\documentclass[a4paper,11pt]{article}
\usepackage{jheppub} 
\usepackage[dvipsnames]{xcolor}
\usepackage{xcolor}
\usepackage{pgfplots}
\pgfplotsset{compat=1.18}
\definecolor{OIorange}{HTML}{E69F00}
\definecolor{OIskyblue}{HTML}{56B4E9}
\definecolor{OIgreen}{HTML}{009E73}
\definecolor{OIyellow}{HTML}{F0E442}
\definecolor{OIblue}{HTML}{0072B2}
\definecolor{OIvermilion}{HTML}{D55E00}
\definecolor{OImagenta}{HTML}{CC79A7}
\definecolor{OIblack}{HTML}{000000}
\usepackage{tikz}
\usepackage{subfigure}
\usepackage{amsmath}
\preprint{IPPP/26/42}
\newcommand{\Fl}{\Phi}
\newcommand{\ktwo}{k_2}
\newcommand{\kthree}{k_3}
\renewcommand{\ktwo}{\phi_2}
\renewcommand{\kthree}{\phi_3}
\newcommand{\Qmin}{Q_{\textrm{min}}}
\newcommand{\mhex}{n_6^m}
\newcommand{\hex}{n_6}
\newcommand{\thedyb}{\theta_{\star}}
\newcommand{\vthedyb}{\vec{\theta}_{\star}}
\newcommand{\thedy}[2]{\theta_{\star}^{(#1),#2}}
\newcommand{\thedyp}[2]{\theta_{\star}^{\prime(#1),#2}}
\newcommand{\latdyB}{L_{\star}}
\newcommand{\vlatdyB}{\vec{L}_{\star}}
\newcommand{\latdy}[2]{L_{\star}^{(#1),#2}}

\newcommand{\latCPp}[1]{L_{\ddagger(#1)}}
\newcommand{\latD}{\Delta L}
\newcommand{\vlatD}{\Delta \vec L}
\newcommand{\theCP}{\theta_{\ddagger}}
\newcommand{\theCPp}[1]{\theta_{\ddagger(#1)}}
\newcommand{\Qem}{Q_{\textrm{em}}}
\newcommand{\Qemin}{Q_{\textrm{em}}^{\textrm{min}}}
\newcommand{\Qq}{Q_{q_L}}

\newcommand{\tHft}{\mathcal{T}}

\title{\boldmath  Dyonic lattices, $\theta$-angles and axions in the Standard Model
}
\author{Rodrigo Alonso, Francesca Chadha-Day, Despoina Dimakou, Yunji Ha, and Valentin V. Khoze}

\medskip
\medskip

\affiliation{Institute for Particle Physics Phenomenology, Department of Physics, Durham University, \newline Durham DH1~3LE, U.K.}

\bigskip

\emailAdd{rodrigo.alonso-de-pablo@durham.ac.uk, 
francesca.chadha-day@durham.ac.uk, despoina.dimakou@durham.ac.uk, yunji.ha@durham.ac.uk, valya.khoze@durham.ac.uk}

\abstract{We investigate the implications of the Witten effect in the Standard Model with a general global gauge group structure and determine the values of the three $\theta$-parameters that lead to distinct families of allowed spectra of dyons. We construct and classify the corresponding dyonic charge lattices consistent with the Standard Model gauge structure. This approach enables us to re-derive the known global-group--dependent periodicities of the $\theta$ angles and to determine all CP-invariant points in $\theta$-space. The electromagnetic subgroup $U(1)_{\mathrm{em}}$ is shown to arise \emph{prior} to electroweak symmetry breaking by factoring out the effect of the anomalous $B+L$ transformations, which reduces the physical $\theta$-parameter space from a three-torus to a two-torus. Our phenomenological conclusions include an observation that a discovery of a $U(1)_{\mathrm{em}}$ monopole carrying non-zero electric charge would determine the last remaining unknown parameter of the Standard Model. Lastly we study how $\theta$-space shapes axion physics with emphasis on the axion-photon coupling and show that a single axion is insufficient to render the Standard Model vacuum fully CP invariant.
}

\keywords{Monopoles, Dyons, $\theta$-angles, the global structure of the Standard Model Group. }

\begin{document}
\maketitle
\tableofcontents
\flushbottom

\section{Introduction}

The Standard Model (SM), defined as the most general renormalisable Lagrangian given the (gauge and Lorentz) symmetry and matter content, is the crowning achievement of particle physics. 
Its well-established theoretical predictions
have withstood the test of experiment, these tests in turn provided by a 
long-standing and active worldwide network of experimental collaborations.
After such sustained scrutiny, it might come as a surprise that there are still unknown fundamental aspects of this theory, yet there are. These harder-to-reach confines are topological and hence beyond perturbation theory, and with experimental signals equally outside the main experimental stream. This is not to say that we are completely in the dark, since some information is known about them both theoretically and experimentally.

The theory sector we are referring to is the Pontryagin density topological terms of the form $F_{\mu\nu}\tilde F^{\mu\nu}$, which are total derivatives and naively not involved in dynamics. The QCD term in particular is the topological sector that has been explored; the development of our understanding of anomalies and the strong interactions led to the $\theta$ vacua of QCD and the strong CP problem as we know it, made a problem by the experimental bounds set on the angle. The two other $\theta$ angles of the SM, associated with the $SU(2)_L$ and $U(1)_Y$ gauge fields, do not lead to such readily experimentally testable predictions, yet they do have physical consequences, obscured in part by an unphysical direction in theory space, as we will explain.
  
  In the exploration of the topology of the Standard Model, new theory tools have proven both useful and regenerative. These are aimed at capturing as much physics as possible in the framework of symmetry and go by the name of generalised symmetries~\cite{Gaiotto:2014kfa}. This line of research has brought to the forefront the connection between topology and the determination of the global gauge group of the SM, as well as new perspectives on confinement, anomalies and applications in condensed matter, see Refs.~\cite{Brennan:2023mmt,Bhardwaj:2023wzd,Bartsch:2022mpm,Bartsch:2022ytj} for reviews.  
  
 The study of said physics will be greatly aided by new physics, and one of the emerging particles most closely related to topology is the axion, in its various incarnations. This paper will use the insight of new and old methods to systematically chart the topological sector, its connection with the gauge group, the electric-magnetic lattice of allowed operators and consequences for the axion.

\section{Dyons, CP and the Witten effect in the Standard Model}
In order to discuss the topological sector of the SM model, a prerequisite is to review the SM model group globally~\cite{Hucks:1990nw,Tong:2017oea}.
There are four distinct groups that share the Lie algebra of the known gauge bosons,
\begin{align}
\label{eq:Gpdef}
    G_p\,=\,\frac{SU(3)_c\times SU(2)_L\times U(1)_Y}{Z_p}\,, \,\qquad\begin{array}{lcccc}
         Z_6:& 1 &\xi&\dots &\xi^5 \,,\\
         Z_3:&  1&\xi^2 &\xi^4\,,&\\
         Z_2:& 1& \xi^3\,,&&\\
         Z_1:& 1\sim \xi^6 &\,, &&
    \end{array}
\end{align}
given by $G_{p=1,2,3,6}$; the true gauge group could be any of them. These possibilities are characterised by the presence or absence, partial or full, of the centre (i.e. the set of elements that commute with every other element). The centre in the SM is $Z_6$ and its generating element $\xi$ is ({\it cf.}~Ref.~\cite{Alonso:2025rkk})
\begin{align}
\label{eq:Q6def}
   \xi=e^{2\pi i Q_6/6}=e^{2\pi i Q_Y/6\Qmin}e^{2\pi in_c/3}e^{\pi i n_L}\,,\qquad
   Q_6\equiv& \frac{Q_Y}{\Qmin}+3\tilde T_{3L}+2\tilde\lambda_{8}\,,
\end{align}
where we have introduced the operator $Q_6$ defined in terms of the hypercharge $Q_Y$
and the diagonal generators $\tilde T_{3L}$ of $SU(2)_L$ and $\tilde\lambda_{8}$ of $SU(3)_c$.
When acting on fields in the fundamental representation, these tilded generators are  
\begin{align}
\label{eq:tildedef}
   \tilde\lambda_8^F\,=\, {\rm diag}(1,1,-2)
   \,,\qquad 
   \tilde T_{3L}^{F}\,=\, {\rm diag}(1,-1)\,,
\end{align}
so that their eigenvalues can be indeed exchanged in Eq.~\eqref{eq:Q6def} for the $n$-alities for weak isospin $n_L$ and colour $n_c$ of a fundamental, $n_L=1,n_c=1$. For higher dimensional representations the same holds true with possible values for $n$-alities $n_L=0,1$, $n_c=0,1,2$. Finally, $\Qmin$ is the minimum possible hypercharge, which may or may not be that of the LH quark doublet $\Qq$. 

\subsection{Outline of our approach}

Our conventions for theta-terms in the SM Lagrangian and for the gauge couplings are
\begin{align}\label{eqs:2.3}
     \mathcal L\supset \frac{\theta_3 g_3^2}{32\pi^2}\,G^a_{\mu\nu}\tilde G^{a\, \mu\nu}+\frac{\theta_2 g^2}{32\pi^2}W^a_{\mu\nu}\tilde W^{a\,\mu\nu}+\frac{\theta_1(g')^2}{16\pi^2}B_{\mu\nu}\tilde B^{\mu\nu}\,. 
\end{align}
All gauge kinetic terms are assumed to be canonically normalised, 
$\mathcal L_{\rm kin} = -\frac{1}{4} G^a_{\mu\nu} G^{a\, \mu\nu}+\ldots$, and the dual field-strengths are 
$\tilde G_{\mu\nu}=\varepsilon_{\mu\nu\rho\sigma}G^{\rho\sigma}/2$, (similarly for $W$ and $B$).
The covariant derivative is also standard,
$D_\mu=\partial_\mu+ig'Q_YB_\mu+ig T_{L a} W^a_\mu+ig_s \lambda_a G^a_\mu$, with the conventionally normalised generators in the fundamental representation,
${\rm Tr}(T_{La}T_{Lb})=\frac{1}{2} \delta_{ab}$ and 
${\rm Tr}(\lambda_{a}\lambda_{b})=\frac{1}{2} \delta_{ab}$. These are related to the tilded generators as $T=\tilde T_L/2$, $\lambda=\tilde\lambda/\sqrt{12}$, as follows from Eq.~\eqref{eq:tildedef}.
Note that this definition allows for different conventions for the hypercharge normalisation, with some authors favouring integer hypercharges and other fractional (a standard pheno-choice is $\Qq=1/6$); irrespective of this choice the ratios of hypercharges of course are not up for debate and are fixed by anomaly cancellation, so it suffices to refer to the LH quark doublet charge as $\Qq$ as done in Tab.~\ref{Tab:SMn6}
\begin{table}[h]
    \centering
\begin{tabular}{c|c|c|c|c|c|c}
     & $q_L$& $u_R$& $d_R$ & $\ell_L$ & $e_R$ & $H$\\ \hline
    $SU(3)_c$ & $\boldsymbol3$ & $\boldsymbol3$ & $\boldsymbol3$ & $\boldsymbol1$ &$\boldsymbol1$& $\boldsymbol1$\\
    $SU(2)_L$ &$\boldsymbol2$ & $\boldsymbol1$ & $\boldsymbol1$& $\boldsymbol2$ & $\boldsymbol1$&$\boldsymbol2$\\
    $ Q_Y$ & $\Qq$ & $4\Qq$ &$-2\Qq$ &$-3\Qq$ &$-6\Qq$&$3\Qq$ 
    \\\hline
    $n_6$ & 0 & 0 & 0 & 0 & 0 & 0
\end{tabular}
\caption{All known SM fields have $n_6=0$ as follows from their hypercharge, non-Abelian representation assignments and the definition~\eqref{eq:n6def} of 
$n_6\,=\, Q_6\mod 6 \,=\, Q_Y/Q_{\rm min} +2n_c+3n_L \mod 6$. }
    \label{Tab:SMn6}
\end{table}

A pervasive assumption about hypercharge that we will drop is that the LH quark doublet hypercharge is the smallest in Nature. This is a physical assumption and we can relax it by allowing for another, smallest, hypercharge $\Qmin$, of a new possible matter field, whose value depends on the group $G_p$ as
\begin{align}
\frac{\Qmin}{\Qq}=\frac{1}{1+p\cdot k}\,,
\end{align}  
with $k$ the degree of compositeness or `staircase' number introduced in~\cite{Alonso:2024pmq} with $k\in\mathbb{Z}$.

It is useful to write the four groups $G_{1,..,6}$ respectively as
\begin{align}
    &SU(3)_c\times SU(2)_L\times U(1)_Y\,, & &SU(3)_c\times U(2)_{L+Y}\,, &&U(3)_{c+Y}\times SU(2)_L\,, & &S\left[U(3)\times U(2)\right]\,,
\end{align}
which highlights how certain sectors in a given group are disconnected or not from another.

So far no use of generalised symmetries has been required. Indeed one can treat the problem without these methods to a large extent, yet they are useful to understand and describe the physics. In particular, one-form symmetries are specific examples of generalised symmetries we will employ. In order to see that one-form is the appropriate symmetry to characterise different groups, one can point at monopoles as the non-perturbative configurations that know about the group globally (given that they require a `large' gauge transformation). The description and classification of monopoles is based on their magnetic \textit{fluxes} -- topologically conserved quantities that are surface integrals. This is different from volume integrals used for commonplace zero-form symmetries, and so one needs a different type of symmetry, a one-form symmetry. This different dimensionality translates at the quantum level to different dimensionality of the operators that the symmetry acts on; zero-forms act on local operators $\mathcal O(x)$ defined at a spacetime point, whereas one-forms act on line-operators defined  
on a line (for example the worldline of a magnetic monopole in spacetime). In general, there are electric and magnetic one-form symmetries. Electric act on Wilson lines ($\mathcal W$), and magnetic act on 't~Hooft lines. Lastly, we note the connection with the global group: a non-trivial quotient in~\eqref{eq:Gpdef} removes the `denominator' from the allowed group space, and with it its associated one-form discrete electric symmetry, but in its place a magnetic one-form $Z^{\textrm{mag},[1]}_{p}$ arises acting on new monopole solutions.  For the Standard Model case with the gauge group $G_p$, there are two one-form discrete symmetries:
\begin{align}
G_p\,:\qquad Z_{6/p}^{\,\textrm{el},[1]}\,,\quad Z_{p}^{\,\textrm{mag},[1]}\,.
\end{align}

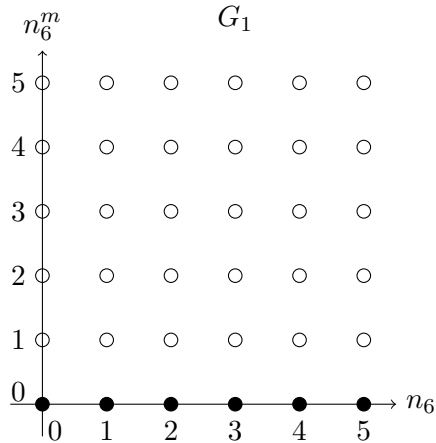
\begin{figure}[h!]
    \centering
        \begin{tikzpicture}[scale=0.85]
  \draw[->] (-0.5,0) -- (5.5,0) node[right] {$n_6$};
  \draw[->] (0,-0.5) -- (0,5.5) node[above] {$n^m_6$};
  \node[left] at (-0.1,0.2) {$0$};
  \node[left] at (-0.1,1) {$1$};
  \node[left] at (-0.1,2) {$2$};
  \node[left] at (-0.1,3) {$3$};
  \node[left] at (-0.1,4) {$4$};
  \node[left] at (-0.1,5) {$5$};
  \foreach \x in {1,...,5}
    \node[below] at (\x,-0.1) {\x};
  \node[below] at (0.2,-0.1) {0};
  \foreach \x in {0,...,5} {
    \foreach \y in {0,...,5} {
      \draw (\x,\y) circle (3pt);
    }
  }
  \foreach \x in {0,...,5}
    \fill (\x,0) circle (3pt);

    \node at (3,6) {$G_1$};
\end{tikzpicture}
    \caption{The lattice looks the same for all values of the $\theta$ angles for the group $G_1$.}
    \label{fig:lattG1}
\end{figure}

A Wilson line is characterised by its electric charge, i.e. gauge group representation, while a '\textit{}t~Hooft line is given by its magnetic fluxes $\Fl$. These two are not independent of one another, but are constrained by  Dirac's quantisation condition, 
which in the SM takes the form,
\begin{align}
    e^{i\Fl_1 Q_Y+i\Fl_2 \tilde T_{3L}+i\Fl_3\tilde \lambda_{8}}=\left[\xi\right]^{n_6^m}\,,\label{eq:DQcond1}
\end{align}
where $n_6^m$ is defined~\cite{Alonso:2025rkk} as an integer mod 6 and can only take values that return elements of the group one takes the quotient of, explicitly
\begin{align}
G_1\,:\,\, n_6^m=0\,, \quad
G_2\,:\,\, n_6^m=0,3\,, \quad
G_3\,:\,\, n_6^m=0,2,4\,, \quad
G_6\,:\,\, n_6^m=0,1,\ldots,5\,.
\end{align}
Indeed, the RHS of the Eq.~\eqref{eq:DQcond1} is the identity, only there are new ways to write the identity once one takes a $Z_p$ quotient in $G_p$. In this way, the connection between the group and the magnetic spectrum is captured by $\mhex$ and there is in turn a correlated connection between $\hex$ in the electric sector defined as
\begin{align}
\label{eq:n6def}
\hex\,\equiv\, q_6\mod 6\,, \qquad Q_6|q_6\rangle&=q_6|q_6\rangle\,,
\end{align}
and the group $G_p$. The possible values for $\hex$ are also different for different $p$ and given by the condition on the spectrum that the quotient acts trivially on the allowed representations. The number of allowed values for $\hex$ is inversely correlated with that of $\mhex$ and this is diagrammatically captured in electric-magnetic or weight lattices,  first presented in the $\hex-\mhex$ plane in Ref.~\cite{Alonso:2025rkk} and here reproduced in the first graphs of each of Figs.~\ref{fig:lattG1},\ref{fig:lattG2},\ref{fig:lattG3},\ref{fig:lattG6}.

\medskip

The next ingredient we need to introduce is the Witten effect \cite{Witten:1979ey}.
For a $U(1)$ theory with non-vanishing $\theta$-parameter, magnetic monopoles acquire an electric charge proportional to $-\theta/2\pi$ (in units of $e$) (see Appendix~\ref{App:WU1} for an overview). 

In the case of the Standard Model,
magnetic monopole fluxes (now labelled by $\mhex$) also source an electric field for $\theta\neq 0$ whose integral defines an electric $n$-ality $\hex$ (in place of the $U(1)$ electric charge) and along with it an effective transformation rule under electric symmetry with $\hex\propto\theta\,\mhex$. Such a transformation is not, in general, that of a true electric representation, just like in the Abelian case, an effective electric charge $-\theta /2\pi$ is not an integer multiple of the fundamental charge. We should note that this effective electric charge does not open the door to extra solutions but merely tilts the electric-magnetic lattice of physical states present in the theory.
For the reader's convenience, we review applications of the Witten effect to $SU(N)$, $SU(N)/Z_p$ and $U(N)$ theories in Appendices~\ref{App:WSUN}, ~\ref{sec:UN}.

In general, for non-vanishing $\theta$-angles, one can choose between two ways of labelling states in the dyonic lattice of the theory. One option is to label states using their effective electric and magnetic charges. These effective charges are physical observables, but as we have already noted, the effective electric charge of $-\theta /2\pi$ is not integer-valued. Alternatively, one can continue labelling the dyonic states by their integer-valued electric and magnetic charges, or more precisely, in the case of the Standard Model, by their electric $n$-ality $\hex$ and the magnetic flux number $\mhex$. Being integer-valued for any $\theta$, the $n$-ality $\hex$ is not a physically measurable analogue of the electric charge, but it can be associated with a representation of the (electric) gauge group $G_p$. 
In general, the pair $(\hex,\mhex)$ assigns well-defined electric and magnetic representations to all dyonic states independently of any of the external parameters such as coupling constants and $\theta$-angles.
The resulting dyonic lattices have been well-explored in the literature for a range of individual non-Abelian groups in the pioneering papers on line operators and generalised symmetries~\cite{Aharony:2013hda,Gaiotto:2014kfa}.
However, to the best of our knowledge, no detailed investigation of dyonic lattices (or, equivalently, of the admissible spectra of all line operators for $\theta\neq 0$ has so far been carried out for the Standard Model. Addressing this question is one of the main goals of the present paper. 

In particular, we will specify and rely upon three key characteristics of the $\theta$-spaces of different $G_p$ incarnations of the Standard Model: 
\begin{enumerate}
\item $\theta$-periodicities $\Delta\vec{\theta}^{\,(p)}$
\item integer-valued-$n_6$ points $\vthedyb^{\,\,(p)}$
\item CP-invariant points $\vec{\theCP}^{(p)}$
\end{enumerate}
where the superscript${}^{\,(p)}$ refers to each particular SM gauge group $G_p$.

For the task of surveying $\theta$-space, the Witten effect will provide the full $\theta$-periodicities, i.e. the equivalence relations between different $\theta$ points that define the $\theta$-manifold. In our notation, these $\theta$-periods or $\theta$-cycles are,
\begin{align}
\label{eq:Dtheta-p}
  \Delta\vec{\theta}^{\,(p)}\,:\qquad   \vec\theta\sim\vec\theta+ \Delta\vec{\theta}^{\,(p)}\,,
\end{align}
where there are as many $\Delta\vec{\theta}^{(p)}$ vectors as the dimensionality of $\vec\theta$ and will be labelled by roman numerals when having to distinguish between them, $\Delta\vec{\theta}^{\,(p)}_I,\Delta\vec{\theta}^{\,(p)}_{II}$ etc. Topologically, this space will always be a hypertorus, but the periodicities do depend on the group through $p$ and this has physical consequences. 

Exploring $\theta$ values starting from the known $\vec\theta=0$ case and varying $\theta$-parameters continuously will allow us to identify non-trivial special values of $\vec\theta=\vthedyb$ which correspond to new integer values of $(\hex,\mhex)$ that one might otherwise miss\footnote{One should note that $\theta$ does span the whole range of consistent electric-magnetic assignments for $SU(N)$ gauge theories. This, however, is no longer the case for other models, such as $SO(N)$ with $N\geq 5$, as was shown in~\cite{Aharony:2013hda}.}
We thus define the set of special points $\vthedyb \in$ $\theta$-space,  
\begin{equation}
\label{eq:thetastar-p}
 \vthedyb^{\,\,(p)}\,:\qquad   n_6(\vthedyb^{\,\,(p)})\equiv\mathbb{Z}.
\end{equation}

Since all $F \tilde F$ operators are CP-odd it follows from the invariance of the action that $\vec\theta$ is also odd under CP, so a generic point violates this discrete symmetry. Other than the trivial $\theta=0$ point, the rest of CP invariant points can be found by requiring
\begin{align}
\label{eq:CPinvthe}
    {\vec{\theCP}}^{(p)}\equiv-\vec{\theCP}^{(p)} \mod  \Delta\vec\theta^{\,(p)}\,.
\end{align}
Once a fundamental domain with its periodicity is obtained, solutions are halfway to the origin along the periodicity vectors and sums of them so that if our $\theta$ space is $n$-dimensional there are $2^n$ solutions in total. 

In the following sections, for each of the $G_p$ groups we will determine the $\theta$-space periods $\Delta\vec{\theta}^{\,(p)}$, as well as the two special sets of points: $\vthedyb^{\,\,(p)}$ and ${\vec{\theCP}}^{(p)}$. Each of these special sets of points serves different purposes. When the value of a continuous $\vec\theta$ parameter coincides with one of the points from the set 
$\vthedyb^{\,\,(p)}$, the corresponding spectrum of the theory is described by a dyonic lattice with integer-valued $(\hex,\mhex)$, but is not guaranteed to be CP-invariant. On the other hand, at the points ${\vec{\theCP}}^{(p)}$, the dyonic lattice gives a CP-invariant spectrum, but can have, as we will see below, half-integer values of $\hex$. Finally, depending on the choice of $G_p$, some of the points from these two sets may coincide; for these coincident points, the dyonic lattice will be both, integer-valued and CP-invariant.

\medskip

In what follows, for each of these special $\theta$-point sets we will reconstruct the spectrum of quantum numbers $\hex$ and $\mhex$ and present the corresponding dyonic lattices on $(\hex,\mhex)$-plane. They are in one-to-one correlation with the group $G_p$ and so are best suited to tell the SM group apart, but they do not capture all the topology of the SM. In particular, the $U(1)$  sector has a conserved integer-valued magnetic charge and for certain groups the triality or $n$-ality are disconnected from $n_6$. When necessary for the study of $\theta$-space, we will inspect these other quantum numbers. 

\subsection{The pure-gauge Standard Model}

Let us study $\theta$ space first neglecting the effect of fermions. The consequences of including fermions are quite dramatic and will be worked out in the next section, yet the effect is a projection into a lower dimensional $\theta$ space so that our starting point loses no information. Our set up is fully general; in order to separate different aspects, the following notation is useful
\begin{align}
    \vec\theta=2\pi\left(p\Qmin^2L_1,L_2,L_3\right)\,,
\end{align}
so that the choice of minimum $U(1)$ charge is factored out and the $L$ three-dimensional torus captures the group $G_p$ dependence.

\subsubsection{The $\mathbf{G_1}$ group $\mathbf{SU(3)_c\times SU(2)_L\times U(1)_Y}$}

We examine first the case of the universal cover $G_1$. While there is no Witten effect in $\hex-\mhex$, there are effects in the extended lattice of magnetic fluxes and electric charges as we will show. The Dirac quantisation condition of Eq.~\eqref{eq:DQcond1} reads
\begin{align}
    e^{i\Fl_1 Q_Y+i\Fl_2 \tilde T_{3L}+i\Fl_3\tilde \lambda_8}=\left[\xi\right]^{\mhex}=1
    \,, \qquad\mhex=0\,,
    \label{eq:FlG1}
\end{align}
given the full centre is present in $G_1$ and $\xi$, ..., $\xi^5$ are group elements distinct from the identity.
 Possible fluxes compatible with this equation are
\begin{align}
\label{eq:fluxesG6}
   \vec{\Fl}\,=\,\left(\frac{2\pi q_m}{6\Qmin}\,,\,\frac{2\pi \ktwo}{2}\,,\,\frac{2\pi \kthree}{3}\right)\,,  \qquad\left(q_m\in 6\mathbb{Z}\,,\,\ktwo\in2\mathbb{Z}\,,\,\kthree\in3\mathbb{Z}\right)\,,
\end{align}
where here and throughout,
\begin{align}
\label{eq:qm_mod_6}
n_6^m=q_m \mod 6\,.
\end{align}
 As alluded to, there is no Witten effect in $\hex,\mhex$ since there is no magnetic one-form symmetry in $G_1$ that might give monopoles an electric charge. 
 Indeed, for the monopole flux in~\eqref{eq:fluxesG6} let us constrain $q_m$ to lie in the interval $0\leq q_m<6$ in which case $q_m=n_6^m$ and the Witten effect induces the effective hypercharge on the monopole ({\it cf.}~Eq.~\eqref{eq:appWit} of Appendix~\ref{App:WU1}),
 \begin{align}
  0\leq q_m&<6\,, \qquad   Q_Y \,=\, -\, \frac{n_6^m}{6 Q_{\rm min}}\,\, \frac{\theta_1}{2\pi}.
 \end{align}
 Since $\mhex=0$ in the $G_6$ theory, this expression is trivially vanishing for any value of $\theta_1$. Nevertheless, let us examine the effect of an innocuous modding of $\mhex$ by units of 6 on the identity 
 \eqref{eq:FlG1} written as $\xi^6$ in the presence of $\theta_1 \neq 0$. We act with $\xi^6=1$, in the form~\eqref{eq:FlG1}, on a monopole solution $\tHft_{\mhex}$ and write,
\begin{align}
\label{eq:xi6T}
    \xi^6\tHft_{\mhex}=\exp\left(\frac{2\pi i Q_Y}{\Qmin}\right)\tHft_{\mhex}=\exp\left(-i\,\frac{n_6^m\theta_1}{6\Qmin^2}\right)\tHft_{\mhex}=\exp\left(-i\,\frac{\theta_1(q_m\mod 6)}{6 \Qmin^2}\right)\tHft_{\mhex}=\tHft_{\mhex}\,.
\end{align}
The outcome is still an identity, with the intermediate mod factor that assigns a unique label to the set of monopoles. However, whereas substituting $n_6^m\to q_m$ is innocuous in Eq.~\eqref{eq:FlG1}, here it does not yield the same identity but extra powers of $e^{-i\theta_1/\Qmin^2}$. This factor is the Abelian $U(1)_Y$ electric phase that the $q_m=6$ monopole would pick via the Witten effect, with other monopoles picking integer powers of this unit. For $\theta=0$, all monopoles in a given $\mhex$ class have the same, if trivial, electric transformation. 
Demanding that this electric transformation is well-defined (i.e. that the last equality in~\eqref{eq:xi6T} holds without the mod) for all elements in each $\mhex$ class, we find specific values of
the $\theta_1$-angle for which this is the case,  
\begin{equation}(\thedyb)_1\,=\,2\pi\, \Qmin^2\mathbb{Z}
\end{equation}
in addition to the trivial value $\theta_1=0$. All these values of $\theta_1$ are 
degenerate in the sense that they result in identical spectra, i.e. produce an electric-magnetic lattice identical to that of $\theta_1=0$. Hence $\theta_1$ is periodic with the period 
$2\pi\, \Qmin^2$.
This periodicity condition can also be derived from demanding a change of charge equal to the smallest possible quantum, $\Delta Q_Y=-\theta_{1}/(2\pi\Qmin)=\Qmin$, see~\eqref{eq:A7}-\eqref{eq:A8} in Appendix~\ref{App:WU1}.

One can similarly extract the periodicity in the $\theta_2$ and $\theta_3$ angles by inspecting $n_L$ and $n_c$ 
\begin{align}
    e^{2\pi in_L/2}\tHft_{\mhex}&=\exp\left(\frac{i\theta_2}{2}(\ktwo\mod 2)\right)\tHft_{\mhex}=\tHft_{\mhex}\,,\\
    e^{2\pi in_c/3}\tHft_{\mhex}&=\exp\left(\frac{i\theta_3}{3}(\kthree\mod 3)\right)\tHft_{\mhex}=\tHft_{\mhex}\,,
\end{align}
where we have taken the result for the $\theta$-shifts of the n-alitites, $n=\theta N (\Phi\mod 2\pi)/(2\pi)^2$ for $SU(N)$ derived in Appendix~\ref{App:WSUN}, Eqs.~\eqref{eq:A.25},\eqref{eq:Flqm} and applied it to the fluxes of Eq.~\eqref{eq:fluxesG6}.
Again, considering the extended lattice of magnetic fluxes one finds the exponential does a consistent mapping only for $\theta_2,\theta_3=2\pi\mathbb{Z}$, and the period of each angle is given by a change of $n$-ality for the first non-trivial magnetic monopole as $\Delta n_c=3\mathbb{Z}$ and $\Delta n_L=2\mathbb{Z}$. 

To summarise our discussion up to this point: for the gauge group $G_1$ we have reproduced the expected result for the periods of the three $\theta$ variables and the cycles of the 3-torus, which in our notation read:
\begin{align}
      G_1:& &\left\{\Delta \vec \theta^{\,(1)}\right\}&=\left\{\begin{array}{lcc}
        (\,2\pi\Qmin^2\,, & 0\,, & 0\,)\\
        (\, 0\,,& 2\pi\,, &0\,)\\
        (\, 0\,,& 0\,, &2\pi\,)
    \end{array}\right\}\,,&
    \left\{\Delta \vec L^{\,(1)}\right\}&=\left\{\begin{array}{lcc}
        (\,1\,, & 0\,, & 0\,)\\
        (\, 0\,,& 1\,, &0\,)\\
        (\, 0\,,& 0\,, &1\,)
    \end{array}\right\}\,.\label{eq:PerG1}
\end{align}

\noindent With the fundamental periods one can now identify the physical $\theta$ space where different values lead to different theories. Selecting this domain is somewhat arbitrary, some people favour $0\leq\theta_3<2\pi$, others $-\pi\leq \theta_3<\pi$, with the choice not making any physical difference. For concreteness here we choose
\begin{align}
\label{eq:thetarG1}
    0\,\leq&\,\theta_1<2\pi \Qmin^2\,, & 
    0\leq&\,\theta_2<2\pi\,, &
    0\leq&\,\theta_3<2\pi\,, \\
\label{eq:LrG1}    
    0\leq&\,L_1<1\,, & 
    0\leq&\,L_2<1\,, &
    0\leq&\,L_3<1 \,.
\end{align}
Within this physical range, we can now search for special CP-invariant points.
 Using either the criterium in Eq.~\eqref{eq:CPinvthe} or the equivalent mirror-reflected-symmetric lattice, leads to $\theta_{2,3}=0,\pi$ and $\theta_1=0,\pi \Qmin^2$. 
 
To present the CP-invariant points within the physical domain~\eqref{eq:thetarG1}-\eqref{eq:LrG1}, it  will be convenient to introduce a universal notation applicable to all gauge groups $G_p$ as follows,
\begin{align}
\label{eq:thetaijk-1}
\theCPp{p}^{(i,j,k)}&\,=\, i\theCPp{p}^{(1,0,0)}+j\theCPp{p}^{(0,1,0)}+k\theCPp{p}^{(0,0,1)}\,,& 0\leq&i\,,j\,,k\leq 1\,,
\end{align}
with $\theCPp{p}^{(0,0,0)}=0$ for all $p$. 

\medskip

In the $G_1$ case we have,
\begin{align}
\theCPp{1}^{(1,0,0)}=&(\pi\Qmin^2,0,0)\,, &\theCPp{1}^{(0,1,0)}=&(0,\pi,0) \,,& \theCPp{1}^{(0,0,1)}=&(0,0,\pi) \,.\end{align}
To condense this information, we will use the schematic form
\begin{align}
    \latCPp{1}=&\left\{(1/2,0,0)\oplus(0,1/2,0)\oplus(0,0,1/2)\right\}\,,\\
    \theCPp{1}=&\left\{(\pi\Qmin^2,0,0)\oplus(0,\pi,0)\oplus(0,0,\pi)\right\}\,.
\end{align}

\medskip

\subsubsection{The $\mathbf{G_2}$ group $\mathbf{SU(3)_c\times U(2)_{L+Y}}$ }

The group $G_{2}$ contains a disconnected $SU(3)_c$ sector with $\theta_3$-periodicity and properties identical to $G_1$. Fluxes in the electroweak sector, however, present new solutions. To find these we again consider the Dirac quantisation condition for the $G_3$-compatible values of $\mhex=0,3$,
\begin{align}
     &e^{i\Fl_1 Q_Y+i\Fl_2 \tilde T_{3L}+i\Fl_3\tilde \lambda_8}=\left\{\begin{array}{c}
          1  \\
          \xi^3 
     \end{array}\right\}=\left[\xi^3\right]^{n^{m}_6/3}=\left[e^{\pi i \,(Q_Y/Q_{\textrm{min}}-n_L)}\right]^{n^{m}_6/3}\,, & &\mhex=0,3\,,
\end{align}
where in the last equality we have used the expression~\eqref{eq:Q6def} for $\xi$ to write $\xi^3$ in the form,
\begin{equation}
\xi^3\,=\, e^{2\pi i Q_Y/(2\Qmin)}\,e^{3\pi i n_L}\,e^{2\pi in_c}\,=\,
e^{\pi i Q_Y/\Qmin}\,e^{-\pi i n_L}\,.
\end{equation}

\noindent Monopole solutions in $G_2$ theory are characterised by magnetic flux~\cite{Alonso:2025rkk},
\begin{align}
    \vec\Fl\,=\,\left(\frac{2\pi q_m}{6\Qmin}\,,\,-\,\frac{2\pi (q_m/3+\ktwo)}{2}\,, \frac{2\pi \kthree}{3}\right), \qquad \left(q_m\in 3\mathbb{Z}\,,\,\ktwo\in 2\mathbb{Z}\,,\,\kthree\in3\mathbb{Z}\right)\,.\label{eq:FlG2}
\end{align}
There is now a magnetic $Z_2^{\textrm{mag},[1]}$ one-form symmetry and hence a non-trivial Witten effect in the $(\hex,\mhex)$ plane. This effect induces $\theta$-dependent (electric) charges on the monopole, and shows itself when we let the quotiented group act on the 't~Hooft lines
\begin{align}\nonumber
    \xi^3\tHft_{\mhex}\,=&\,\exp\left(\pi i \left(\frac{Q_Y}{\Qmin}-n_L\right)\right)\tHft_{\mhex}\,\\
    =&\,\exp\left(i\theta_2\,\frac{(q_m/3 \mod 2)}{2}\,-\,\frac{i\theta_1}{2\Qmin^2}\,\frac{(q_m/3 +\ktwo\mod 2)}{2}\right)\tHft_{\mhex}\label{eq:ElG2T-2}    \\
    =&\,\exp\left(i\frac{n_6^m}{6}\left[\theta_2-\frac{\theta_1}{2\Qmin^2}\right]\right)\tHft_{\mhex}\,,\label{eq:ElG2T}
\end{align}
Using this relation, we can define the parity of the $Z_2$ group
\begin{align}
    \xi^3\tHft_{\mhex}\,=\, e^{2\pi i n_2/2}\tHft_{\mhex}\,,\qquad n_2\,=\,\frac{n_6^m}{3(2\pi)}\left[\theta_2-\frac{\theta_1}{2\Qmin^2}\right]=\hex\mod 2\,.
\end{align}

To find the periodicity of theta-angles for the $G_2$ theory we are discussing here, we need to impose two conditions, which are accidentally the same as in the case of $G_1$ in the previous section. The first condition is that the exponential mapping of Eq.~\eqref{eq:ElG2T} returns the same for $\mhex$ substituted by $q_m,\ktwo$ as in Eq.~\eqref{eq:ElG2T-2} without mod terms, and analogously for $n_c$ in the QCD sector, 
\begin{align}\label{eq:WelG2}
&e^{i(\thedyb)_2}=e^{i(\thedyb)_1/2\Qmin^2}=e^{i(\thedyb)_3}=1\,, 
& \vthedyb&=2\pi (2\Qmin^2(\latdyB)_1, (\latdyB)_2,(\latdyB)_3)\,,\qquad \vlatdyB \in\vec{\mathbb{Z}}\,,
\end{align}
which is a condition shared with dyonic lattices. The second condition is a variation of $\Delta n_2=0 \mod 2$ for all magnetic lines; for this it suffices to impose it for the smallest non-trivial magnetic flux
\begin{align}
\label{eq:n6min-2}
   (n_2)_{\min}= \frac{(n_6^m)_{\min}}{3}\,
   \frac{1}{2\pi}\left[\Delta \theta_2-\frac{\Delta\theta_1}{2\Qmin^2}\right]=
   \frac{1}{2\pi}\left[\Delta\theta_2-\frac{\Delta\theta_1}{2\Qmin^2}\right]=0 \mod 2\,.
\end{align}
Substituting the solution of Eq.~\eqref{eq:WelG2} this reads
\begin{align}
\label{eq:DL-even}
    (\latD^{(2)})_1-(\latD^{(2)})_2=0\mod 2\,.
\end{align}
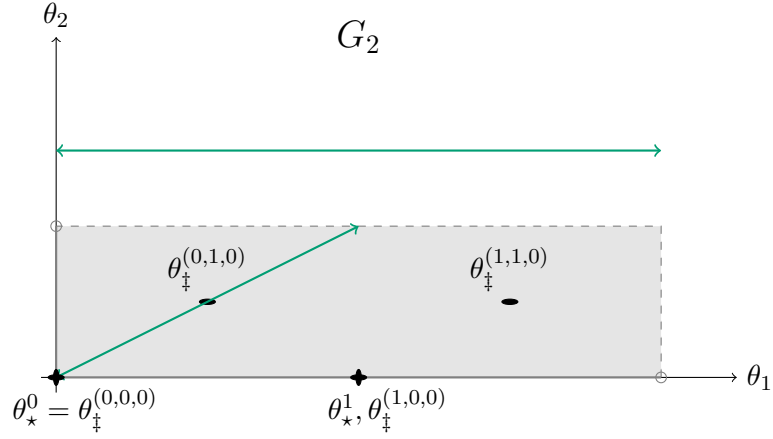
\begin{figure}[h!]
    \centering
\begin{tikzpicture}
        \draw [->] (0,-0.2)--(0,4.5) node[above] {$\theta_2$};
        \draw [->] (-0.2,0)--(9,0) node[right] {$\theta_1$};
             \filldraw [gray,dashed,opacity=0.2] (0,0) -- (8,0) -- (8,2) -- (0,2) -- (0,0);
          \draw [gray,dashed] (8,0) -- (8,2) -- (0,2);
          \draw [gray,thick] (0,2) -- (0,0) -- (8,0);
       \filldraw (4,0) ellipse (3pt and 1pt);
      \filldraw(4,0) ellipse (1pt and 3pt)  node[below] {$\qquad\thedyb^{1},\theCP^{(1,0,0)}$}--(4,0.1);
        \filldraw (2,1) ellipse (3pt and 1pt) node[above] {$\theCP^{(0,1,0)}$};
        \filldraw (6,1) ellipse (3pt and 1pt) node[above] {$\theCP^{(1,1,0)}$};
        \draw [OIgreen, <->,thick] (0,0) -- (4,2);
        \draw [OIgreen, <->,thick] (0,3) -- (8,3);
         \draw[gray] (8,0) circle (2pt);
         \draw[gray] (0,2) circle (2pt);
         \node at (4,4.5) {\Large $G_2$};
\filldraw (0,0) ellipse (3pt and 1pt);
\filldraw (0,0) ellipse (1pt and 3pt) node[below]
{$\qquad\thedyb^{0}=\theCP^{(0,0,0)}$};
    \end{tikzpicture}
    \caption{$\theta_3=0$ section of the $\theta$-space of the $G_2$ gauge theory.
    Fundamental domain on the $(\theta_1,\theta_2)$ plane is the gray rectangle of dimensions  ($2\pi (2\Qmin)^2$\,,\,$2\pi$) including the lower and left sides and origin but not the rest of the boundary (i.e. not the other two sides and three vertices). The equivalence relation that defines the two torus is given by the green double-sided arrows. CP-invariant points $\theCP^{(i,j,k)}$ are marked with a horizontal ellipse. We also show the point $\thedyb^{0}=0=\theCP^{(0,0,0)}$ and the point $\thedyb^{1}=\theCP^{(1,0,0)}$ for which $n_6$ is shifted by 1 unit. Such points $\thedyb$ corresponding to integer shifts in $n_6$ are indicated by vertical ellipses.}
    \label{fig:G2domain}
\end{figure}
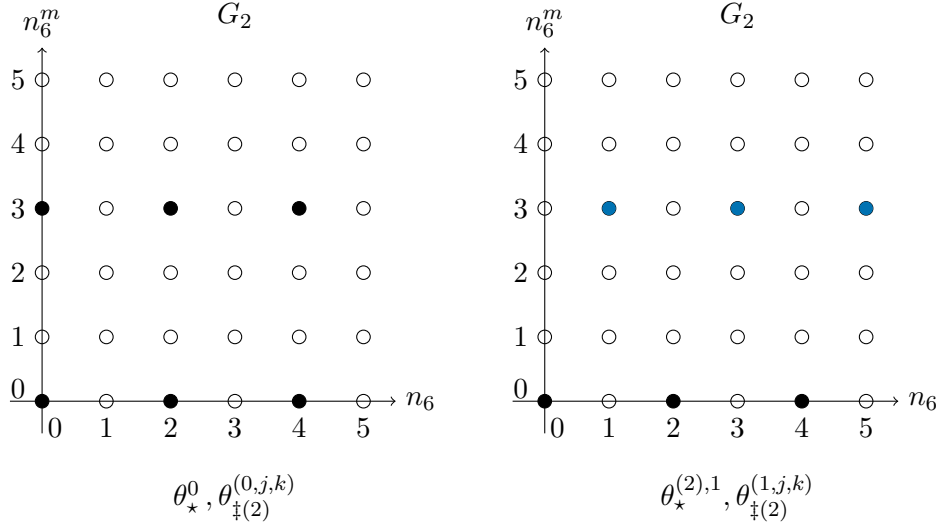
\begin{figure}
    \centering\begin{tikzpicture}[scale=0.85]
  \draw[->] (-0.5,0) -- (5.5,0) node[right] {$n_6$};
  \draw[->] (0,-0.5) -- (0,5.5) node[above] {$n^m_6$};
  \node[left] at (-0.1,0.2) {$0$};
  \node[left] at (-0.1,1) {$1$};
  \node[left] at (-0.1,2) {$2$};
  \node[left] at (-0.1,3) {$3$};
  \node[left] at (-0.1,4) {$4$};
  \node[left] at (-0.1,5) {$5$};
  \foreach \x in {1,...,5}
    \node[below] at (\x,-0.1) {\x};
  \node[below] at (0.2,-0.1) {0};
  \foreach \x in {0,...,5} {
    \foreach \y in {0,...,5} {
      \draw (\x,\y) circle (3pt);
    }
  }
  \fill (0,0) circle (3pt);
  \fill (2,0) circle (3pt);
  \fill (4,0) circle (3pt);
\fill (0,3) circle (3pt);
  \fill (2,3) circle (3pt);
  \fill (4,3) circle (3pt);
     

    \node at (3,6) {$G_2$};
  \node at (3,-1.5) {$\thedyb^{0}\,,\theCPp{2}^{(0,j,k)}$};
\end{tikzpicture}\qquad
\begin{tikzpicture}[scale=0.85]
  \draw[->] (-0.5,0) -- (5.5,0) node[right] {$n_6$};
  \draw[->] (0,-0.5) -- (0,5.5) node[above] {$n^m_6$};
  \node[left] at (-0.1,0.2) {$0$};
  \node[left] at (-0.1,1) {$1$};
  \node[left] at (-0.1,2) {$2$};
  \node[left] at (-0.1,3) {$3$};
  \node[left] at (-0.1,4) {$4$};
  \node[left] at (-0.1,5) {$5$};
  \foreach \x in {1,...,5}
    \node[below] at (\x,-0.1) {\x};
  \node[below] at (0.2,-0.1) {0};
  \foreach \x in {0,...,5} {
    \foreach \y in {0,...,5} {
      \draw (\x,\y) circle (3pt);
    }
  }
  \fill (0,0) circle (3pt);
  \fill (2,0) circle (3pt);
  \fill (4,0) circle (3pt);
\fill[OIblue] (1,3) circle (3pt);
  \fill[OIblue] (3,3) circle (3pt);
  \fill[OIblue] (5,3) circle (3pt);
  \node at (3,6) {$G_2$};
    \node at (3,-1.5) {$\thedy{2}{1},\theCPp{2}^{(1,j,k)}$};
\end{tikzpicture}
    \caption{Two distinct dyonic lattices in $G_2$ theory. LHS: Electric-magnetic lattice for $\thedyb^{0}=0$ and the CP invariant points $\theCPp{2}^{(0,j,k)}$. RHS:  Spectral lattice for dyonic states in the models with $\theta$-parameters fixed at $\thedy{2}{1}$ and $\theCPp{2}^{(1,j,k)}$. Both dyonic lattices give CP-invariant spectra and have integer-valued indices $(n_6,n_6^m)$.}
    \label{fig:lattG2}
\end{figure}
The two conditions combined return the periodicity structure 
({\it cf.}~Eq.~\eqref{eq:Dtheta-p}),
\begin{align}
         G_2:& & \left\{\Delta\vec \theta^{\,(2)}\right\}&=\left\{\begin{array}{lcc}
        (\,4\pi\Qmin^2\,, & 2\pi\,, & 0\,)\\
        (\, 8\pi\Qmin^2\,,& 0\,, &0\,)\\
        (\, 0\,,& 0\,, &2\pi\,)
    \end{array}\right\}\,,& \left\{ \vlatD^{(2)} \right\}&=\left\{\begin{array}{lcc}
        (\,1\,, & 1\,, & 0\,)\\
        (\, 2\,,& 0\,, &0\,)\\
        (\, 0\,,& 0\,, &1\,)
    \end{array}\right\}\,.\label{eq:PerG2}
\end{align}
For concreteness, we choose the fundamental domain consistent with these conditions as follows
\begin{align}
\label{eq:theta-per}
    0\leq&\,\theta_1<8\pi \Qmin^2\,, & 
    0\leq&\,\theta_2<2\pi\,, &
    0\leq&\,\theta_3<2\pi\,, \\
\label{eq:L-per}    
        0\leq&\,L_1<2\,, & 0\leq&\,L_2<1\,, & 0\leq&\,L_3<1\,.
\end{align}
In Fig.~\ref{fig:G2domain} we show this fundamental domain of the $\theta$-space of the $G_2$ theory on the $(\theta_1,\theta_2)$-plane for the $\theta_3=0$ section. 
The figure also shows non-trivial $\theta$-points which result in CP-invariant theories with mass spectra characterised by integer-valued $(n_6,n_6^m)$. We will soon see that there are two such dyonic lattices in the $G_2$ theory; they are shown in Fig.~\ref{fig:lattG2}.

We shall now first pinpoint locations of special points $\thedyb^{(2)}$ 
(as in~Eq.~\eqref{eq:thetastar-p}) for which the mass-spectrum is characterised by integer-valued $(n_6,n_6^m)$  in our $G_2$ model, and then we will address CP-invariance following Eq.~\eqref{eq:CPinvthe}. Notably, the points in $\theta$-space that satisfy the first equation in Eq.~\eqref{eq:WelG2} and lead to integer-valued $n_6$ charges, extend beyond just those that live at the ends of the periodicity intervals in~\eqref{eq:theta-per}. In particular, while the difference $L_1-L_2$ between the interval ends in~\eqref{eq:L-per}  is always an integer it does not have to be an even number (as it was in~\eqref{eq:DL-even}), it could be odd as well in our case,
\begin{align}
    (\latdy{2}{1})_1-(\latdy{2}{1})_2&=1 \mod 2
\end{align}
with solution, within the fundamental domain,\footnote{The superscript $1$ in $\latdy{2}{1}$ and $\thedy{2}{1}$ is introduced merely as a label to specify this particular solution.}
\begin{align}
\label{eq:thetastar2}
\latdy{2}{1}\,=\,(1,0,0)\,, \quad
    \thedy{2}{1}\,=\,(4\pi\Qmin^2,0,0).
\end{align}
It is easy to verify using Eq.~\eqref{eq:n6min-2} that the minimal shift in $n_6$ induced by the solution
$\thedy{2}{1}$ in~\eqref{eq:thetastar2} is an integer, $\Delta n_6^{\rm min} =-1$, as it should be.

Other solutions can  be found by shifting 
$\latdy{2}{1}$ and $\thedy{2}{1}$ by the full periods, but they take us out of the domain of Fig.~\ref{fig:G2domain} and hence there is a single physically distinct point $\thedy{2}{1}$ that leads to the dyonic lattice shown in the diagram on the right in Fig.~\ref{fig:lattG2}. The diagram on the left in Fig.~\ref{fig:lattG2} depicts the dyonic lattice of the original $G_2$ theory with all trivial $\theta_i^{(2)}=0$. 

\medskip

Our final task in this section is to determine CP-invariant points $\theCP$ that should satisfy Eq.~\eqref{eq:CPinvthe} or equivalently, give mirror-symmetric dyonic lattices. They can be identified as points at a half-distance from the origin along any of the three periodicities of the $G_2$ theory and read
\begin{align}
\latCPp{2}&=\left\{(1,0,0)\oplus(1/2,1/2,0)\oplus(0,0,1/2)\right\}\,,\\
    \theCPp{2} &=\left\{(4\pi\Qmin^2,0,0)\oplus(2\pi\Qmin^2,\pi,0)\oplus(0,0,\pi)\right\}.\label{eq:CPG2}
\end{align}
Or, equivalently, in notations of Eq.~\eqref{eq:thetaijk-1}, 
\begin{align}
\label{eq:thetaijk-2}
\theCPp{p}^{(i,j,k)}&\,=\, i\theCPp{p}^{(1,0,0)}+j\theCPp{p}^{(0,1,0)}+k\theCPp{p}^{(0,0,1)}\,,& 0\leq&i\,,j\,,k\leq 1\,,
\end{align}
in our $p=2$ case we have,
\begin{align}
\theCPp{2}^{(0,0,0)}=(0,0,0)\,,\quad
\theCPp{2}^{(1,0,0)}=(4\pi\Qmin^2,0,0)\,, \quad
\theCPp{2}^{(0,1,0)}=(2\pi\Qmin^2,\pi,0) \,, \quad
\theCPp{2}^{(0,0,1)}=(0,0,\pi) \,.
\end{align}
In this notation, the CP-invariant points that lie within the $\theta_3=0$ section of the fundamental domain in Fig.~\ref{fig:G2domain}  are $\theCPp{2}^{(1,0,0)}$, $\theCPp{2}^{(0,1,0)}$ and $\theCPp{2}^{(1,1,0)}=(6\pi\Qmin^2,\pi,0)$.
The first of the CP-invariant points coincides with the solution $\thedy{2}{1}$ and, hence, gives an integer-valued $\Delta n_6$ shift in the dyonic lattice, as we already noted. The other two CP-invariant points, result in same dyonic lattice as the original one for vanishing $\theta$-angles,
\begin{eqnarray}
\thedy{2}{1}\,=\, &\theCPp{2}^{(1,0,0)}\,=\, (4\pi\Qmin^2,0,0) \quad \Rightarrow &\quad \Delta n_6^{(\rm min)}=-1\,, \\
&\theCPp{2}^{(0,1,0)}\,=\, (2\pi\Qmin^2,\pi,0) \quad \Rightarrow &\quad \Delta n_6^{(\rm min)}=\,0\,, \\
&\theCPp{2}^{(1,1,0)}\,=\, (6\pi\Qmin^2,\pi,0) \quad \Rightarrow &\quad \Delta n_6^{(\rm min)}=\,0\,.
\end{eqnarray}
These points are shown in Fig.~\ref{fig:G2domain} and the corresponding dyonic lattices are presented in Fig.~\ref{fig:lattG2}. There are only two distinct dyonic lattices.
For each of these two lattices of Fig.~\ref{fig:lattG2} one has several $\theta$ values, in particular the same lattice on the left is characterised by points $\theta=0$ and $\theCPp{2}^{(0,j,k)}$ for any integer $j$ and $k$; these can be disentangled by looking at quantum numbers outside of  the $\hex-\mhex$ subset.
Similarly, the dyonic lattice on the right hand side of Fig.~\ref{fig:lattG2} arises from $\thedy{2}{1}$ and any of points $\theCPp{2}^{(1,j,k)}$, as indicated in the figure.

\subsubsection{The $\mathbf{G_3}$ group $\mathbf{U(3)_{c+Y}\times SU(2)_L}$}

\begin{figure}[h]
    \centering
    \begin{tikzpicture}[scale=1.2]
    \draw [OIgreen,<->] (0,2.5) -- (6,2.5);
         \draw [OIgreen,<->] (0,0) -- (2,2);
        \draw[->] (-0.1,0) -- (6.25,0) node[right] {$\theta_{1}$};
        \draw[->] (0,-0.1) -- (0,2.75) node[above] {$\theta_3$};
        \filldraw (2,0)  ellipse (1pt and 3pt) node[below] {$\thedyb^{2}$};
        \filldraw (4,0)  ellipse (1pt and 3pt) node[below] {$\thedyb^{1}$};
        \filldraw (3,0)  ellipse (3pt and 1pt) node[above] {$\theCP^{(1,0,0)}$};
        \filldraw (1,1)  ellipse (3pt and 1pt) node[above] {$\theCP^{(0,1,0)}$};
        \filldraw (4,1)  ellipse (3pt and 1pt) node[above] {$\theCP^{(1,1,0)}$};
        \filldraw [gray,dashed,opacity=0.2] (0,0) -- (6,0) -- (6,2) -- (0,2) -- (0,0);
        \draw [gray,thick] (0,2) -- (0,0) -- (6,0);
        \draw[gray] (0,2) circle (2pt);
        \draw[gray] (6,0) circle (2pt);
        \node at (3,3.5) {\Large $G_3$};
        \filldraw (0,0) ellipse (1pt and 3pt);
        \filldraw (0,0) ellipse (3pt and 1pt) node[below]
{$\qquad\thedyb^{0}=\theCP^{(0,0,0)}$};
    \end{tikzpicture}
    \caption{$\theta_2=0$ section of the $\theta$-space of the $G_3$ gauge theory.
    Fundamental domain on the $(\theta_1,\theta_3)$ plane is the gray rectangle of dimensions ($2\pi (3\Qmin)^2$,$2\pi$). Our notation is the same as in Fig.~\ref{fig:G2domain}.}
    \label{fig:domainG3}
\end{figure}
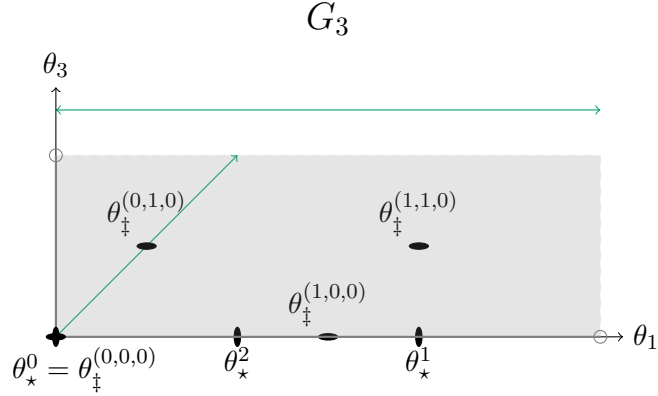
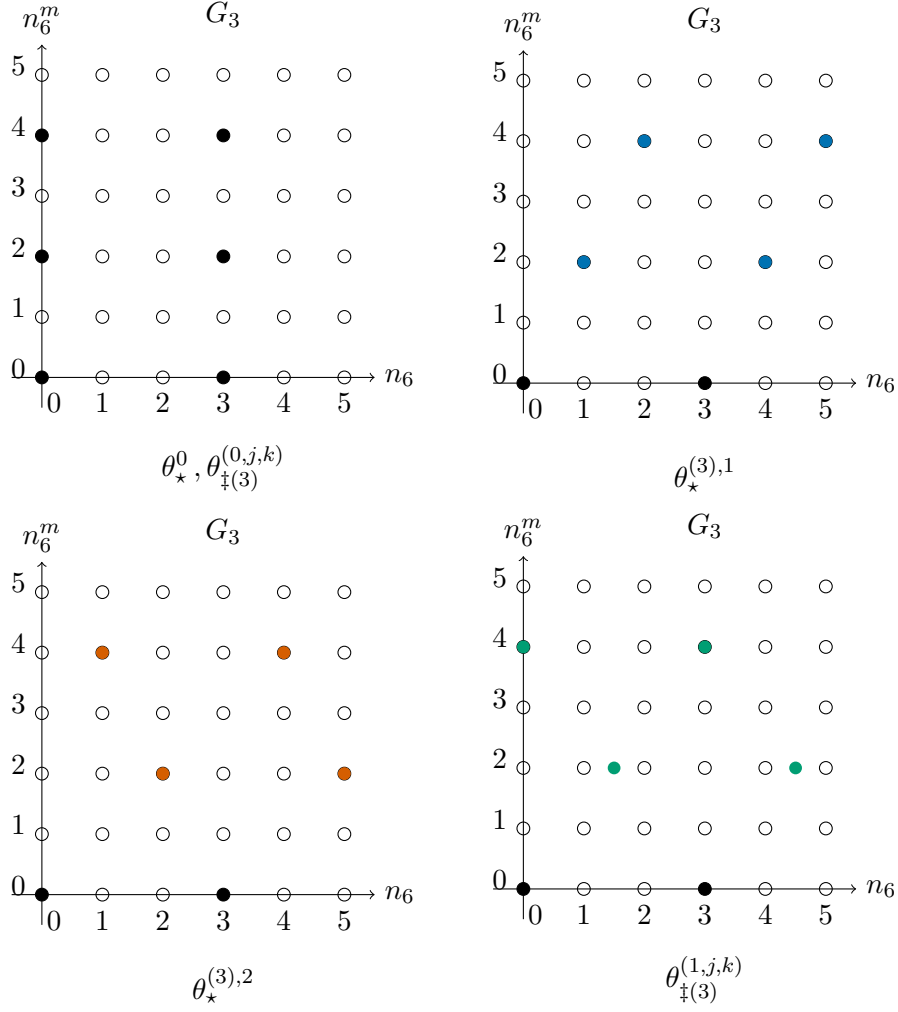
\begin{figure}[h!]
    \centering
    \begin{tikzpicture}[scale=0.8]
      \draw[->] (-0.5,0) -- (5.5,0) node[right] {$n_6$};
      \draw[->] (0,-0.5) -- (0,5.5) node[above] {$n^m_6$};
      \foreach \y in {0,...,5} \node[left] at (-0.1,\y+0.15) {$\y$};
      \foreach \x in {1,...,5} \node[below] at (\x,-0.1) {$\x$};
      \node[below] at (0.2,-0.1) {$0$};
      \foreach \x in {0,...,5} \foreach \y in {0,...,5}
        \draw (\x,\y) circle (3pt);
      \fill (0,0) circle (3pt);
      \fill (3,0) circle (3pt);
      \fill (0,2) circle (3pt);
      \fill (3,2) circle (3pt);
      \fill (0,4) circle (3pt);
      \fill (3,4) circle (3pt);
      \node at (3,6) {$G_3$};
            \node at (3,-1.5) {$\thedyb^{0}\,,\theCPp{3}^{(0,j,k)}$};
    \end{tikzpicture}\qquad
    \begin{tikzpicture}[scale=0.8]
      \draw[->] (-0.5,0) -- (5.5,0) node[right] {$n_6$};
      \draw[->] (0,-0.5) -- (0,5.5) node[above] {$n^m_6$};
      \foreach \y in {0,...,5} \node[left] at (-0.1,\y+0.15) {$\y$};
      \foreach \x in {1,...,5} \node[below] at (\x,-0.1) {$\x$};
      \node[below] at (0.2,-0.1) {$0$};
      \foreach \x in {0,...,5} \foreach \y in {0,...,5}
        \draw (\x,\y) circle (3pt);
      \fill (0,0) circle (3pt);
      \fill (3,0) circle (3pt);
      \fill[OIblue] (1,2) circle (3pt);
      \fill[OIblue] (2,4) circle (3pt);
      \fill[OIblue] (4,2) circle (3pt);
      \fill[OIblue] (5,4) circle (3pt);
      \node at (3,6) {$G_3$};
            \node at (3,-1.5) {$\thedy{3}{1}$};   
            \end{tikzpicture}\\
     \begin{tikzpicture}[scale=0.8]
      \draw[->] (-0.5,0) -- (5.5,0) node[right] {$n_6$};
      \draw[->] (0,-0.5) -- (0,5.5) node[above] {$n^m_6$};
      \foreach \y in {0,...,5} \node[left] at (-0.1,\y+0.15) {$\y$};
      \foreach \x in {1,...,5} \node[below] at (\x,-0.1) {$\x$};
      \node[below] at (0.2,-0.1) {$0$};
      \foreach \x in {0,...,5} \foreach \y in {0,...,5}
        \draw (\x,\y) circle (3pt);
      \fill (0,0) circle (3pt);
      \fill (3,0) circle (3pt);
      \fill[OIvermilion] (2,2) circle (3pt);
      \fill[OIvermilion] (5,2) circle (3pt);
      \fill[OIvermilion] (4,4) circle (3pt);
      \fill[OIvermilion] (1,4) circle (3pt);
      \node at (3,6) {$G_3$};
            \node at (3,-1.5) {$\thedy{3}{2}$};
    \end{tikzpicture}\qquad
    \begin{tikzpicture}[scale=0.8]
      \draw[->] (-0.5,0) -- (5.5,0) node[right] {$n_6$};
      \draw[->] (0,-0.5) -- (0,5.5) node[above] {$n^m_6$};
      \foreach \y in {0,...,5} \node[left] at (-0.1,\y+0.15) {$\y$};
      \foreach \x in {1,...,5} \node[below] at (\x,-0.1) {$\x$};
      \node[below] at (0.2,-0.1) {$0$};
      \foreach \x in {0,...,5} \foreach \y in {0,...,5}
        \draw (\x,\y) circle (3pt);
      \fill (0,0) circle (3pt);
      \fill (3,0) circle (3pt);
      \fill[OIgreen] (1.5,2) circle (3pt);
      \fill[OIgreen] (4.5,2) circle (3pt);
      \fill[OIgreen] (0,4) circle (3pt);
      \fill[OIgreen] (3,4) circle (3pt);
      \node at (3,6) {$G_3$};
            \node at (3,-1.5) {$\theCPp{3}^{(1,j,k)}$};
    \end{tikzpicture}
    \caption{Dyonic lattices for different choices of special $\theta$-points in $G_3$ theory. The first three are dyonic lattices with integer values of $n_6$. The fourth lattice includes half-integer $n_6$ values, it corresponds to a perfectly valid theory but it does not have a well-defined electric symmetry. The first and the last of the lattices (those labelled by $\theCP$ points) respect $n_6\to -n_6$ axial symmetry and give CP-invariant spectra; the other two (which arise solely from $\thedyb$ points) do not.}
    \label{fig:lattG3}
\end{figure}

The case of $G_3$ has quantisation condition
\begin{align}
   &e^{i\Fl_1 Q_Y+i\Fl_2 \tilde T_{3L}+i\Fl_3\tilde \lambda_8}=\left\{\begin{array}{c}
          1  \\
          \xi^2\\
          \xi^4
     \end{array}\right\}=\left[\xi^2\right]^{n_{6}^m/2}=\left[e^{2\pi i/3 (Q/Q_{\textrm{min}}-n_c)}\right]^{n_{6}^m/2}\,, & \mhex&=(0,2,4)\,,
\end{align}
with magnetic fluxes solving this equation as
\begin{align}
     \vec\Fl\,=\,\left(\frac{2\pi q_m}{6\Qmin},\frac{2\pi \ktwo}{2},-\frac{2\pi (q_m/2+\kthree)}{3}\right), \qquad\left(q_m\in 2\mathbb{Z}\,,\,\ktwo\in2\mathbb{Z}\,,\,\kthree\in 3\mathbb{Z}\right),
\end{align}
so that the action of the electric symmetry on 't Hooft lines
\begin{align}
    \xi^2\tHft_{\mhex}=&e^{2\pi i/3 (Q_Y/Q_{\textrm{min}}-n_c)}\tHft_{\mhex}
    \,
    =\exp\left(i\theta_3\,\frac{q_m/2+\kthree\mod 3}{3}\,-\,
    \frac{i\theta_1}{3\Qmin^2}\frac{q_m/2\mod 3}{3}\right)\tHft_{\mhex}\,,\nonumber\\
    =&\, \exp\left(\frac{i n^m_6}{6}\left[\theta_3-\frac{\theta_1}{3\Qmin^2}\right]\right)\tHft_{\mhex} 
\label{eq:255}    
 \end{align}
which leads to
\begin{align}
    \xi^2\tHft_{\mhex}\,=\,e^{2\pi n_3 i/3}\tHft_{\mhex}\,, \qquad n_3&=\frac{n^m_6}{2}\left[\frac{\theta_3}{2\pi}-\frac{\theta_1}{6\pi\Qmin^2}\right]=n_6\mod 3.
\label{eq:256}
\end{align}
The condition of a well-defined electric symmetry 
\begin{align}
    &e^{i(\thedyb)_3}=e^{i(\thedyb)_2}=e^{i(\thedyb)_1/3\Qmin^2}=1\,, & \thedyb&=2\pi (3\Qmin^2(\latdyB)_1,(\latdyB)_2, (\latdyB)_3)\,,\qquad \vec\latdyB\in\mathbb{Z}\,.
\end{align} An increase of $n$-ality by 3 units for $n_3$ for the smallest flux is on the other hand
\begin{align}
    (n_3)_\textrm{min}=\frac{(n^m_6)_{\min}}{2\times2\pi}\left[\Delta\theta_3-\frac{\Delta\theta_1}{3\Qmin^2}\right]=(\latD^{(3)})_3-(\latD^{(3)})_1=0\mod 3\,,
\end{align}
which returns the  periods $\Delta\vec\theta$ of Eq.~\eqref{eq:Dtheta-p} with the values given by,
\begin{align}
    G_3:&  &\left\{\Delta\vec\theta^{\,(3)}\right\}&=\left\{\begin{array}{lcc}
        (\,6\pi\Qmin^2\,, & 0\,, & 2\pi\,)\\
        (\, 18\pi\Qmin^2\,,& 0\,, &0\,)\\
        (\, 0\,,& 2\pi\,, &0\,)
    \end{array}\right\}\,,& \left\{\Delta\vec L^{\,(3)}\right\}=&\left\{\begin{array}{lcc}
        (\,1\,, & 0\,, & 1\,)\\
        (\, 3\,,& 0\,, &0\,)\\
        (\, 0\,,& 1\,, &0\,)
    \end{array}\right\}\,,\label{eq:PerG3}
\end{align}
and a valid choice of domain is
\begin{align}
        0\leq&\theta_1<18\pi \Qmin^2\,, & 
    0\leq&\theta_2<2\pi\,, &
    0\leq&\theta_3<2\pi\,, \\
    0\leq&L_1<3\,, & 0\leq&L_2<1\,, & 0\leq&L_3<1\,.
\end{align}

As in the case for $G_2$, now there are also more possibilities for well-defined electric symmetry which we can characterise as 
\begin{align}
    (\latdy{3}{\ell})_3-(\latdy{3}{\ell})_1=\ell\mod 3\,,
\end{align}
and $\ell=1,2$ return two distinct dyonic lattices
\begin{align}
\thedy{3}{1} =&2\pi\left(6\Qmin^2,0,0\right)\,, &
\thedy{3}{2}=&2\pi\left(3\Qmin^2,0,0\right)\,, \\
\latdy{3}{1}=&(2,0,0) & \latdy{3}{2}=&(1,0,0)\,.
\end{align}
The set of CP invariant points $\theCPp{3}$ in this case is non-overlapping with the $\vthedyb^{\,\,(3)}$ points and reads
\begin{align}
    \theCPp{3}=&\left\{(9\pi\Qmin^2\,,\,0\,,\,0)\oplus (3\pi\Qmin^2\,,\,0\,,\,\pi)\oplus (0\,,\,\pi\,,\,0)\right\}\,,\\
    \latCPp{3}=&\left\{(3/2,0,0)\oplus(1/2,0,1/2)\oplus (0,1/2,0)\right\}\,.
\end{align}
Location of these points in the fundamental $\theta$-space domain of the $G_3$ theory is given in Fig.~\ref{fig:domainG3}.
The corresponding electric-magnetic lattices that these points lead to are shown in Fig.~\ref{fig:lattG3}.

\subsubsection{The $\mathbf{G_6}$ group $\mathbf{S\left[U(3)\times U(2)\right]}$}

\begin{figure}[h!]
    \centering
    \begin{tikzpicture}[scale=0.8]
  \draw[->] (-0.5,0) -- (5.5,0) node[right] {$n_6$};
  \draw[->] (0,-0.5) -- (0,5.5) node[above] {$n^m_6$};
  \node[left] at (-0.1,0.2) {$0$};
  \node[left] at (-0.1,1) {$1$};
  \node[left] at (-0.1,2) {$2$};
  \node[left] at (-0.1,3) {$3$};
  \node[left] at (-0.1,4) {$4$};
  \node[left] at (-0.1,5) {$5$};
  \foreach \x in {1,...,5}
    \node[below] at (\x,-0.1) {\x};
  \node[below] at (0.2,-0.1) {0};
  \foreach \x in {0,...,5} {
    \foreach \y in {0,...,5} {
      \draw (\x,\y) circle (3pt);
    }
  }
  \fill (0,0) circle (3pt);
  \fill (0,1) circle (3pt);
  \fill (0,2) circle (3pt);
\fill (0,3) circle (3pt);
     
    \fill (0,4) circle (3pt);
  \fill (0,5) circle (3pt);
  \node at (3,6) {$G_6$};
    \node at (3,-1.5) {$\thedyb^{0}\,,\theCPp{6}^{(0,j,k)}$};
\end{tikzpicture}
\begin{tikzpicture}[scale=0.8]
  \draw[->] (-0.5,0) -- (5.5,0) node[right] {$n_6$};
  \draw[->] (0,-0.5) -- (0,5.5) node[above] {$n^m_6$};
  \node[left] at (-0.1,0.2) {$0$};
  \node[left] at (-0.1,1) {$1$};
  \node[left] at (-0.1,2) {$2$};
  \node[left] at (-0.1,3) {$3$};
  \node[left] at (-0.1,4) {$4$};
  \node[left] at (-0.1,5) {$5$};
  \foreach \x in {1,...,5}
    \node[below] at (\x,-0.1) {\x};
  \node[below] at (0.2,-0.1) {0};
  \foreach \x in {0,...,5} {
    \foreach \y in {0,...,5} {
      \draw (\x,\y) circle (3pt);
    }
  }
  \fill (0,0) circle (3pt);
  \fill[OIgreen] (1,1) circle (3pt);
  \fill[OIgreen] (2,2) circle (3pt);
\fill[OIgreen] (3,3) circle (3pt);
     
    \fill[OIgreen] (4,4) circle (3pt);
  \fill[OIgreen] (5,5) circle (3pt);

  \node at (3,6) {$G_6$};
    \node at (3,-1.5) {$\thedy{6}{1}$};
\end{tikzpicture}
\begin{tikzpicture}[scale=0.8]
  \draw[->] (-0.5,0) -- (5.5,0) node[right] {$n_6$};
  \draw[->] (0,-0.5) -- (0,5.5) node[above] {$n^m_6$};
  \node[left] at (-0.1,0.2) {$0$};
  \node[left] at (-0.1,1) {$1$};
  \node[left] at (-0.1,2) {$2$};
  \node[left] at (-0.1,3) {$3$};
  \node[left] at (-0.1,4) {$4$};
  \node[left] at (-0.1,5) {$5$};
  \foreach \x in {1,...,5}
    \node[below] at (\x,-0.1) {\x};
  \node[below] at (0.2,-0.1) {0};
  \foreach \x in {0,...,5} {
    \foreach \y in {0,...,5} {
      \draw (\x,\y) circle (3pt);
    }
  }
  \fill (0,0) circle (3pt);
  \fill[OIblue] (2,1) circle (3pt);
  \fill[OIblue] (4,2) circle (3pt);
\fill[OIblue] (0,3) circle (3pt);
    \fill[OIblue] (2,4) circle (3pt);
  \fill[OIblue] (4,5) circle (3pt);
  \node at (3,6) {$G_6$};
    \node at (3,-1.5) {$\thedy{6}{2}$};
\end{tikzpicture}
\begin{tikzpicture}[scale=0.8]
  \draw[->] (-0.5,0) -- (5.5,0) node[right] {$n_6$};
  \draw[->] (0,-0.5) -- (0,5.5) node[above] {$n^m_6$};
  \node[left] at (-0.1,0.2) {$0$};
  \node[left] at (-0.1,1) {$1$};
  \node[left] at (-0.1,2) {$2$};
  \node[left] at (-0.1,3) {$3$};
  \node[left] at (-0.1,4) {$4$};
  \node[left] at (-0.1,5) {$5$};
  \foreach \x in {1,...,5}
    \node[below] at (\x,-0.1) {\x};
  \node[below] at (0.2,-0.1) {0};
  \foreach \x in {0,...,5} {
    \foreach \y in {0,...,5} {
      \draw (\x,\y) circle (3pt);
    }
  }
  \fill (0,0) circle (3pt);
  \fill[OIorange] (3,1) circle (3pt);
  \fill[OIorange] (0,2) circle (3pt);
\fill[OIorange] (3,3) circle (3pt);
    \fill[OIorange] (0,4) circle (3pt);
  \fill[OIorange] (3,5) circle (3pt);
  \node at (3,6) {$G_6$};
    \node at (3,-1.5) {$\thedy{6}{3},\theCPp{6}^{(1,j,k)}$};\end{tikzpicture}
\begin{tikzpicture}[scale=0.8]
  \draw[->] (-0.5,0) -- (5.5,0) node[right] {$n_6$};
  \draw[->] (0,-0.5) -- (0,5.5) node[above] {$n^m_6$};
  \node[left] at (-0.1,0.2) {$0$};
  \node[left] at (-0.1,1) {$1$};
  \node[left] at (-0.1,2) {$2$};
  \node[left] at (-0.1,3) {$3$};
  \node[left] at (-0.1,4) {$4$};
  \node[left] at (-0.1,5) {$5$};
  \foreach \x in {1,...,5}
    \node[below] at (\x,-0.1) {\x};
  \node[below] at (0.2,-0.1) {0};
  \foreach \x in {0,...,5} {
    \foreach \y in {0,...,5} {
      \draw (\x,\y) circle (3pt);
    }
  }
  \fill (0,0) circle (3pt);
  \fill[OIvermilion] (4,1) circle (3pt);
  \fill[OIvermilion] (2,2) circle (3pt);
\fill[OIvermilion] (0,3) circle (3pt);
    \fill[OIvermilion] (4,4) circle (3pt);
  \fill[OIvermilion] (2,5) circle (3pt);
  \node at (3,6) {$G_6$};
    \node at (3,-1.5) {$\thedy{6}{4}$};
\end{tikzpicture}
\begin{tikzpicture}[scale=0.8]
  \draw[->] (-0.5,0) -- (5.5,0) node[right] {$n_6$};
  \draw[->] (0,-0.5) -- (0,5.5) node[above] {$n^m_6$};
  \node[left] at (-0.1,0.2) {$0$};
  \node[left] at (-0.1,1) {$1$};
  \node[left] at (-0.1,2) {$2$};
  \node[left] at (-0.1,3) {$3$};
  \node[left] at (-0.1,4) {$4$};
  \node[left] at (-0.1,5) {$5$};
  \foreach \x in {1,...,5}
    \node[below] at (\x,-0.1) {\x};
  \node[below] at (0.2,-0.1) {0};
  \foreach \x in {0,...,5} {
    \foreach \y in {0,...,5} {
      \draw (\x,\y) circle (3pt);
    }
  }
  \fill (0,0) circle (3pt);
  \fill[OImagenta] (5,1) circle (3pt);
  \fill[OImagenta] (4,2) circle (3pt);
\fill[OImagenta] (3,3) circle (3pt);
     
    \fill[OImagenta] (2,4) circle (3pt);
  \fill[OImagenta] (1,5) circle (3pt);
  \node at (3,6) {$G_6$};
    \node at (3,-1.5) {$\thedy{6}{5}$};
\end{tikzpicture}
    \caption{Dyonic lattices for different special values of $\theta$; the top and bottom left lattices are CP-invariant. The four remaining lattices (those  marked solely by $\thedyb^{(6)}$ points) do not produce CP-invariant spectrum.}
    \label{fig:lattG6}
\end{figure}
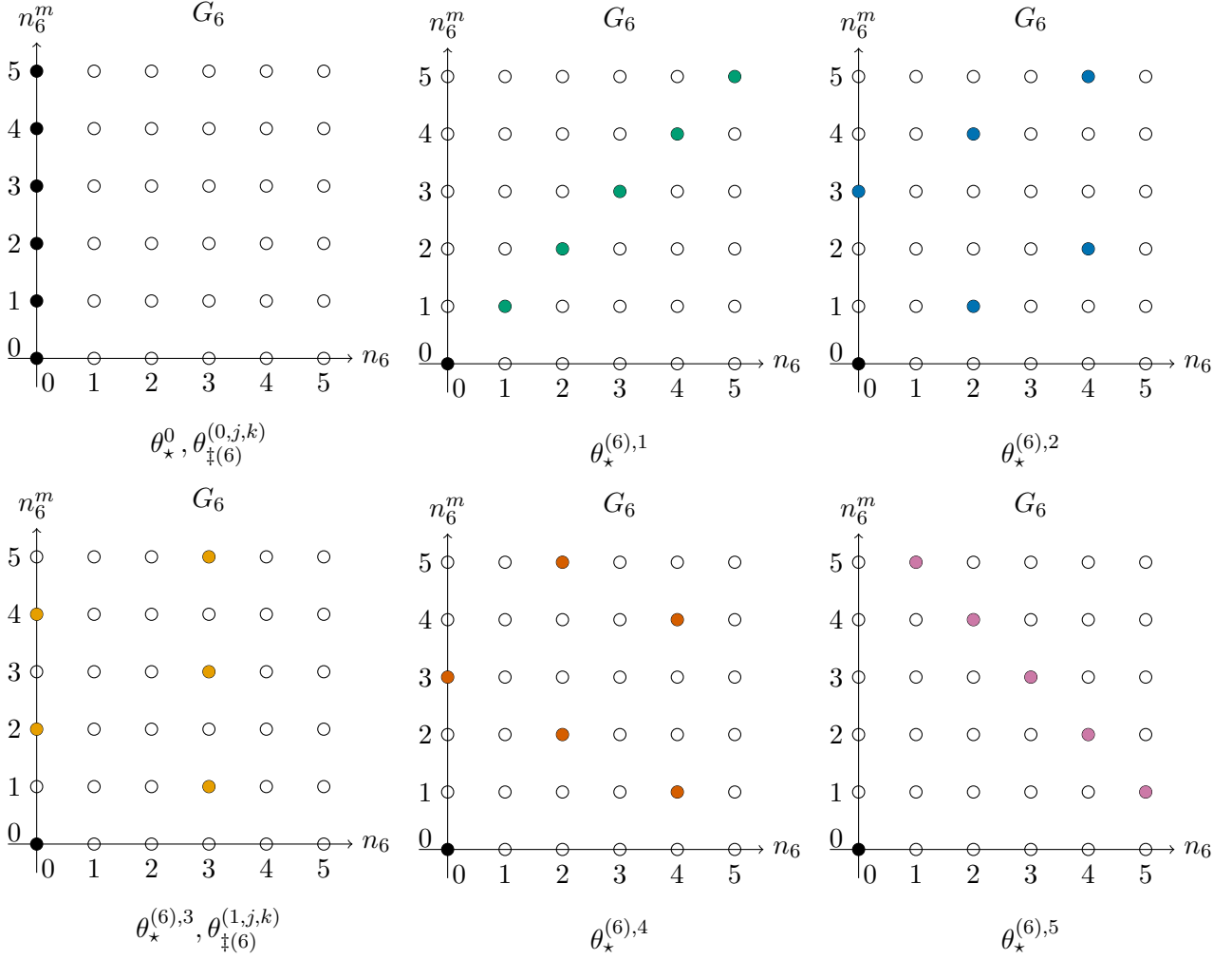

The smallest group, $G_6$, has the entire $Z_6$ centre quotiented out and allows for all elements of $Z_6$ on the RHS of Eq.~\eqref{eq:DQcond1}
\begin{align}
    e^{i\Fl_1 Q_Y+i\Fl_2 \tilde T_{3L}+i\Fl_3\tilde \lambda_8}=\left[e^{2\pi i Q_6/6}\right]^{\mhex}=\left[\xi\right]^{\mhex}=\left[e^{2\pi i Q_Y/6\Qmin}e^{2\pi in_c/3}e^{\pi n_Li}\right]^{\mhex}\,,
\end{align}
so that allowed fluxes read
\begin{align}
    \vec\Fl\,=\,\left(\frac{2\pi q_m}{6\Qmin}\,,\,\frac{2\pi (q_m+\ktwo)}{2}\,,\,\frac{2\pi(q_m+ \kthree)}{3}\right),
    \qquad\left(q_m\in \mathbb{Z}\,,\,\ktwo\in2\mathbb{Z}\,,\,\kthree\in3\mathbb{Z}\right)\,,
\end{align}
and the electric symmetry action on monopoles
\begin{align}\nonumber
    \xi \tHft_{\mhex}=&e^{2\pi i Q_6/6}T_{n_6^m}\,
    =\,\exp\left(\left[\frac{-i \theta_1}{6\Qmin^2}\right] \frac{q_m\mod 6}{6}
\,+\,i\theta_2\,\frac{(q_m\mod 2)}{2}\,+\,
i\theta_3\,\frac{(q_m\mod 3)}{3}\right)\tHft_{\mhex}\,,\\
=&\,
    \exp\left(\frac{i}6\left[2(n^m_6\mod 3)\theta_3+3(n^m_6\mod 2)\theta_2-\frac{n_6^m\theta_1}{6\Qmin^2}\right]\right)\tHft_{\mhex}
\end{align}
with
\begin{align}
     \xi \tHft_{\mhex}\,=\,e^{2\pi n_6i/6}\tHft_{\mhex}\,, \qquad n_6=\frac{1}{2\pi}\left[2(n^m_6\mod 3)\theta_3+3(n^m_6\mod 2)\theta_2-\frac{n_6^m\theta_1}{6\Qmin^2}\right]\,.
\end{align}
\begin{figure}[h!]
    \centering
    \begin{tikzpicture}
\draw [->] (0,0) -- (0,2.5*1.1) node[above] {$\theta_2$};
\draw [->] (0,0) -- (-2.2*1.1,-1.1) node[left] {$\theta_3$};
\draw [->] (0,0) -- (6,-3) node[right] {$\theta_1$};
\filldraw (6/6,-3/6) ellipse (1pt and 3pt);     
\filldraw (2*6/6,-2*3/6) ellipse (1pt and 3pt); 
\filldraw (3*6/6,-3*3/6) ellipse (1pt and 3pt);
\filldraw (3*6/6,-3*3/6)  ellipse (3pt and 1pt); 
\filldraw (4*6/6,-4*3/6) ellipse (1pt and 3pt);  
\filldraw (5*6/6,-5*3/6) ellipse (1pt and 3pt);  
\draw[<->,OIgreen] (0,0) -- (2*6/6-2.2,-2*3/6-1);
\draw[<->,OIgreen] (0,0) --(3*6/6,-3*3/6+2.5);
\filldraw (2*6/12-2.2/2,-2*3/12-1/2)  ellipse (3pt and 1pt) node [left] {$\theCP^{(0,1,0)}$};
\filldraw (3*6/12,-3*3/12+2.5/2)  ellipse (3pt and 1pt) node [left] {$\theCP^{(0,0,1)}$};
\filldraw (3*6/12+3*6/6,-3*3/12+2.5/2-3*3/6)  ellipse (3pt and 1pt) node [right] {$\theCP^{(1,0,1)}$};
\filldraw ((2*6/12-2.2/2+3*6/6,-2*3/12-1/2-3*3/6)  ellipse (3pt and 1pt) node [left] {$\theCP^{(1,1,0)}$};
\filldraw (3*6/12+2*6/12-2.2/2,-2*3/12-1/2-3*3/12+2.5/2)  ellipse (3pt and 1pt) node [above] {$\theCP^{(0,1,1)}$};
\draw[dashed,OIgreen] (3*6/12+2*6/12-2.2/2+1.1,-2*3/12-1/2-3*3/12+2.5/2+0.5-1.25) --(3*6/12+2*6/12-2.2/2+1.1,-2*3/12-1/2-3*3/12+2.5/2+0.5)--(3*6/12+2*6/12-2.2/2,-2*3/12-1/2-3*3/12+2.5/2) --(3*6/12+2*6/12-2.2/2,-2*3/12-1/2-3*3/12)--(3*6/12+2*6/12-2.2/2+1.1,-2*3/12-1/2-3*3/12+2.5/2+0.5-1.25);
\draw [dashed,OIgreen] (3*6/12,-3*3/12+2.5/2) -- (3*6/12+2*6/12-2.2/2+3*6/6+1.1,-3*3/6-2*3/12-1/2-3*3/12+2.5/2+0.5);
\draw [dashed,OIgreen] ((2*6/12-2.2/2,-2*3/12-1/2) -- (3*6/12+2*6/12-2.2/2+3*6/6,-3*3/6-2*3/12-1/2-3*3/12);
\filldraw (3*6/12+2*6/12-2.2/2+3*6/6,-3*3/6-2*3/12-1/2-3*3/12+2.5/2)  ellipse (3pt and 1pt) node [right] {$\theCP^{(1,1,1)}$};
\draw [dashed,OIgreen] (3*6/12+2*6/12-2.2/2,-2*3/12-1/2-3*3/12+2.5/2) -- (3*6/12+2*6/12-2.2/2+3*6/6,-3*3/6-2*3/12-1/2-3*3/12+2.5/2);
\filldraw[gray,opacity=0.2] (-2.2,-1) -- (-2.2+6,-1-3) -- (6,-3) -- (6,-3+2.5) -- (0,2.5) -- (-2.2,-1+2.5) --(-2.2,-1);
\draw[gray]     (-2.2,-1) -- (-2.2+6,-1-3) -- (6,-3) -- (6,-3+2.5) -- (0,2.5) -- (-2.2,-1+2.5) --(-2.2,-1);
\draw [gray] (-2.2,-1+2.5) -- (-2.2+6,-1-3+2.5) -- (-2.2+6,-1-3);
\draw [gray] (-2.2+6,-1-3+2.5) -- (6,-3+2.5);
\draw[dashed,OIgreen]  (3*6/12+2*6/12-2.2/2+3*6/6,-3*3/6-2*3/12-1/2-3*3/12)--(3*6/12+2*6/12-2.2/2+3*6/6,-3*3/6-2*3/12-1/2-3*3/12+2.5/2) -- (3*6/12+2*6/12-2.2/2+3*6/6+1.1,-3*3/6-2*3/12-1/2-3*3/12+2.5/2+0.5)--(3*6/12+2*6/12-2.2/2+3*6/6+1.1,-3*3/6-2*3/12-1/2-3*3/12+2.5/2+0.5-1.25) 
--(3*6/12+2*6/12-2.2/2+3*6/6,-3*3/6-2*3/12-1/2-3*3/12);
\draw[dashed,OIgreen]  (3*6/12+2*6/12-2.2/2+3*6/6-5.5,-3*3/6-2*3/12-1/2-3*3/12+2.75)--(3*6/12+2*6/12-2.2/2+3*6/6-5.5,-3*3/6-2*3/12-1/2-3*3/12+2.5/2+2.75) --  (3*6/12+2*6/12-2.2/2+3*6/6+1.1-5.5,-3*3/6-2*3/12-1/2-3*3/12+2.5/2+0.5+2.75)--(3*6/12+2*6/12-2.2/2+3*6/6+1.1-5.5,-3*3/6-2*3/12-1/2-3*3/12+2.5/2+0.5-1.25+2.75) 
--(3*6/12+2*6/12-2.2/2+3*6/6-5.5,-3*3/6-2*3/12-1/2-3*3/12+2.75);
\draw[dashed,OIgreen] (2*6/12-2.2/2,-2*3/12-1/2) --(3*6/12+2*6/12-2.2/2+3*6/6-5.5,-3*3/6-2*3/12-1/2-3*3/12+2.75);
\draw[dashed,OIgreen] (3*6/12+2*6/12-2.2/2,-2*3/12-1/2-3*3/12+2.5/2)-- (3*6/12+2*6/12-2.2/2+3*6/6-5.5,-3*3/6-2*3/12-1/2-3*3/12+2.5/2+2.75);
\draw[dashed,OIgreen] (3*6/12,-3*3/12+2.5/2)  --(3*6/12+2*6/12-2.2/2+3*6/6+1.1-5.5,-3*3/6-2*3/12-1/2-3*3/12+2.5/2+0.5+2.75);
\filldraw (0,0) ellipse (1pt and 3pt);
\filldraw (0,0) ellipse (3pt and 1pt);
\node at (3,3) {\Large $G_6$};
\end{tikzpicture}
    \caption{Fundamental domain in the $G_6$ theory as the gray cuboid of dimensions  ($2\pi (6\Qmin)^2$\,,\,$2\pi$\,,\,$2\pi$). The domain includes the three sides defined by the axes, the edges that overlap with said axes and the origin, but not the rest of the boundary given by the other three sides, nine edges and seven vertexes. The equivalence relation that defines the three-torus is given by the green double-sided arrows and CP-invariant points  are marked with ellipses.
    }
    \label{fig:DomainG6}
\end{figure}
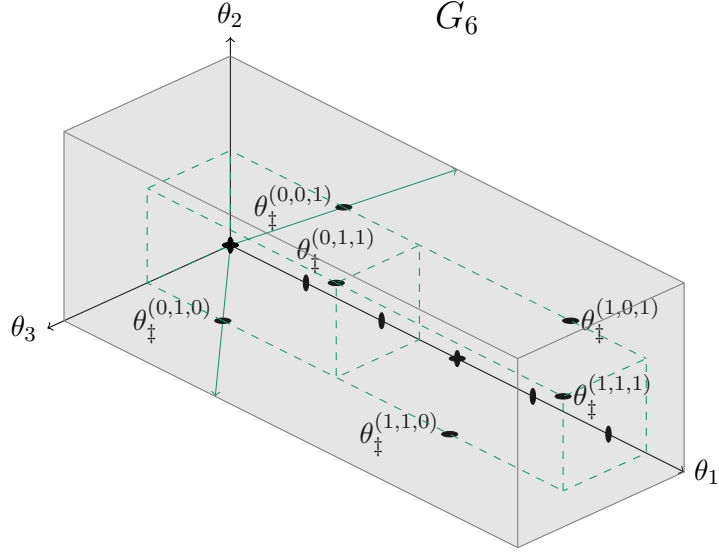
A well-defined electric symmetry demands
\begin{align}
    e^{i(\thedyb)_3}=e^{i(\thedyb)_2}=e^{i(\thedyb)_1/6\Qmin^2}=1\,,& & \thedyb&=2\pi\left(6\Qmin^2(\latdyB)_1,(\latdyB)_2,(\latdyB)_3\right)\,,\quad \vec\latdyB\in \vec{\mathbb{Z}}\,.
\end{align}
One can classify these according to the $n$-ality shift they produce for the minimum $n_6$
\begin{align}
    &\frac{1}{2\pi}\left[2(\thedyb)_3+3(\thedyb)_2-\frac{(\thedyb)_1}{6\Qmin^2}\right]=2(\latdyB)_3+3(\latdyB)_2-(\latdyB)_1=\ell \mod 6\,.
\end{align}
Where $\ell=0$ gives the full periodicity ({\it cf.}~Eq.~\eqref{eq:Dtheta-p}),
\begin{align}
      G_6:& & \left\{\Delta\theta^{(6)}\right\}&=\left\{\begin{array}{lcc}
        (\,24\pi\Qmin^2\,, & 0\,, & 2\pi\,)\\
        (\, 36\pi\Qmin^2\,,& 2\pi\,, &0\,)\\
        (\, 72\pi\Qmin^2\,,& 0\,, &0\,)
    \end{array}\right\}\,,& \left\{\latD^{(6)}\right\}&=\left\{\begin{array}{lcc}
        (\,2\,, & 0\,, & 1\,)\\
        (\, 3\,,& 1\,, &0\,)\\
        (\, 6\,,& 0\,, &0\,)
    \end{array}\right\}\,.\label{eq:PerG6}
\end{align}
The minimum domain can now be chosen as
\begin{align}
    0\leq&\,\theta_1<72\pi \Qmin^2 \,,& 
    0\leq&\,\theta_2<2\pi\,, &
    0\leq&\,\theta_3<2\pi\,, \\
        0\leq&\,L_1<6\,, & 0\leq&\,L_2<1\,, & 0\leq&\,L_3<1\,,
\end{align}
and within this domain the points $ \vthedyb^{\,\,(6)}$ that lead to integer-$n_6$ dyonic lattices with shift $\ell$ read
\begin{align}
    \thedy{6}{\ell}\,=\,2\pi((6-\ell) 6\Qmin^2\,,\,0\,,\,0\,)\,, \qquad \latdy{6}{\ell}&=((6-\ell) \,,\,0\,,\,0\,)\,.
\end{align}
Lastly, the CP invariant points ({\it cf.}~Eq.~\eqref{eq:CPinvthe}) are
\begin{align}
    \latCPp{6}&=\left\{(3,0,0)\oplus(1,0,1/2)\oplus(3/2,1/2,0)\right\}\,,\\
    \theCPp{6}=&\left\{ (36\pi\Qmin^2\,,\,0\,,\,0)\oplus(12\pi\Qmin\,,\,0\,,\,\pi)\oplus (18\pi\Qmin^2\,,\,\pi\,,\,0)\right\}\,,
\end{align}
where now some degeneracy is present with dyonic lattices, see Fig.~\ref{fig:lattG6}, while the location of points is shown in Fig.~\ref{fig:DomainG6}.


\subsection{The effect of fermions in the full SM}\label{sec:BL}

Consider now the full Standard Model and its fermion content. The charge assignment of fermions is such that gauge anomalies cancel, but  global symmetries can suffer from anomalies and contribute to the theta angles. The well-known case is the axial anomaly in the quark sector, the phase of this axial rotation contributes to $\theta_3$ and makes it by itself unphysical. There is, however, a term in the action that breaks this symmetry at the classical level, Yukawa couplings that give rise to masses as
\footnote{For simplicity we consider one generation only; the $\theta$ projection we are after remains the same if three are considered.}
\begin{align}
    \mathcal L \supset \bar u_R Y_u \tilde H^\dagger q_L+\bar d_R Y_dH^\dagger q_L+h.c.\,,
\end{align}
which one can use to define the properly weighted difference between $\theta_3$ and the phase in the Yukawas that is invariant under the action of this symmetry
\begin{align}
    \bar\theta_3&=\theta_3-\arg(Y_u Y_d).\label{eq:the3phys}
\end{align}
In the following, we will drop the bar in $\theta$ with the understanding that it is the physical combination that is discussed. There is another global symmetry with an anomaly yet exact at the classical level: $B+L$. To be precise, a $U(1)_{B+L}$ rotation leads to a shift in $(\theta_1,\theta_2)$-angles.
For one family we have
\begin{align}
    &\begin{array}{c}
         (q_L,u_R,d_R)\to e^{i\alpha/3}(q_L,u_R,d_R)  \\
          (\ell_L,e_R)\to e^{i\alpha}(\ell_L,e_R)
    \end{array},& (\theta_1,\theta_2)\to (\theta_1,\theta_2)+(-36 \Qq^2,2)\alpha\equiv (\theta_1,\theta_2)+\vec{u}_{B+L}\alpha\,,
    \label{eq:BLshift}
\end{align}
while in the SM there is now no term in the Lagrangian that shifts under this symmetry. Before we move on to examine the consequences of this let us note that, in a SM extension with a $B+L$ breaking operator  such as
\begin{align}
    \mathcal L_{SMEFT}\supset C d_R u_R u_R e_R\,,
\end{align}
one could define physical $\bar \theta_{1,2}$ angles as
\begin{align} 
\bar\theta_1\,=\,\theta_1+18 \Qq^2 \arg (C)\,,
    \qquad\bar\theta_2&=\theta_2-\arg(C)\,,
\label{the12SMEFT}
\end{align}
and the discussion of the preceding section would apply as it is with the understanding that $\theta_{i}\to \bar\theta_i$.

\medskip

Let us nevertheless restrict to the original SM with no explicit $(B+L)$-violating terms in the Lagrangian and deal with the unphysical $\vec u_{B+L}$ direction. As with any other fermion rotation that affects $\vec\theta$, 
there must be a consistency between the period of the rotation angle given by $\alpha=6\pi$ in~\eqref{eq:BLshift}, and the period of $\vec\theta$ angles in $\theta$-space,
\begin{align}
6\pi\left(-36Q_{q_L}^2,2\,,\,0\right)\mod\left\{ \Delta\vec\theta^{(p)}\right\}=0\,,
\end{align}
which is true for all $p$ by virtue of $Q_{q_L}/\Qmin=\mathbb{Z}$.

The consequence of this shift is that the distinction between any two points connected by a trajectory along $u_{B+L}$ is unphysical. To avoid making unphysical predictions one can take an orthogonal direction, i.e. a linear combination of $\theta_1,\theta_2$ invariant under a $B+L$ shift, this would be
\begin{align}
    \theta_\perp& = \frac{\theta_1}{(6\Qq)^2}+\frac{\theta_2}{2}\,,
    \label{eq:thetaperp}
\end{align} 
as was already noted in Ref.~\cite{Tong:2017oea}.

Selecting such a direction clashes with the Witten effect prediction of the previous section and naively renders it  unphysical.
Take the $U(2)\times SU(3)$ case for illustration, with Witten effect~\eqref{eq:255}-\eqref{eq:256},
\begin{align}
    n_2&=\frac{1}{ 2\pi}\left((q_m/3+\ktwo\mod 2)\theta_2-\frac{\theta_{1}}{2 Q_{\mathrm{min}}^2}(q_m/3 \mod 2)\right)=\frac{\mhex}{3(2\pi)}\left[\theta_2-\frac{\theta_1}{2\Qmin^2} \right].
\end{align}
This expression is not invariant under the $B+L$ shift of Eq.~\eqref{eq:BLshift}, yet it is suggestive of a modified version that could be invariant; if instead of taking the mod of the fluxes one could select an appropriate $SU(2)_L$ and $U(1)_Y$ flux pair the resulting Witten effect could be independent of a $B+L$ shift.

To pursue this line of thought while having a physical interpretation we begin by treating the $\tilde T_{3L}$ direction as Abelian, since for an Abelian gauge symmetry different fluxes are topologically distinct as opposed to $SU(N)$, where they are only different mod $N$. In particular, since our normalisation for covariant derivative is
\begin{align}
    D_\mu\,=\,\partial_\mu+i\frac{g}{2}\tilde T_{aL} W^a_\mu +ig'Q_Y B_\mu\,, \qquad \mathcal L\supset&\frac{\theta_2g^2}{32\pi^2}W^a_{\mu\nu}\tilde W^{a\,\mu\nu}+\frac{\theta_1(g')^2}{16\pi^2}B_{\mu\nu}\tilde B^{\mu\nu}\,,
\end{align}
we can borrow the Abelian result identifying a second charge $Q'=\tilde T_{3L}/2$ with Abelian angle $\theta_2'=\theta_2/2$. We parametrise the group direction with $\gamma$ as $Q_\perp=Q_Y/\Qmin+\tan\gamma (\tilde T_{3L}/2)$. One has a quantisation condition for fluxes in this $Q_\perp$ direction
\begin{align}
    &e^{i\Fl_1 Q_Y+i\Fl_2 (\tilde T_{3L}/2)}= e^{2\pi q_m^\perp i (Q/\Qmin+\tan\gamma (\tilde T_{3L}/2))}=1\,, &    &\Fl=2\pi q_m^\perp\left(\frac1{\Qmin}\,,\,\tan\gamma\right)\,,\label{eq:QcondPerp}
\end{align}
and Witten effect on a 't Hooft line $\tHft_{q^\perp_m}$ of flux parametrized by $q_m^\perp$ as above,
\begin{align}
    \left(\frac{Q_Y}{\Qmin}+\tan\gamma\frac{\tilde T_{3L}}{2}\right)\tHft_{q^\perp_m}=q_m^\perp\left(-\frac{\theta_1}{2\pi \Qmin^2}-\tan\gamma^2\frac{\theta_2'}{2\pi}\right)\tHft_{q_m^\perp}=\frac{-q_m^\perp}{2\pi\Qmin^2}\left(\theta_1+(\Qmin\tan\gamma)^2\frac{\theta_2}{2}\right)\tHft_{q_m^\perp}\,.
\end{align}
For this linear combination to be proportional to $\theta_\perp=(\theta_1+(6Q_{q_L})^2\theta_2/2)/(6Q_{q_L}^2)$ the required $\gamma$ is
\begin{align}
    (\tan\gamma)=\pm\left(\frac{6Q_{q_L}}{\Qmin}\right)\,.
\end{align}
The sign ambiguity can be fixed by realising that $\tilde T_{3L}$ has as many positive as negative entries and so the sign simply relabels states, which leads us to a charge
\begin{align}
    Q_\perp=\frac{Q_Y}{\Qmin}+\frac{6Q_{q_L}}{\Qmin}\frac{\tilde T_{3L}}{2}\,,\label{eq:perpem}
\end{align}
consistent with the quantisation conditions of Eq.~\eqref{eq:QcondPerp}. 

The charge we have solved for is in fact none other than the QED electric charge rescaled,
\begin{align}
   Q_\perp= \frac{6Q_{q_L}}{\Qmin}\Qem\,,
\end{align}
as we will show in the next section. From now on we will also identify the 
physical direction $\theta_\perp$ with the QED $\theta$-parameter $\theta_{\textrm{em}}$ (we will return to this point in the discussion leading to Eq.~\eqref{eq:perp=em-2}),
\begin{align}
\label{eq:perp=em}
    \theta_\perp=\theta_{\textrm{em}}\,.
\end{align}

This is a remarkable result, so let us outline how we arrived here. Given only the fermionic content of the SM one identifies $U(1)_{B+L}$ as an anomalous symmetry respected at the classical level which makes a combination of $\theta_1,\theta_2$ unphysical; the orthogonal physical direction in $\theta$ space, through the Witten effect, selects electromagnetic charge as the physical charge in the electroweak sector. The role of the Higgs is to realise one compatible theory by providing masses for fermions, but one could envision a theory without a Higgs or where the Higgs does not trigger EWSB in which electromagnetism is nevertheless singled out as the physical charge. 

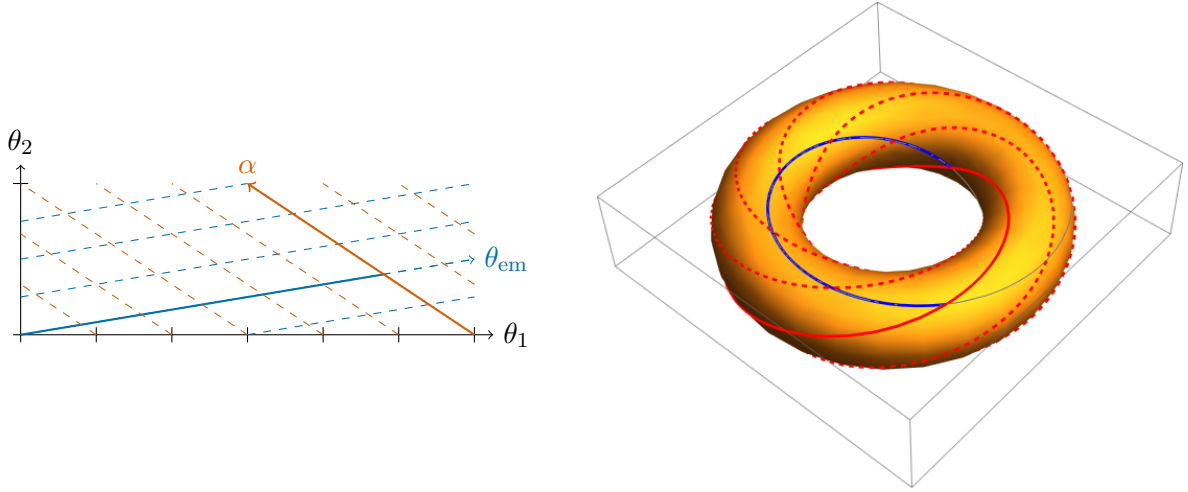
\begin{figure}[h!]
    \centering
     \raisebox{50pt}{\begin{tikzpicture}
        \draw[->] (-0.1,0) -- (6.25,0)node[right] {$\theta_1$};
         \foreach \x in {1,...,6}
        \draw (\x,0.1) -- (\x,-0.1);
        \draw [OIblue,thick] (0,0) -- (4.81,0.8);
        \draw[->] (0,-0.1) -- (-0,2.25) node[above] {$\theta_2$};
        \draw (0.1,2) -- (-0.1,2);
         \draw [OIblue,dashed] (3,0) -- (6,0.5);
         \draw [OIblue,dashed,->] (0,0) -- (6,1) node[right]{$\theta_{\textrm{em}}$};
         \draw [OIblue,dashed] (0,0.5) -- (6,1.5);
         \draw [OIblue,dashed] (0,1.5) -- (3,2);
        \foreach \x in {1,...,2}
         \draw [OIblue,dashed] (0,2*\x/2-1) -- (6,2*\x/2);
         \draw [OIvermilion,thick,->] (6,0) -- (3,2) node[above] {$\alpha$};
         \draw [OIvermilion,dashed] (3,0) -- (0,2);
         \draw [OIvermilion,dashed] (1,0) -- (0,2/3);
         \draw [OIvermilion,dashed] (2,0) -- (0,4/3);
         \draw [OIvermilion,dashed] (4,0) -- (1,2);
         \draw [OIvermilion,dashed] (5,0) -- (2,2);
         \draw [OIvermilion,dashed] (6,2/3) -- (4,2);
         \draw [OIvermilion,dashed] (6,4/3) -- (5,2);
    \end{tikzpicture}}\qquad
    \includegraphics[width=0.45\textwidth]{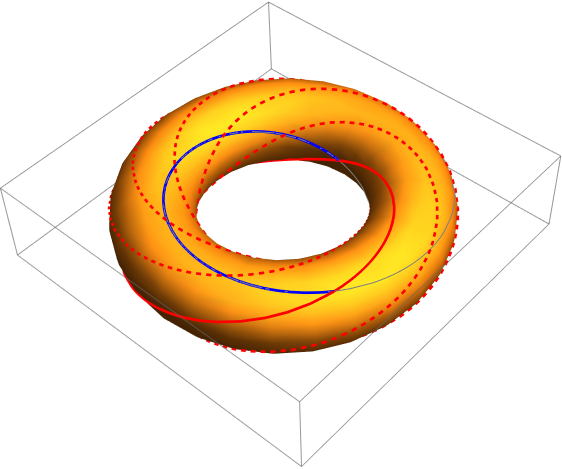}
    \caption{LHS: Fundamental (defined as in Fig.~\ref{fig:DomainG6}) and its projection $(\theta_1,\theta_2)\to\theta_{\textrm{em}}$ for $G_6$ and $\Qmin/\Qq=1$. The thick blue line is the range of $\theta_{\textrm{em}}$ one projects onto and its defined by the intersection with the thick red line which takes all points along it to the origin; the dashed blue line would continue the $\theta_{\textrm{em}}$ line past its true period and wrap back to the beginning. Analogously all points along a red dashed line would be projected to their intersection with the blue line, noting that lines right of the red solid line continue on to the left side given the periodicities. RHS: The same set up now visualised as a 2-torus in 3d.}
    \label{fig:torustrip}
\end{figure}

We now turn to the practical problem of the projection onto $(\theta_{\textrm{em}}, \theta_3)$ space from $(\theta_1,\theta_2,\theta_3)$.
The new angle $\theta_{\textrm{em}}$ is itself periodic, and one can solve for its period. Decompose the plane into the physical and $B+L$ direction
\begin{align}
    (\theta_1,\theta_2)\,=\,\vec{u}_{B+L}\,\alpha\,+\,\frac{\vec{u}_\perp}{u_\perp^2}\,\theta_{\textrm{em}}\,, \qquad
\vec u_\perp \,=\,\left(\frac{1}{(6\Qq)^2}\,,\,\frac12\right)\,,
\end{align}
where $\theta_{\textrm{em}}=\vec\theta\cdot\vec u_\perp$, the vector $\vec u_{B+L}$ is defined in Eq.~\eqref{eq:BLshift} and $\vec u_{B+L}\cdot\vec u_\perp=0$. These two directions have periodicities that can be found given the periodicities of $\theta_1$, $\theta_2$. Visually, $(\theta_1,\theta_2)\sim T^2$ so taking a trajectory with say $\alpha=0$ and changing $\theta_{\textrm{em}}$ traces a line around the torus which will eventually come back to the starting point. The same is true for $\alpha$, whose periodicity we know beforehand, but this direction is unphysical:
any point connected to $(\theta_1,\theta_2)=(0,0)$ by a trajectory in $\alpha$ only is equivalent to the origin (which is to say that the Witten effect cancels for such sets of points). The directions of $\theta_{\textrm{em}}$ and $\alpha$ are defined to be orthogonal so one might be tempted to conclude that all points along $\alpha=0$, $\theta_{\textrm{em}}\neq 0$ are physically distinct  points, up until the period of $\theta_{\textrm{em}}$ found as described above. On the torus, however, two orthogonal lines intersect at more than one point, unlike on the plane. This means that if the $\alpha$-axis intersects the $\theta_{\textrm{em}}$ axis before its stand-alone period, the true period of $\theta_{\textrm{em}}$ will be shorter and given by said intersection. This is always the case and it is this intersection of axes that gives the period, with a visualisation of the projection in Fig.~\ref{fig:torustrip}.
One can solve for this periodicity  as
\begin{align}\label{eq:graydot}
    \frac{\vec{u}_\perp}{u_\perp^2}\,\Delta\theta_{\textrm{em}}\equiv\vec{u}_{B+L}\alpha+(2\pi(p\Qmin)^2,0)\,,
\end{align}
with solution
\begin{align}
    \Delta\theta_{\textrm{em}}=2\pi\frac{(p \Qmin)^2}{(6\Qq)^2}\,.
\end{align}
This leads us to the relation that defines the projection as
\begin{align}
    \theta_{\textrm{em}}(\theta_1,\theta_2)\equiv\frac{\theta_1}{(6\Qq)^2}+\frac{\theta_2}{2}\mod 2\pi\frac{(p \Qmin)^2}{(6\Qq)^2}\,.\label{eq:Project}
\end{align}
Deriving this relation has made use of the periodicity in the $\theta_1$ direction only. One might wonder if the periodicities of certain groups contained in the $(\theta_1,\theta_2)$ plane will introduce an even smaller period. Let us check $G_2$ and the vector $\Delta\vec\theta^{(2)}_I=(4\pi\Qmin^2,2\pi,0)$
\begin{align}
    \theta_{\textrm{em}}(\Delta\vec\theta^{(2)}_I)=2\pi\left(\frac{2\Qmin^2}{(6\Qq)^2}+\frac{1}{2}\mod \frac{(\Qmin)^2}{(3\Qq)^2}\right)=\frac{2\pi \Qmin^2}{(3\Qq)^2}\left(\frac{1+(1+2k)^2}{2}\mod 1\right)=0\,,
\end{align}
where we used $k\in\mathbb{Z}$ and the fact that the square of an odd number plus one is even (but not a multiple of four, a result necessary to map the CP point $\theta^{(0,1,0)}$ of Eq.~\eqref{eq:CPG2}). The fact that we obtain a null result means this period vector is redundant and degenerate with $\Delta\vec\theta^{(2)}_{II}=(8\pi\Qmin^2,0,0)$ after the projection. To collect the rest of the results here while displaying the lattice structure, let us introduce
\begin{align}
\left(\theta_{\textrm{em}}\,,\,\theta_3\right)=&\left(2\pi\frac{p\Qmin^2}{(6\Qq)^2}L_{\perp}\,,\,2\pi L_3\right)\,,
\end{align}
with
\begin{align}
L_\perp&=L_1+\frac{18\Qq^2}{p\Qmin^2}L_2 \mod p    =L_1+\frac{18(1+p\,k)^2}{p}L_2 \mod p\,.
\end{align}
Where to avoid confusion with the 3d $\theta$ space we will use a prime when using vector notation
\begin{align}
    \vec\theta^{\,\prime}\,=\,(\theta_{\textrm{em}},\theta_3)\,, \qquad \vec{L}'&=(L_\perp,L_3)\,.
\end{align}
We note that $L_i$, which were introduced to capture the group structure independent of the $\Qmin$ choice, now seem to have a $\Qmin$ dependence; this is not so and due to the mod factor, as we will show.
The periodicities are now, for each group,
\begin{align}
G_1: & & \left\{\Delta\theta^{\prime(1)}\right\}&=\left\{\begin{array}{lc} 
(\pi \Qmin^2/(18\Qq^2) & 0) \\
(0 & 2\pi)
\end{array}\right\}\,,
& \left\{\latD^{\prime(1)}\right\}&=\left\{\begin{array}{cc} 
(1 & 0) \\
(0 & 1)
\end{array}\right\}\,,
\\
G_2:&  & \left\{\Delta\theta^{\prime(2)}\right\}&=\left\{\begin{array}{lc}
(2\pi \Qmin^2/(3\Qq)^2 & 0) \\
(0 & 2\pi)
\end{array}\right\}\,,& \left\{\latD^{\prime(2)}\right\}&=\left\{\begin{array}{cc}
(2 & 0) \\
(0 & 1)
\end{array}\right\}\,,\\
G_3: & & \left\{\Delta\theta^{\prime(3)}\right\}&=\left\{\begin{array}{cc}
(\pi\Qmin^2/(6\Qq^2) & 2\pi) \\
(\pi\Qmin^2/(2\Qq^2) & 0)
\end{array}\right\}\,,& \left\{\latD^{\prime(3)}\right\}&=\left\{\begin{array}{cc}
(1 & 1) \\
(3 & 0)
\end{array}\right\}\,,\\
G_6: & & \left\{\Delta\theta^{\prime(6)}\right\}&=\left\{\begin{array}{cc}
(2\pi\Qmin^2/(3\Qq^2) & 2\pi) \\
(2\pi\Qmin^2/\Qq^2 & 0)
\end{array}\right\}\,,& \left\{\latD^{\prime(6)}\right\}&=\left\{\begin{array}{cc}
(2 & 1) \\
(6 & 0)
\end{array}\right\}\,,
\end{align}
where one can observe that $G_2$ ($G_6$) is a $L_\perp$-rescaled version of $G_1$ ($G_3$).
The dyonic lattices map one-to-one in this smaller space
\begin{align}
   \vthedyb^{\,\, \prime (p), \ell} 
   \to\, 2\pi\frac{p\Qmin^2}{(6\Qq)^2}\left(p-\ell \mod p\,,\,0\right)\,,
\end{align}
yet as we will see not all of them are dyonic lattices in the reduced group $U(1)_\perp\times SU(3)_c$ and so they lose their singular status.

The CP invariant points do become degenerate, unavoidably so, since one goes from $2^{3}$ to $2^2$ such points. In the projected space, the CP invariant points can be found from the periodicities, in the order $\theCP^{(1,0)},\theCP^{(0,1)}$, their basis is
\begin{align}\label{eq:CPproj}
G_1:& & \latCPp{1}^{\prime}&=\left\{(1/2,0)\oplus (0,1/2)\right\}\,, &\theCPp{1}^{\prime}&=\left\{(\pi \Qmin^2/(36\Qq^2),0)\oplus (0,\pi)\right\}\,,\\
G_2:& & \latCPp{2}^{\prime}&=\left\{(1,0)\oplus (0,1/2)\right\}\,,&\theCPp{2}^{\prime}&=\left\{(\pi \Qmin^2/(3\Qq)^2,0)\oplus (0,\pi)\right\}\,,\\
G_3:& & \latCPp{3}^{\prime}&=\left\{(3/2,0)\oplus (1/2,1/2)\right\}\,,&\theCPp{3}^{\prime}&=\left\{ (\pi\Qmin^2/(4\Qq^2),0)\oplus (\pi\Qmin^2/(12\Qq^2),\pi)\right\}\,,\\
G_6:& & \latCPp{6}^{\prime}&=\left\{(3,0)\oplus (1,1/2)\right\}\,,&\theCPp{6}^{\prime}&=\left\{(\pi\Qmin^2/\Qq^2,0)\oplus (\pi\Qmin^2/(3\Qq^2),\pi)\right\}\,,
\end{align}
with the 4 possibilities marked by $\theCP^{(i,j)}$ with $i,j=0,1$.
The correspondence with CP points pre-projection is
\begin{align}\label{eq:CPconnect}
G_1: & & \latCPp{1}^{(1,0,0)}&\to \latCPp{1}^{\prime(1,0)}\,, & \latCPp{1}^{(0,0,1)}&\to \latCPp{1}^{\prime(0,1)}\,,\\
& & \left[(1/2,0,0)\right]&\to (1/2,0) \,,& \left[(0,0,1/2)\right]&\to (0,1/2)\,,\\
G_2:& & \latCPp{2}^{(1,0,0)}\,,\latCPp{2}^{(0,1,0)}\,&\to \latCPp{2}^{\prime(1,0)} \,,& \latCPp{2}^{(0,0,1)}&\to \latCPp{2}^{\prime(0,1)}\,,\\
& & \left[(1,0,0)\,,(1/2,1/2,0)\right]&\to (1,0)\,, & \left[(0,0,1/2)\right]&\to (0,1/2)\,,\\
G_3:& & \latCPp{3}^{(1,0,0)}\,&\to \latCPp{3}^{\prime(1,0)}\,, & \latCPp{3}^{(0,1,0)}&\to \latCPp{3}^{\prime(0,1)}\,,\\
& & \left[(3/2,0,0)\right]&\to (3/2,0)\,, & \left[(1/2,0,1/2)\right]&\to (1/2,1/2)\,,\\
G_6:& &\latCPp{6}^{(1,0,0)}\,,\latCPp{6}^{(0,0,1)}\,&\to \latCPp{6}^{\prime(1,0)} \,,& \latCPp{6}^{(0,1,0)}&\to \latCPp{6}^{\prime(0,1)}\,,\\
& & \left[(3,0,0)\,,(3/2,1/2,0)\right]&\to (3,0)\,, & \left[(1,0,1/2)\right]&\to (1,1/2)\,,
\end{align}
with the case of $\latCPp{3}^{(0,0,1)}=(0,1/2,0)$ in $G_3$ mapped to $(0,0)$ and $\latCPp{1}^{(0,1,0)}=(0,1/2,0)$ for $G_1$ going to $(0,0)$. In this projection no dependence on $k$ (i.e. on $\Qmin$) is left and we have that $L_\perp,L_3$ do describe the group structure, independently of $\Qmin$.

While this was largely a mapping exercise to track the original special values of $\theta$, one general feature that can be extracted is that CP invariant points remain so after the projection but the integer dyonic lattices might not.

\begin{figure}[h!]
    \centering
    \begin{tikzpicture}
     \draw [->] (-0.1,0) -- (3.25,0) node[right] {$\theta_{\textrm{em}}$} ;
         \draw [->] (0,-0.1) -- (0,3.25) node[above] {$\theta_3$};
         \filldraw [gray,dashed,opacity=0.2] (0,0) -- (3,0) -- (3,3) -- (0,3) -- (0,0);
         \draw [OIgreen,<->] (0,.5) -- (3,0.5);
         \draw [OIgreen,<->] (.5,0) -- (0.5,3);
         \filldraw (0,0) ellipse (1pt and 3pt) ;
        \filldraw (0,0) ellipse (3pt and 1pt) ;
        \node at (1.5,3.5) {\Large $G'_1$};
         \filldraw (1.5,0)  ellipse (3pt and 1pt) node[below] {$\theCP^{\prime(1,0)}$};
         \filldraw (0,1.5) ellipse  (3pt and 1pt) node[left] {$\theCP^{\prime(0,1)}$};
         \filldraw (1.5,1.5) ellipse (3pt and 1pt)node[above] {$\theCP^{\prime(1,1)}$};
    \end{tikzpicture}\qquad
    \begin{tikzpicture}
     \draw [OIgreen,<->] (0,2.5) -- (6,2.5);
         \draw [OIgreen,<->] (0,0) -- (2,2);
        \draw[->] (-0.1,0) -- (6.25,0) node[right] {$\theta_{\textrm{em}}$};
        \draw[->] (0,-0.1) -- (0,2.75) node[above] {$\theta_3$};
        \filldraw [gray,dashed,opacity=0.2] (0,0) -- (6,0) -- (6,2) -- (0,2) -- (0,0);
        \filldraw (0,0) ellipse (1pt and 3pt) ;
        \filldraw (0,0) ellipse (3pt and 1pt) ;
        \node at (3,3.5) {\Large $G'_3$};
        \filldraw (2,0) ellipse (1pt and 3pt) node[below] {$\thedyp{3'}{2}$};
        \filldraw (4,0) ellipse (1pt and 3pt) node[below] {$\thedyp{3'}{1}$};
         \filldraw (1,1) ellipse (3pt and 1pt) node[right] {$\theCP^{\prime(0,1)}$};
         \filldraw (3,0)  ellipse (3pt and 1pt)node[below] {$\theCP^{\prime(1,0)}$};
         \filldraw (4,1)  ellipse (3pt and 1pt)node[above] {$\theCP^{\prime(1,1)}$};
    \end{tikzpicture}
    \caption{Fundamental domain in $\theta$ space for $G^\prime_1$ (LHS) and $G^\prime_3$ (RHS) in gray, with special points signposted and in accordance with the notation of the main text. The dimensions of the gray rectangle marking the domain are $(2\pi(\Qemin)^2,2\pi)$ for $G'_1$ and $(2\pi(3\Qemin)^2,2\pi)$ for $G'_3$.}
    \label{fig:EMQCDtheta}
\end{figure}
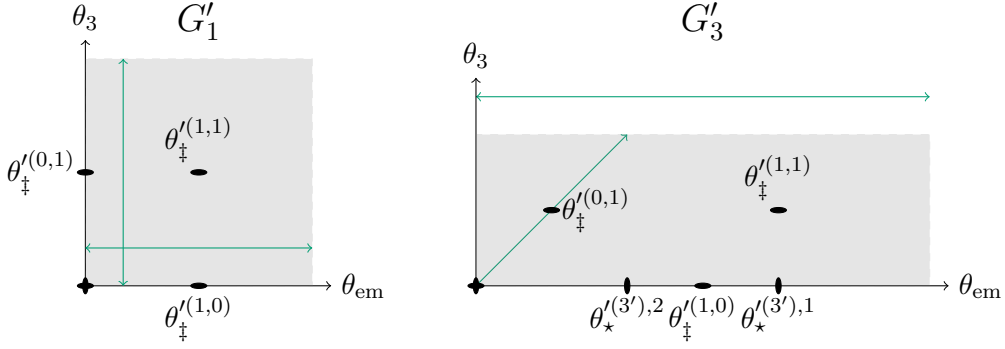



\subsection{Low energy theory and electro-weak symmetry breaking}\label{sec:EWSB}

Let us analyse the low energy theory with gauge symmetry
\begin{align}G'_{p'}\equiv \frac{U(1)_{\textrm{em}}\times SU(3)_c}{Z_{p'}}\,,
\qquad p'=1,3\,,
\end{align}
wilfully ignoring the connection with the SM group for now. The analysis of $\theta$ space in this group can be carried out simply knowing the group and its centre, which is $Z_3$ and is generated by
\begin{align}
    \xi^\prime= \exp\left(\frac{2\pi i}{3}\left[\frac{\Qem}{\Qemin}-\tilde \lambda_8\right]\right).
\end{align}
The two theta angles of this theory are physical, the $\theta_3$ angle in its subtracted form of Eq.~\eqref{eq:the3phys}, and $\theta_{\textrm{em}}$, which is unaffected by a $B+L$ shift given both $U(1)_{\textrm{em}}$ and $U(1)_{B+L}$ symmetries are non-chiral and the triangle diagram vanishes. Our conventions are as in previous sections, i.e. 
\begin{align}
    \mathcal L&\supset\, \frac{\theta_3g_s^2}{32\pi^2}\,G^a_{\mu\nu}\tilde G^{a\,\mu\nu}\,+\,\frac{\theta_{em}e^2}{16\pi^2}\,F_{\mu\nu}\tilde F^{\mu\nu}\,, & D_\mu&=\partial_\mu+ie\Qem A_\mu+ig_s\lambda G_\mu\,,
\end{align}
with $\lambda=\tilde\lambda/\sqrt{12}$.
Analysis of the period, CP invariant points and dyonic lattice points follows the same method of the previous subsection so let us here simply collect the results. 

In analogy with the previous subsection we define
\begin{align}
    \theta_{\textrm{em}}=2\pi p'(\Qemin)^2 L_{\textrm{em}}\,.
\end{align}

\begin{figure}[h!]
\centering
\hspace{5mm}
\begin{tikzpicture}[scale=0.85]
  \node at (1.6,3.2) {$G'_1$};
  \draw[->] (-0.5,0) -- (3,0) node[right] {$n$};
  \draw[->] (0,-0.5) -- (0,3) node[above] {$n_m$};
  \node[left] at (-0.1,0.2) {$0$};
  \node[left] at (-0.1,1) {$1$};
  \node[left] at (-0.1,2) {$2$};
  \foreach \x in {1,...,2}
    \node[below] at (\x,-0.1) {\x};
  \node[below] at (0.2,-0.1) {0};
  \foreach \x in {0,...,2} {
    \foreach \y in {0,...,2} {
      \draw (\x,\y) circle (3pt);
    }
  }
  \fill (0,0) circle (3pt);
  \fill (1,0) circle (3pt);
  \fill (2,0) circle (3pt);
\end{tikzpicture}
\\
\begin{tikzpicture}[scale=0.85]
  \node at (1.6,3.2) {$G'_3$};
  \draw[->] (-0.5,0) -- (3,0) node[right] {$n$};
  \draw[->] (0,-0.5) -- (0,3) node[above] {$n_m$};
  \node[left] at (-0.1,0.2) {$0$};
  \node[left] at (-0.1,1) {$1$};
  \node[left] at (-0.1,2) {$2$};
  \foreach \x in {1,...,2}
    \node[below] at (\x,-0.1) {\x};
  \node[below] at (0.2,-0.1) {0};
  \foreach \x in {0,...,2} {
    \foreach \y in {0,...,2} {
      \draw (\x,\y) circle (3pt);
    }
  }
  \fill (0,0) circle (3pt);
  \fill (0,1) circle (3pt);
  \fill (0,2) circle (3pt);
  \node at (1.2,-1.2) {$\theCPp{3'}^{\prime(0,i)}$};
\end{tikzpicture}
\hspace{5mm}
\begin{tikzpicture}[scale=0.85]  
\node at (1.6,3.2) {$G'_3$};
  \draw[->] (-0.5,0) -- (3,0) node[right] {$n$};
  \draw[->] (0,-0.5) -- (0,3) node[above] {$n_m$};
  \node[left] at (-0.1,0.2) {$0$};
  \node[left] at (-0.1,1) {$1$};
  \node[left] at (-0.1,2) {$2$};
  \foreach \x in {1,...,2}
    \node[below] at (\x,-0.1) {\x};
  \node[below] at (0.2,-0.1) {0};
  \foreach \x in {0,...,2} {
    \foreach \y in {0,...,2} {
      \draw (\x,\y) circle (3pt);
    }
  }
  \fill[OIblue] (0,0) circle (3pt);
  \fill[OIblue] (1,1) circle (3pt);
  \fill[OIblue] (2,2) circle (3pt);
  \node at (1.2,-1.2) {$\thedyp{3'}{1}$};
\end{tikzpicture}
\hspace{5mm}
\begin{tikzpicture}[scale=0.85]
\node at (1.6,3.2) {$G'_3$};
\draw[->] (-0.5,0) -- (3,0) node[right] {$n$};
  \draw[->] (0,-0.5) -- (0,3) node[above] {$n_m$};
  \node[left] at (-0.1,0.2) {$0$};
  \node[left] at (-0.1,1) {$1$};
  \node[left] at (-0.1,2) {$2$};
  \foreach \x in {1,...,2}
    \node[below] at (\x,-0.1) {\x};
  \node[below] at (0.2,-0.1) {0};
  \foreach \x in {0,...,2} {
    \foreach \y in {0,...,2} {
      \draw (\x,\y) circle (3pt);
    }
  }
  \fill[OIvermilion] (0,0) circle (3pt);
  \fill[OIvermilion] (2,1) circle (3pt);
  \fill[OIvermilion] (1,2) circle (3pt);
  \node at (1.2,-1.2) {$\thedyp{3'}{2}$};
\end{tikzpicture}
\hspace{5mm}
\begin{tikzpicture}[scale=0.85]
\node at (1.6,3.2) {$G'_3$};
\draw[->] (-0.5,0) -- (3,0) node[right] {$n$};
  \draw[->] (0,-0.5) -- (0,3) node[above] {$n_m$};
  \node[left] at (-0.1,0.2) {$0$};
  \node[left] at (-0.1,1) {$1$};
  \node[left] at (-0.1,2) {$2$};
  \foreach \x in {1,...,2}
    \node[below] at (\x,-0.1) {\x};
  \node[below] at (0.2,-0.1) {0};
  \foreach \x in {0,...,2} {
    \foreach \y in {0,...,2} {
      \draw (\x,\y) circle (3pt);
    }
  }
  \fill[OIgreen] (0,0) circle (3pt);
  \fill[OIgreen] (1.5,1) circle (3pt);
  \fill[OIgreen] (0,2) circle (3pt);
  \node at (1.2,-1.2) {$\theCP^{\prime(1,i)}$};
\end{tikzpicture}
\caption{Dyonic lattices for $G'_1$ and $G'_3$ theories that follow from the special theta-values collected in Fig.~\ref{fig:EMQCDtheta}\label{fig:CPEWSB}. The first, second and fourth of the spectral lattices presented are CP-invariant theories. The fourth lattice includes half-integer values of $n$.}
\end{figure}
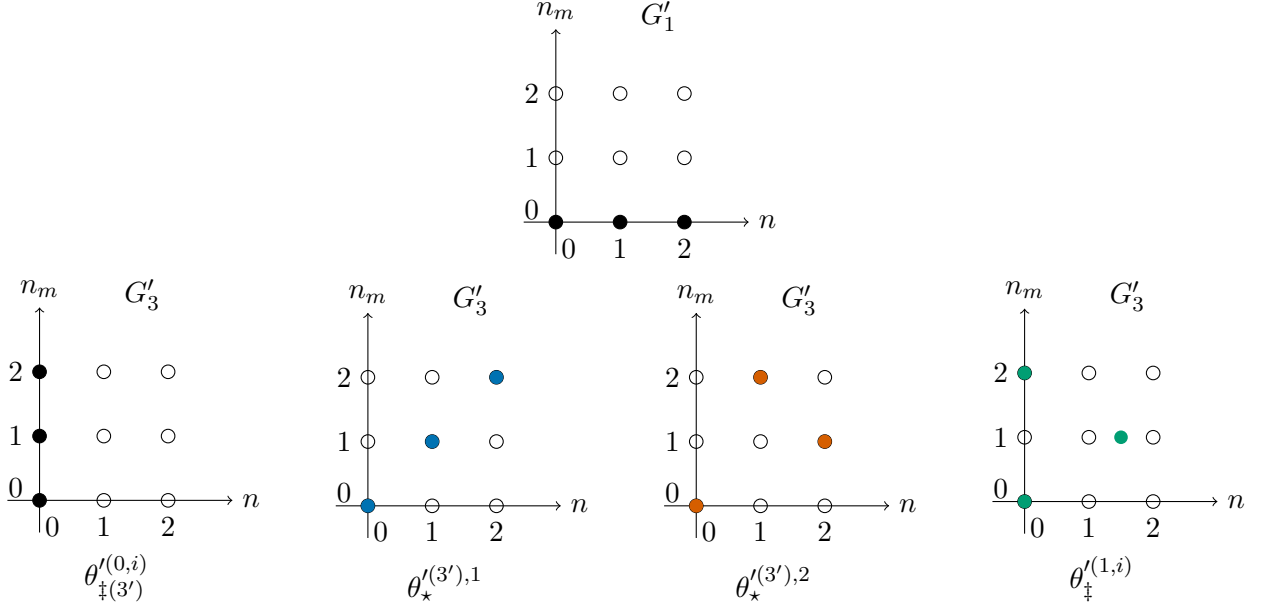

The case of $U(3)$ has flux quantisation condition as
\begin{align}
    e^{i\Fl_{\textrm{em}} Q_{\textrm{em}}+i\Fl_3 \tilde \lambda_8}=\left[e^{2\pi i/3(\Qem/\Qemin-\tilde \lambda_8)}\right]^{n_3^{m\prime}}\,,
\end{align}
with $n_3^{m\prime}=0,1,2$, solutions as
\begin{align}
    \vec \Fl\,=\,2\pi\left(\frac{q_m'}{3\Qemin},-\frac{q_m'+\kthree}{3}\right) \,,\qquad q'_m\in\mathbb{Z}\,,\,\kthree\in 3 \mathbb{Z}\,,
\end{align}
and Witten effect
\begin{align}
    \xi'\tHft_{q_m'}=\exp\left(\frac{q_m\mod 3}{3}\left[\theta_3-\frac{\theta_{em}}{3 (\Qemin)^2}\right]\right)\tHft_{q_m'}\equiv e^{2\pi in_3'/3}\tHft_{q_m'}.
\end{align}
From the Witten effect we extract the periodicity, 
\begin{align}
\label{eq:PerG3p}
    G'_{3}:& &\left\{\Delta\theta^{'(3')}\right\}&=\left\{\begin{array}{lc}
        (\,6\pi (\Qemin)^2\,,  &\,2\pi\,)\\
        (\,18\pi (\Qemin)^2\,,&0\,)
        \end{array}\right\}\,,&\left\{\latD^{'(3')}\right\}&=\left\{\begin{array}{lc}
        (\,1\,,  &\,1\,)\\
        (\,3\,,&0\,)
        \end{array}\right\}\,,
\end{align}
where we use $(3')$, $(1')$ as index for $G'_3,G'_1$ in order to avoid confusion with $G_3,G_1$ in the previous subsection. Our choice of domain
\begin{align}
    0&\leq\theta_{\textrm{em}}<2\pi(3\Qemin)^2\,, & 0&\leq \theta_3<2\pi\,,\\
    0&\leq L_{\textrm{em}}<3 \,,& 0&\leq L_3<1\,,
\end{align}
while CP invariant points are, in the order $\{\latCPp{3'}^{(1,0)},\latCPp{3'}^{(1,0)}\}$
\begin{align}
    G'_{3}:& &\theCPp{3'}^\prime&=\left\{
       (\,9\pi (\Qemin)^2\,,0\,) \oplus (\,3\pi (\Qemin)^2\,,  \,\pi\,)
        \right\}\,,&\latCPp{3'}^{\prime}&=\left\{
        (\,3/2\,,0\,)\oplus
        (\,1/2\,,  \,1/2\,)
        \right\}\,,
\end{align}
and dyonic lattices with $\ell=1,2$
\begin{align}
    G_{3}':& & \thedyp{3'}{\ell}&=6\pi(\Qemin)^2(3-\ell,0)\,, & L^{\prime(3')\ell}_{\star}&=(3-\ell,0)\,.\label{eq:thestG3p}
\end{align}
The case of the universal cover $SU(3)_c\times U(1)_{\textrm{em}}$ reads
\begin{align}
    e^{i\Fl Q_{\textrm{em}}+i\Fl_3 \tilde \lambda_8}=\left[e^{2\pi i\Qem/\Qemin}\right]^{n_3^{m\prime}/3}\,,
\end{align}
with $n_3^{m\prime }=0$ and solutions
\begin{align}
    \vec \Fl\,=\,2\pi\left(\frac{q_m'}{3\Qemin},\frac{\kthree}{3}\right)\,, \qquad &q'_m\in3\mathbb{Z}\,,\,\kthree\in 3 \mathbb{Z}\,,
\end{align}
and trivial Witten effect
\begin{align}
    (\xi')^3\tHft_{q_m'}=\exp\left(\frac{q_m\mod 3}{3}\left[-\frac{\theta_{em}}{ (\Qemin)^2}\right]\right)\tHft_{q_m'}\equiv e^{2\pi in_1'}\tHft_{q_m'}=\tHft_{q_m'}\,.
\end{align}
As for the $G_1$ case one can extract the periods
\begin{align}
\label{eq:PerG1p}
    G'_{1}:& & \left\{\Delta\theta^{'(1')}\right\}&=\left\{\begin{array}{lc}
        (\,2\pi(\Qemin)^2\,,  &\,0\,)\\
        (\,0\,,&2\pi\,)
        \end{array}\right\}\,,& \left\{\latD^{'(1')}\right\}&=\left\{\begin{array}{lc}
        (\,1\,,  &\,0\,)\\
        (\,0\,,&1\,)
        \end{array}\right\}\,,
\end{align}
and domain
\begin{align}
    0&\leq\theta_{\textrm{em}}<2\pi(\Qemin)^2\,, & 0&\leq \theta_3<2\pi\,,\\
    0&\leq L_{\textrm{em}}<1\,, & 0&\leq L_3<1\,.
\end{align}
The CP invariant points are given by the basis
\begin{align}
    G'_{1}:& &\theCPp{1'}^\prime&=\left\{
        (\,\pi (\Qemin)^2\,,  \,0\,)\oplus
        (\,0\,,\pi\,)
        \right\}\,,&\latCPp{1'}^\prime&=\left\{
        (\,1/2\,,  \,0\,)\oplus
        (\,0\,,1/2\,)
        \right\}\,.
\end{align}
There are no dyonic lattices in this case. Both $G'_{1,3}$ groups have $\theta$-space characterised in Fig.~\ref{fig:EMQCDtheta} and electric-magnetic lattices as in Fig.~\ref{fig:CPEWSB}.

At this point let us remove our blindfold to the SM and lay out the connection. One has the photon singled out as the direction orthogonal to the massive $Z$ boson, that is
\begin{align}
    \Qem\langle H\rangle=0\,,\qquad \Qem=\frac{Q_Y}{2Q_H}+\frac{1}{2}\tilde T_{3L}=\frac{Q_Y}{6\Qq}+\frac{1}{2}\tilde T_{3L}\,,
\end{align}
from which it follows the connection with $Q_\perp$ of Eq.~\eqref{eq:perpem}.
The rotation to the mass basis and relation between couplings
\begin{align}
    \tan \theta_W\,=\,\frac{6Q_L g'}{g}\,, \qquad e\, =\,\frac{6\Qq gg'}{\sqrt{g^2+(6\Qq g')^2}}\,,
\end{align}
lead to
\begin{align}
    \frac{\theta_2g^2}{32\pi^2}W_{\mu\nu}\tilde W^{\mu\nu}+\frac{\theta_1(g')^2}{16\pi^2}B_{\mu\nu}\tilde B^{\mu\nu} \to&\frac{1}{16\pi^2}\left(\theta_1(g' c_W)^2+\frac{\theta_2}{2}g^2s_{W}^2\right) A_{\mu\nu}\tilde A^{\mu\nu}\\=&\frac{e^2}{16\pi^2}\left(\frac{\theta_1}{(6\Qq)^2}+\frac{\theta_{2}}{2}\right)A_{\mu\nu}\tilde A^{\mu\nu}\,,
\end{align}
so that the two angles are indeed one and the same (as we have already anticipated),
\begin{align}
\label{eq:perp=em-2}
    \theta_\perp=\theta_{\textrm{em}}\,.
\end{align}

This means that we have explored the same $\theta$ space with two approaches: the projection of sec.~\ref{sec:BL}, and the direct analysis of $SU(3)_c\times U(1)_{\textrm{em}}/Z_{p'}$. One can use these two viewpoints both as a check on our derivation and to obtain connections. First, the structure of the periods $\Delta L$ in Eqs.~\eqref{eq:PerG3p}, \eqref{eq:PerG1p} for $G'$ and Eqs.~\eqref{eq:PerG6}, \eqref{eq:PerG3}, \eqref{eq:PerG2}, \eqref{eq:PerG1} for $G$ signals the correspondence $G_{6,3}\to G_3'$, $G_{2,1}\to G_1'$, whereas the length in the $\theta_\textrm{em},\theta_\perp$ direction leads to the relations between minimum charges
\begin{align}
\label{eq:MinimumElectricCharge}
G_6:&&|\Qemin|&=\left|\frac{\Qmin}{3\Qq}\right|=\left|\frac{1}{3(1+6k)}\right|\,,
 &G_3:& &|\Qemin|&=\left|\frac{\Qmin}{6\Qq}\right|=\left|\frac{1}{6(1+3k)}\right|\,,
\\
    G_2:& 
     &|\Qemin|&=\left|\frac{\Qmin}{3\Qq}\right|=\left|\frac{1}{3(1+2k)}\right|\,, & G_1:& & |\Qemin|&=\left|\frac{\Qmin}{6\Qq}\right|=\left|\frac{1}{6(1+k)}\right|\,.\label{eq:MinimumElectricCharge2}
\end{align}
The smallest electromagnetic charge as a function of the group $G_p$ and $\Qmin$ was first given in~\cite{Alonso:2024pmq} and reproduced here in Eqs.~\eqref{eq:MinimumElectricCharge}-\eqref{eq:MinimumElectricCharge2}, and we find that it agrees with the derivation presented here based on the Witten effect and $\theta$-space analysis.

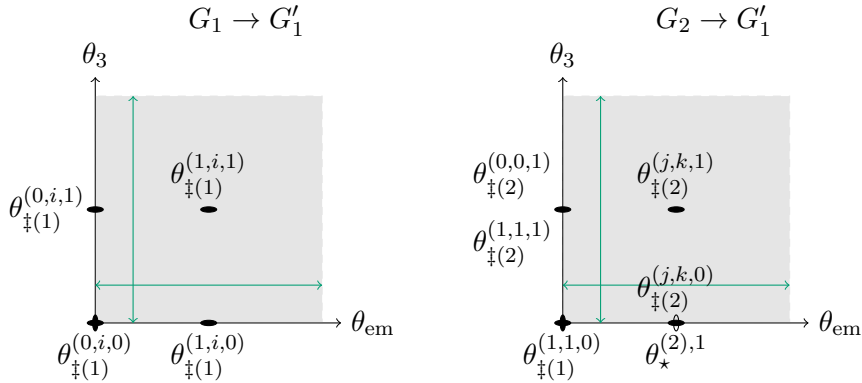
\begin{figure}[h!]
    \centering 
    \begin{tikzpicture}
     \draw [->] (-0.1,0) -- (3.25,0) node[right] {$\theta_{\textrm{em}}$} ;
         \draw [->] (0,-0.1) -- (0,3.25) node[above] {$\theta_3$};
         \filldraw [gray,dashed,opacity=0.2] (0,0) -- (3,0) -- (3,3) -- (0,3) -- (0,0);
         \draw [OIgreen,<->] (0,.5) -- (3,0.5);
         \draw [OIgreen,<->] (.5,0) -- (0.5,3);
         \filldraw (0,0) ellipse (1pt and 3pt) ;
         \filldraw (0,0) ellipse (3pt and 1pt)node[below] {$\theCPp{1}^{(0,i,0)}$};
         \filldraw (1.5,0)  ellipse (3pt and 1pt) node[below] {$\theCPp{1}^{(1,i,0)}$};
         \filldraw (0,1.5) ellipse  (3pt and 1pt) node[left] {$\theCPp{1}^{(0,i,1)}$};
         \filldraw (1.5,1.5) ellipse (3pt and 1pt)node[above] {$\theCPp{1}^{(1,i,1)}$};
         \node at (2,4) {$G_1\to G_1^\prime$};
    \end{tikzpicture}\qquad
     \begin{tikzpicture}
     \draw [->] (-0.1,0) -- (3.25,0) node[right] {$\theta_{\textrm{em}}$} ;
         \draw [->] (0,-0.1) -- (0,3.25) node[above] {$\theta_3$};
         \filldraw [gray,dashed,opacity=0.2] (0,0) -- (3,0) -- (3,3) -- (0,3) -- (0,0);
         \draw [OIgreen,<->] (0,.5) -- (3,0.5);
         \draw [OIgreen,<->] (.5,0) -- (0.5,3);
         \filldraw (0,0) ellipse (1pt and 3pt);
         \filldraw (0,0) ellipse (3pt and 1pt) node[below] {$\theCPp{1}^{(1,1,0)}$};
         \filldraw (1.5,0)  ellipse (3pt and 1pt) node[below] {$\thedy{2}{1}$} node[above] {$\theCPp{2}^{(j,k,0)}$};
         \draw (1.5,0) ellipse (1pt and 3pt);
         \filldraw (0,1.5) ellipse  (3pt and 1pt) node[anchor=south east] {$\theCPp{2}^{(0,0,1)}$}node[anchor=north east] {$\theCPp{2}^{(1,1,1)}$};
         \filldraw (1.5,1.5) ellipse (3pt and 1pt)node[above] {$\theCPp{2}^{(j,k,1)}$};
         \node at (2,4) {  $G_2\to G_1^\prime$};
    \end{tikzpicture}
    \caption{Correspondence of $\theta$ points in the original 3d $\theta$ space to the 2d $\theta$ space after EWSB for $G_1^\prime$. The scale is as in Fig.~\ref{fig:EMQCDtheta} and points that lose their defining property after EWSB are marked as emptied out ellipses, while $i=0,1$ and $(j,k)=(1,0),(0,1)$.}
    \label{fig:ConnectGp1}
\end{figure}

\begin{figure}[h!]
    \centering 
 \begin{tikzpicture}
    \node at (3,3.5) {$G_3\to G_3^\prime$};
    \draw [OIgreen,<->] (0,2.5) -- (6,2.5);
         \draw [OIgreen,<->] (0,0) -- (2,2);
        \draw[->] (-0.1,0) -- (6.25,0) node[right] {$\theta_{\textrm{em}}$};
        \draw[->] (0,-0.1) -- (0,2.75) node[above] {$\theta_3$};
        \filldraw [gray,dashed,opacity=0.2] (0,0) -- (6,0) -- (6,2) -- (0,2) -- (0,0);
        \filldraw (0,0) ellipse (1pt and 3pt);
         \filldraw (0,0) ellipse (3pt and 1pt) node[below] {$\theCPp{3}^{(0,0,i)}$} ;
        \filldraw (2,0) ellipse (1pt and 3pt) node[below] {$\thedy{3}{2}$};
        \filldraw (4,0) ellipse (1pt and 3pt) node[below] {$\thedy{3}{1}$};
         \filldraw (1,1) ellipse (3pt and 1pt) node[right] {$\theCPp{3}^{(0,1,i)}$};
         \filldraw (3,0)  ellipse (3pt and 1pt)node[below] {$\theCPp{3}^{(1,0,i)}$};
         \filldraw (4,1)  ellipse (3pt and 1pt)node[above] {$\theCPp{3}^{(1,1,i)}$};
    \end{tikzpicture}\quad
     \begin{tikzpicture}
        \node at (3,3.5) { $G_6\to G_3^\prime$};
      \draw [OIgreen,<->] (0,2.5) -- (6,2.5);
         \draw [OIgreen,<->] (0,0) -- (2,2);
        \draw[->] (-0.1,0) -- (6.25,0) node[right] {$\theta_{\textrm{em}}$};
        \draw[->] (0,-0.1) -- (0,2.75) node[above] {$\theta_3$};
        \filldraw [gray,dashed,opacity=0.2] (0,0) -- (6,0) -- (6,2) -- (0,2) -- (0,0);
        \filldraw (0,0) ellipse (1pt and 3pt);
         \filldraw (0,0) ellipse (3pt and 1pt) node[below] {$\theCPp{6}^{(1,0,1)}$};
        \draw (1,0) ellipse (1pt and 3pt) node[below] {$\thedy{6}{5}$};
        \filldraw (2,0) ellipse (1pt and 3pt) node[below] {$\thedy{6}{4}$};
        \filldraw (4,0) ellipse (1pt and 3pt)node[below] {$\thedy{6}{2}$};
        \draw (5,0) ellipse (1pt and 3pt) node[below] {$\thedy{6}{1}$};
         \filldraw (1,1) ellipse (3pt and 1pt) node[right] {$\theCPp{6}^{(0,1,0)}$} node[left] {$\theCPp{6}^{(1,1,1)}$};
         \filldraw (3,0)  ellipse (3pt and 1pt) node[below] {$\theCPp{6}^{(j,0,k)}$} node[above] {$\thedy{6}{3}$};
         \draw (3,0)  ellipse (1pt and 3pt);
         \filldraw (4,1)  ellipse (3pt and 1pt)node[above] {$\theCPp{6}^{(j,1,k)}$};
    \end{tikzpicture}
    \caption{Correspondence of $\theta$ points in the original 3d $\theta$ space to the 2d $\theta$ space after EWSB for $G_3^\prime$. The scale is as in Fig.~\ref{fig:EMQCDtheta} and points that lose their defining property after EWSB are marked as emptied out ellipses, while $i=0,1$ and $(j,k)=(1,0),(0,1)$.}
    \label{fig:ConnectGp3}
\end{figure}
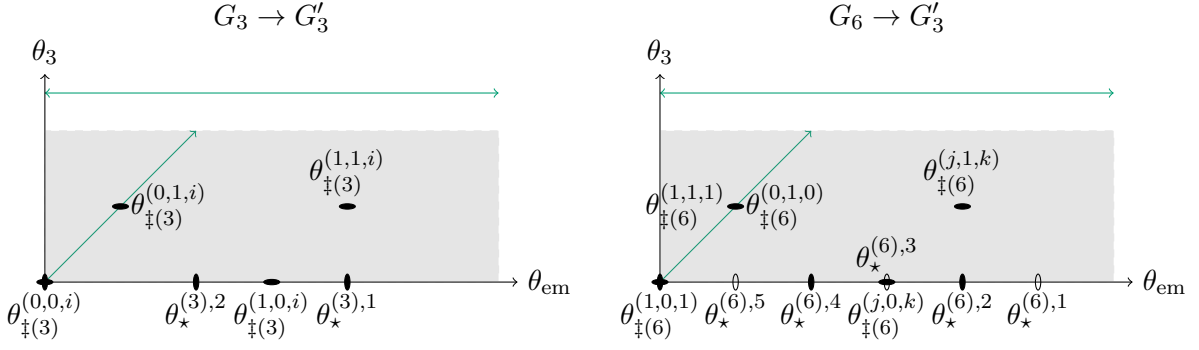

Finally one can compare Fig.~\ref{fig:EMQCDtheta} displaying the special $\theta$ points extracted from the analysis of $G_{3,1}'$ with the relevant $\theta$ points obtained for $G_{1,2,3,6}$ and put through the projection of Eq.~\eqref{eq:Project} in Figs.~\ref{fig:ConnectGp1} and~\ref{fig:ConnectGp3}. This exercise shows explicitly how some dyonic-lattice points $\vec\thedyb$ of $G_{1,2,3,6}$ end up mapped to CP invariant points $\vec\theCP^\prime$ as in the RHS panel of Fig.~\ref{fig:ConnectGp1} or they completely lose their special status as in the RHS panel of Fig.~\ref{fig:ConnectGp3}.

\section{Consequences for the QCD axion}

The physical $\theta$ angles of the Standard Model span a torus whose length depends on the two possible group choices at low energy: $G_3^{\prime}$ or $G_1^{\prime}$, each in turn related to the SM groups $G_{1,2,3,6}$ as shown in the previous section. Experimental observations constrain the $\theta_3$ angle to be smaller than $10^{-10}$ \cite{Abel:2020pzs}, while there is no bound on $\theta_{\textrm{em}}$ so that the observation-compatible region is a thin strip around the horizontal axis in Fig.~\ref{fig:EMQCDtheta}.  To account for the smallness of $\theta_3$ dynamically, one may add a QCD axion $a$ to the Standard Model \cite{Peccei:1977ur,Wilczek:1977pj,Weinberg:1977ma}, whose role is to set the effective value of $\theta_3$ to zero by moving $(\theta_{\rm em}, \theta_3)$ on the $\theta$-torus. The periodicity of the $\theta$-torus therefore constrains axion couplings to the gluon and electromagnetic field strength tensors defined as

\begin{align}
 \mathcal{L} \supset   \frac{Ng_3^2}{16\pi^2}\frac{a}{f}\,G^a_{\mu\nu}\tilde G^{a\,\mu\nu}\,+\, \frac{Ee^2}{16\pi^2}\frac{a}{f}\,F_{\mu\nu}\tilde F^{\mu\nu},
\end{align}
where $f$ is the axion decay constant and $E$ and $N$ are its anomaly coefficients.
As the Nambu-Goldstone boson of a $U(1)$ symmetry, the axion has a physical field range $0\leq a/f<2\pi$. The quantisation of $E$ and $N$ then follows from the periodicity of $(\theta_{\rm em}, \theta_3)$ by considering the full $\theta$ terms
\begin{align}
    \mathcal{L} \, \supset\, &\frac{g_3^2}{32\pi^2}\left(\theta_3+2N\frac{a}{f}\right)G^a_{\mu\nu}\tilde G^{a\,\mu\nu} \,+\, \frac{e^2}{16\pi^2}\left(\theta_{\rm em}+E\frac{a}{f}\right)F_{\mu\nu}\tilde F^{\mu\nu}\,.
\end{align}
It is convenient to define  the rescaled magnitudes
\begin{align}
    \bar a\,\equiv\,\frac{a}{2\pi f}, \qquad N_{DW}\,=\,2N, \qquad E_{DW}\,=\,\frac{E}{p'(\Qemin)^2},
\end{align}
while considering the effective $\theta$ angles including the axion as
\begin{align}
    \vec \theta^{\prime \rm eff} &= (\theta_{\rm em}^{\rm eff}, \theta_3^{\rm eff})\equiv \left(\theta_{\rm em}+E\frac{a}{f}\,,\,\theta_3+2N\frac{a}{f}\right) \label{eq:thetaEff},\\    
    \vec L^{\prime \rm eff} &= (L_{\rm em}^{\rm eff}\,,\, L_3^{\rm eff})\equiv \left(\frac{\theta_{\rm em}^{\rm eff}}{2\pi p'(\Qemin)^2}, \frac{\theta_3^{\rm eff}}{2\pi}\right)= \left(L_{\textrm{em}}+E_{DW}\bar a\,,\, L_3+N_{DW}\bar a \right).
\end{align}
We must thread the axion round the $\theta$-torus so that the periodicities agree, that is
\begin{align}
    (E,2N)&=\mathbb{Z}\frac{\Delta\vec{\theta}_{I}^{(p')}}{2\pi}+\mathbb{Z}'\frac{\Delta\vec{\theta}_{II}^{(p')}}{2\pi} \,,
\end{align}
where the subscripts $I$ and $II$ label the two solutions in 
Eqs.~\eqref{eq:PerG1p} and~\eqref{eq:PerG3p}. 
This is the condition that $S^1\to T^2$ maps  satisfy, and
translated to the rescaled variables reads
\begin{align}
    (E_{DW}\,,\,N_{DW})&=\mathbb{Z}\Delta L^{\prime(p')}_{I}+ \mathbb{Z'}\Delta L^{\prime(p')}_{II}\,.
\end{align}
We thus find for each of the two group realisations after EWSB, 
\begin{eqnarray}
    G_1^\prime: \qquad &(E,2N)\,=\, (\mathbb{Z}(\Qemin)^2,\,\mathbb{Z}'), \qquad &(E_{DW},N_{DW})\,=\,(\mathbb{Z},\mathbb{Z}')\,,\\
    G_3^\prime:\qquad &(E,2N)\,=\, \left(3(\mathbb{Z}+3\mathbb{Z}')(\Qemin)^2,\,\mathbb{Z}\right), \qquad
    &(E_{DW},N_{DW})\,=\,(\mathbb{Z}+3\mathbb{Z}',\mathbb{Z})\,.
\end{eqnarray}
For $G'_3$ there is a correlation between the anomaly coefficients
\begin{align}
    G'_3: \qquad\frac{E}{3(\Qemin)^2}+4N&=E_{DW}+2N_{DW}=3\mathbb{Z}\,,\label{eq:ENquant}
\end{align}
and in turn $\Qemin$ depends on the full SM group $G_p$ and the compositeness degree $k$ as in Eqs.~\eqref{eq:MinimumElectricCharge} and \eqref{eq:MinimumElectricCharge2}. The correlation of Eq.~\eqref{eq:ENquant} was first presented in its explicit form in~\cite{Choi:2023pdp} for $k=0$, here it is generalised to $k\neq0$.

The topology of the $\theta$ torus can be further used to obtain the possible frequencies of the axion potential. Consider an axion potential as a sum of cosine potentials, as would follow from the dilute instanton gas approximation
\begin{equation}
    V(a) = \sum_n A_n\cos\left[\vec b_n\cdot\vec\theta^{\prime \rm eff}\right]
\end{equation}
with $\vec b_n$ a 2-vector which is itself quantised since every term in the potential should satisfy
\begin{align}
    \cos\left[\vec b_n\cdot\vec\theta^{\prime \rm eff}\right]=\cos\left[\vec b_n\cdot\vec\theta^{\prime \rm eff}+\vec b_n \cdot\left(
\mathbb{Z}\Delta\vec{\theta}_I^{(p')}+\mathbb{Z}'\Delta\vec{\theta}_{II}^{(p')}\right)\right],
\end{align}
which implies
\begin{align}
    G_3^{\prime}: \qquad \vec{b} = &\left( \frac{\mathbb{Z}}{9 (\Qemin)^2}, \mathbb{Z}^{\prime } - \frac{\mathbb{Z}}{3} \right) \\
    G_1^{\prime}: \qquad  \vec{b} =& \left( \frac{\mathbb{Z}}{(\Qemin)^2}, \mathbb{Z^{\prime}} \right).
\end{align}
and one has that the smallest frequencies return potential terms as
\begin{align}
G_3': & & \cos(\theta_3^{\rm eff}/3-\theta_{\rm em}^{\rm eff}/(3\Qemin)^2) && \cos(\theta_3^{\rm eff}) &&\cos(\theta_{\rm em}^{\rm eff}/3(\Qemin)^2) \label{eq:potentials}\\
G_1':&& \cos(\theta_3^{\rm eff}) && \cos(\theta_{\rm em}^{\rm eff}/(\Qemin)^2) && \dots
\label{eq:potentials-1}
\end{align}
which depend on the axion $a$ through Eq.~\eqref{eq:thetaEff}.

There is generically a strong hierarchy between different $A_n$ coefficients in the potential since what we referred to as frequencies are in one-to-one relation to instanton number, and higher numbers are suppressed by higher powers of non-perturbative factors $e^{-4\pi^2/g_i^2}$. In particular, potential terms that depend only on $\theta_{\textrm{em}}^{\textrm{eff}}$ would be negligible against $\theta_{3}^{\textrm{eff}}$-only dependent ones, reflecting the hierarchy of couplings in QCD and QED, whose non-perturbative dynamics would generate each corresponding term. 
For $G_3^\prime$ the first term in Eq.~\eqref{eq:potentials} depends on both angles and would correspond to the smallest instanton number; in particular this number in the QCD sector is $1/3$ of the minimum QCD-only instanton number. 
Instantons with fractional $k_{\rm inst}=1/N_c$ instanton numbers are formally incompatible with the usual topological classification $\pi_3(SU(N_c)) =\mathbb{Z}$
which allows only integer $k_{\rm inst}$. This would imply that the coefficient in front of such a potential must be trivial, $A=0$.
In general, there are two options to relax this constraint and allow either for genuine fractional instantons~\cite{Davies:1999uw,Davies:2000nw,Unsal:2007jx,Csaki:2019vte,tHooft:1981nnx,Anber:2021upc,Anber:2023sjn,Choi:2023pdp,Cordova:2023her}
or for novel instanton solutions~\cite{Gherghetta:2020keg}
monopoles~\cite{Csaki:2024ajo,Fan:2021ntg,GarciaGarcia:2025uub}.
In the latter case, the presence of monopoles introduces non-contractible cycles in spacetime and allows for the existence of purely Abelian $U(1)$ instantons which depend only on the $\theta_{\textrm{em}}^{\textrm{eff}}$ angle.
The second option is to modify the topology of $SU(N_c)$ by considering product groups spontaneously broken to a diagonal subgroup~\cite{Agrawal:2017ksf,Csaki:2019vte}. In all such cases, there are novel semi-classical contributions to the axion potential that have been studied in the literature. We will not pursue this program further in the present paper, leaving a detailed exploration of how to generate such non-standard instanton potentials dynamically for future work.

\subsection{QCD and Electromagnetic Domain Walls}
This qualitative understanding of the potential terms allows for one more application of topology. It follows from the above discussion that the dominant contribution to the axion potential will be 
\begin{align}
    V_{\rm QCD\,inst}(a) \,=\, \Lambda_{\rm QCD}^4(1-\cos(\theta_3+2 Na/f))
  \,=\, 2\Lambda_{\rm QCD}^4\, \sin^2\,\left(\frac{\theta_3 + 2Na/f}{2}\right)
    ,\label{eq:VQCD}
\end{align}
if we rely on the dilute QCD instanton approximation, or
\begin{align}
   V_{\chi{\rm PT}}(a)
\,=\,
m_\pi^2 f_\pi^2\left[1-
\sqrt{
1 - \frac{4z}{(1+z)^2}
\, \sin^2\,\left(\frac{\theta_3 + 2Na/f}{2}\right)
}\right],
\qquad z = \frac{m_u}{m_d},
\label{eq:VchPT}
\end{align}
if we use the phenomenologically robust expression from chiral perturbation theory, see e.g.~\cite{DiLuzio:2020wdo} for a recent review. Both potentials are non-perturbatively generated by the QCD scale; they are functions of the same argument, $\theta_3 + 2Na/f$, and have the same periodicity properties. Beyond that, the actual form of the potential and whether it is single-valued~\eqref{eq:VQCD} or multi-branched~\cite{Witten:1998uka},
will not be relevant to the discussion below.
The number of physically distinct axion minima of the QCD potential~\eqref{eq:VQCD}, or equivalently~\eqref{eq:VchPT}, is $N_{DW}=2N$, which corresponds to the number of axion domain walls. While this is a well-known result, we would like to elucidate it from the perspective of the present work.

\begin{figure}[h!]
    \centering
    \begin{tikzpicture}
    \filldraw[gray,opacity=0.2] (0,0) -- (4.5,0) -- (4.5,1.5)--(0,1.5);
    \foreach \x in {0,...,1} {
    \foreach \y in {0,...,1} {
      \draw (1.5*\x+1,3*1.5*\y+1.5*\x+0.5) circle (2pt);
    }
  }
  \foreach \x in {0,...,1} {
    \foreach \y in {0,...,1} {
      \draw (1.5*\x+4.5+1,3*1.5*\y+1.5*\x+0.5) circle (2pt);
    }
  }
   \draw (1.5*2+1,2*1.5+0.5) circle (2pt);
  \draw[thick,->] (0,0) --(0,7)node[left]  {$L_{3}^{\rm eff}$};
     \draw[thick,->] (0,0) --(7,0) node[below] {$L_{\rm em}^{\rm eff}$};
       \draw[dashed] (1,0) node[below] {$L_{\textrm{em}}$} -- (1,0.5) -- (0,0.5) node[left] {$L_3$};
       \draw[thick] (1,0.5) -- (7,2);
       \foreach \z in {0,...,4} {
    \draw[OIblue] (1.5*\z,0) -- (1.5*\z,7);
    \draw[OIvermilion] (0,1.5*\z) -- (7,1.5*\z);
    }
    \draw [<->] (1,2.1) -- (7,2.1); 
    \node at (4,2.3){\small$E_{DW}$};
    \draw [<->] (7.1,0.5) -- (7.1,2);
    \node at (6.6,1.2) {\small$N_{DW}$};
    \foreach \x in {1,...,4} {
    \draw (1.5*\x,0.1) -- (1.5*\x,-0.1) node[below] {\x};
    \draw (0.1,1.5*\x) -- (-0.1,1.5*\x) node[left] {\x};
    }
    \end{tikzpicture}
    \raisebox{25pt}{\begin{tikzpicture}
     \filldraw[gray,opacity=0.2] (0,0) -- (6,0) -- (6,2)--(0,2);
         \draw[thick,->] (0,0) --(0,5)node[left]  {$L_{3}^{\rm eff}$};
     \draw[thick,->] (0,0) --(7,0) node[below] {$L_{\rm em}^{\rm eff}$};
     \draw (4/3,2/3) circle (2pt);
     \draw [thick](4/3,2/3) -- (4/3+14/3,2/3+14/12);
     \draw [thick,->](4/3,2/3) -- (4/3+10/3,2/3+10/12);
     \draw [thick](0,11/6) -- (2/3,2);
     \draw [thick,->](0,11/6) -- (1/3,11.5/6);
     \draw [thick](14/3,0) -- (6,1/3);
     \draw [thick,<-](16/3,1/6) -- (6,1/3);
     \draw [thick](0,1/3)-- (4/3,2/3);
     \draw [thick,-<](0,1/3)-- (2/3,1/3+1/6);
     \draw[dashed] (4/3,0) node[below] {$L_{\textrm{em}}$} -- (4/3,2/3) -- (0,2/3) node[left] {$L_{3}$};
     \draw[OIvermilion, very thick] (0,0) -- (6,0);
     \draw[OIblue, very thick] (0,0) -- (0,2);
    \end{tikzpicture}}
    \caption{$\vec L^{\prime\textrm{eff}}=(L^{\textrm{eff}}_{\textrm{em}},L^{\textrm{eff}}_{3})=(\theta^{\textrm{eff}}_{\textrm{em}}/(6\pi (\Qemin)^2),\theta^{\textrm{eff}}_{3}/2\pi)$ plane for $G_3'$ with the values spanned by a given axion as the black thick interval, the set of values that minimise the QCD (QED) potential as red (blue) horizontal (vertical) lines both for the fundamental domain on the RHS panel and for the extended plane with an equivalence relation in the LHS panel. The number of axion values that minimise the QCD (QED) potential is given by the intersection of the black and red (blue) horizontal (vertical) lines and equates $N_{DW}$ ($E_{DW}$). After the QCD phase transition, the value of the axion in a given space point will be driven to its closest minimum (i.e. intersection point) following the arrows in the black interval, if $N_{DW}>1$ this can lead to domain walls.}
    \label{fig:G3DomainWalls}
\end{figure}
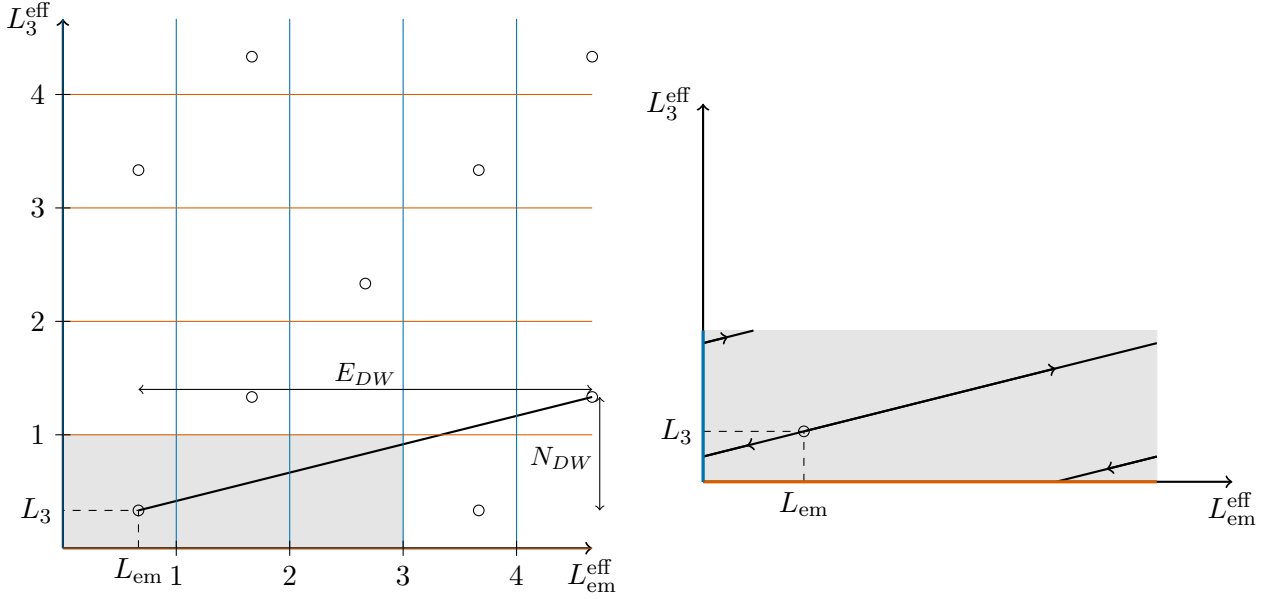

The axion field $0\le a < 2\pi f$ spans a semi-open interval in $\theta$-space following Eq.~\eqref{eq:thetaEff}. At $a=0$ where this interval begins we have $\vec\theta^{\prime \rm eff}=(\theta_{\textrm{em}},\theta_3)$, whereas increasing $a$ until $a=2\pi f$ one traverses a $N_{DW}$-vertical distance and $E_{DW}$ horizontal distance 
on the $(L^{\textrm{eff}}_{\textrm{em}},L^{\textrm{eff}}_{3})$ plane in Fig.~\ref{fig:G1DomainWalls} to reach a point that should be the same as $(\theta_{\textrm{em}},\theta_3)$ modulo the torus periodicities. This is the condition that quantises $E_{DW},N_{DW}$. As shown in  Figs.~\ref{fig:G1DomainWalls}, \ref{fig:G3DomainWalls}, this means that the axion trajectory should end in any of the elements of the lattice of circles. On the other hand, the potential of Eq.~\eqref{eq:VQCD} or \eqref{eq:VchPT} defines a set of $\vec\theta^{\prime \rm eff}$ values that minimise it, this set is simply $\theta_3^{\textrm{eff}}=2\pi\mathbb{Z}$. If constrained to our fundamental domain, this corresponds to $\theta_3^{\textrm{eff}}=0$. It will, however,  be useful to work with the extended plane where the minimum of the potential is a set of lines parallel to $\theta_3^{\textrm{eff}}=0$, shown as red lines in Figs.~\ref{fig:G1DomainWalls} and \ref{fig:G3DomainWalls}. The number of minima of the axion potential in Eq.~\eqref{eq:VQCD} (and equivalently in~\eqref{eq:VchPT}) is given by the number of solutions to $\sin(\theta_3+N_{DW}a/f)=0$, $0\leq a<2\pi f$; visually this is the number of intersections of the axion trajectory with the set of possible minimum values in $\vec\theta^{\prime \rm eff}$ just described. In the extended plane on the left hand side of Figs.~\ref{fig:G1DomainWalls} and \ref{fig:G3DomainWalls} it is simpler to visualise that $N_{DW}$ does indeed count such intersections and one recovers the known result. The same result follows from the fundamental-domain-only analysis on the right hand side of these figures.

The analysis in the $\theta_{\textrm{em}}$ direction is analogous; the leading QED-only potential term\footnote{It is given by $V(\theta_{\rm em}^{\rm eff})\,\sim\,\cos(\theta_{\rm em}^{\rm eff}/p'(\Qemin)^2)$ as follows from the last terms on the RHS of Eqs.~\eqref{eq:potentials}-\eqref{eq:potentials-1} for the $G'_p$ theories under consideration.} is minimised for $\theta_{\textrm{em}}^{\textrm{eff}}=2\pi (\Qemin)^2 p'\mathbb{Z}$ (or equivalently for $L_{\textrm{em}}^\textrm{eff}=\mathbb{Z}$) and $E_{DW}$ counts the number of minima as it returns the number of intersections in Figs.~\ref{fig:G1DomainWalls} and \ref{fig:G3DomainWalls} of the axion trajectory and the QED minima in the form of blue vertical lines. 
The number of domain walls in the purely electromagnetic direction, $E_{DW}$, has not been identified or studied previously, to the best of our knowledge. In practice, the (potential) effect of non-degeneracy in the electromagnetic direction would be caused by a QED-induced potential which is hugely suppressed relative to the QCD effect of~\eqref{eq:VQCD}, \eqref{eq:VchPT}, though it may still be of interest. We also note that for a gluonphobic ($N_{DW}=0$) axion $E_{DW}$ would be the topological magnitude of relevance, alas such an axion would not solve the strong CP problem.

\medskip

Let us now consider the cosmological formation of axion domain walls from the point of view of the $\theta$-torus. Solving the Strong CP problem requires that at least one quark be charged under both the PQ symmetry and under QCD. In DFSZ type axion models \cite{Dine:1981rt,Zhitnitsky:1980tq}, one or more of the SM quarks fulfils this role. This requires extending the Higgs sector and could include shared global structure between the PQ and SM groups, as discussed in \cite{Choi:2026oqz}. Here we will consider the (group-theoretically) simpler KSVZ type axion \cite{Kim:1979if,Shifman:1979if}, in which a new heavy quark carries both QCD and PQ charge while the SM quarks are uncharged under the PQ symmetry. 

Assuming that $f$ is below the scale of inflation, after spontaneous symmetry breaking the massless axion takes values along its range at different spatial points, which leads generically to the formation of strings, around which the axion field winds from $0$ to $2 \pi f$. 
At the QCD phase transition, domain walls form attached to these strings, since in the axion winding $0\leq a<2\pi f$ there will be $N_{DW}$ maxima in $N_{DW}$ directions perpendicular to the string. For $N_{DW}>1$ domain walls can separate physically distinct vacua and be long lived coming to dominate the energy density of the universe. This phenomenology is dictated by $N_{DW}$. Let us cast this same phenomenology visually in the plane of $\vec \theta^{\prime \textrm{eff}}$. Immediately after PQ symmetry breaking, the axion in a given location will take a value and hence correspond to a point in the trajectories of Figs.~\ref{fig:G1DomainWalls} and \ref{fig:G3DomainWalls}. Around a string the whole axion period is sampled continuously filling the whole black interval of Figs.~\ref{fig:G1DomainWalls} and \ref{fig:G3DomainWalls}. After the QCD phase transition, the axion value will be driven to its closest minimum, i.e. the point in $\vec \theta^{\prime \textrm{eff}}$ will be pulled towards its closest intersection with the red lines. Around a string, the axion values near a maximum will be pulled away from it but this will increase the energy in the gradient $(\nabla a)^2$ until it compensates the potential energy and a wall is formed. For $N_{DW}=1$ one domain wall is attached to one string and the surface tension will pull the string towards the wall and make the structure shrink and disappear. If $N_{DW}>1$ there is more than one wall attached to a string and equivalently more than one minimum to choose from. Between two spacetime points with axion values in different minima a topologically stable wall will form. Note that in the two distinct vacua that such a wall would separate, the value of $\theta_3^{\rm eff}$ is the same, but the value of $\theta_{\textrm{em}}^{\rm eff}$ in general is not. 
Information about the vacuum is therefore shifted to $\theta^{\rm eff}_{\rm em}$, explicitly
\begin{align}
    \langle\vec \theta^{\prime \rm eff}\rangle  &= (\langle \theta_{\rm em}^{\rm eff}\rangle, 0)\equiv \left(\theta_{\rm em}-\frac{E}{N_{DW}}\left(\theta_3-2\pi n\right)\,,\,0\right) \mod \Delta \vec{\theta}^{\prime (p')}\,,
\end{align}
where there are $N_{DW}$ possible integer values for $n$ such that $0\leq(-\theta_3 +2\pi n)/N_{DW}<2\pi$ and the mod factor above brings the $\langle \theta_{\rm em}^{\rm eff}\rangle$ value within the fundamental domain. Visually this minimisation will drive points in the axion trajectories of Figs.~\ref{fig:G1DomainWalls} and \ref{fig:G3DomainWalls} along the arrows to the uniquely defined intersection with the red line of the RHS panel.

There is physical information stored in $\theta_{\rm em}^{\rm eff}$ and it could be measured through, for example, the Witten effect in newfound monopoles and used to tell the different vacua apart. As the axion field falls into its potential minimum at the QCD phase transition, it will wind round the $\theta$-torus in the $\theta^{\rm eff}_{\rm em}$ direction up to $\frac{1}{2} \frac{E_{DW}}{N_{DW}}$ times, depending on the initial vacuum angles. An example of such a trajectory is shown in Figs. \ref{fig:G1DomainWalls} and \ref{fig:G3DomainWalls}. This will lead to potentially large oscillations in the physical value $\theta_{\rm em}^{\rm eff}$ within an axion domain wall or during the oscillation of QCD axion dark matter.

\begin{figure}[h!]
    \centering
     \begin{tikzpicture}
     \draw[thick,->] (0,0) --(0,7) node[left]  {$L_{3}^{\rm eff}$};
     \draw[thick,->] (0,0) --(7,0)node[below] {$L_{\rm em}^{\rm eff}$};
\foreach \x in {0,...,4} {
    \foreach \y in {0,...,4} {
      \draw (1.5*\x+0.5,1.5*\y+1) circle (2pt);
    }
  }
    \draw[thick] (0.5,1) --(5,4);
    \filldraw[gray,opacity=0.2] (0,0) -- (0,1.5) -- (1.5,1.5) --(1.5,0);
    \draw[dashed] (0.5,0) node[below] {$L_{\textrm{em}}$} -- (0.5,1) -- (0,1) node[left] {$L_3$};
    \foreach \z in {0,...,4} {
    \draw[OIblue] (1.5*\z,0) -- (1.5*\z,7);
    \draw[OIvermilion] (0,1.5*\z) -- (7,1.5*\z);
    }
    \draw [<->] (0.5,4.1) -- (5,4.1);
    \node at (2.8,4.3) {\small $E_{DW}$};
    \draw [<->] (5.1,1) -- (5.1,4);
    \node at (5.55,2.3) {\small$N_{DW}$};
    \foreach \x in {1,...,4} {
    \draw (1.5*\x,0.1) -- (1.5*\x,-0.1) node[below] {\x};
    \draw (0.1,1.5*\x) -- (-0.1,1.5*\x) node[left] {\x};
    }
\end{tikzpicture}
\begin{tikzpicture}
     \draw[thick,->] (0,0) --(0,7)node[left]  {$L_{3}^{\rm eff}$};
     \draw[thick,->] (0,0) --(7,0) node[below] {$L_{\rm em}^{\rm eff}$};
     \draw (2,4) circle (2pt);
     \draw [dashed] (2,0) node[below] {$L_{\textrm{em}}$} -- (2,4) --(0,4) node[left] {$L_3$};
     \draw [thick] (2,4) -- (5,6);
     \draw [thick,->] (2,4) -- (3.5,5);
     \draw [thick] (5,0) -- (6,2/3);
     \draw [thick,-<] (5,0) -- (5.5,1/3);
     \draw [thick] (0,2/3) -- (6,2/3+4);
     \draw [thick,<->] (1.5,2/3+1.5*2/3) -- (4.5,2/3+4.5*2/3);
     \draw [thick](0,14/3) -- (2,6);
     \draw [thick,>-](1,16/3) -- (2,6);
     \draw [thick](2,0) -- (6,8/3);
     \draw [thick,-<](2,0) -- (4,4/3);
     \draw [thick](0,8/3) -- (2,4);
     \draw [thick,->](0,8/3) -- (1,10/3);
     \filldraw[gray,opacity=0.2] (0,0) -- (6,0) -- (6,6) -- (0,6);
     \draw [OIblue, very thick] (0,0) -- (0,6);
     \draw [OIvermilion, very thick] (0,0) -- (6,0);
\end{tikzpicture}
    \caption{$\vec L^{\prime\textrm{eff}}=(L^{\textrm{eff}}_{\textrm{em}},L^{\textrm{eff}}_{3})=(\theta^{\textrm{eff}}_{\textrm{em}}/(2\pi (\Qemin)^2),\theta^{\textrm{eff}}_{3}/2\pi)$ plane for $G_1'$ with the values spanned by a given axion as the black thick interval, the set of values that minimise the QCD (QED) potential as red (blue) horizontal (vertical) lines both for the fundamental domain on the RHS panel and for the extended plane with an equivalence relation in the LHS panel. The number of axion values that minimise the QCD (QED) potential is given by the intersection of the black and red (blue) horizontal (vertical) lines and equates $N_{DW}$ ($E_{DW}$). After the QCD phase transition the value of the axion in a given space point will be driven to its closest minimum (i.e. intersection point) following the arrows along the black interval, if $N_{DW}>1$ this can lead to domain walls.}
    \label{fig:G1DomainWalls}
\end{figure}
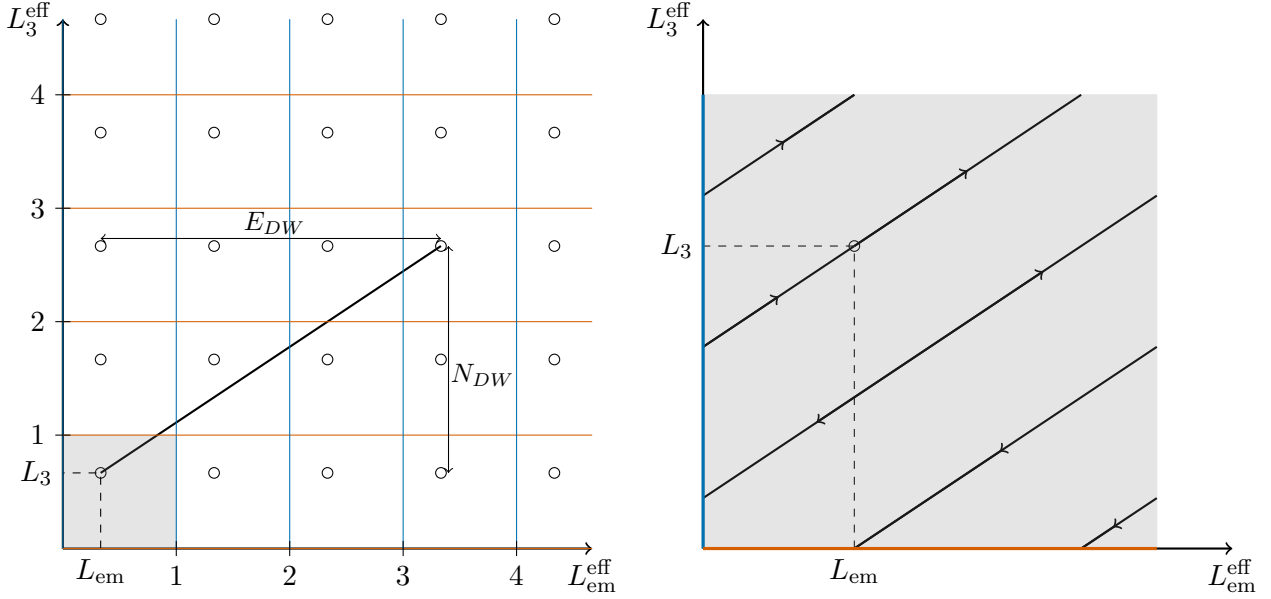

The most straightforward way of addressing cosmological constraints is therefore to set $N_{DW} = 1$ in order to avoid the domain wall problem, yet this implies restrictions on the allowed axion couplings to the Standard Model. For example, for $G'_3$, $(E,2N)=(0,1)$ is not a valid solution of the constraints (unlike for $G_1'$). A purely gluonphilic (at tree level) axion in these models must have $2N = 3 \mathbb{Z}$ and therefore suffer from a QCD domain wall problem.

\subsection{The Axion Photon Coupling}

\begin{figure}[h!]
  \includegraphics[width=0.95\textwidth]{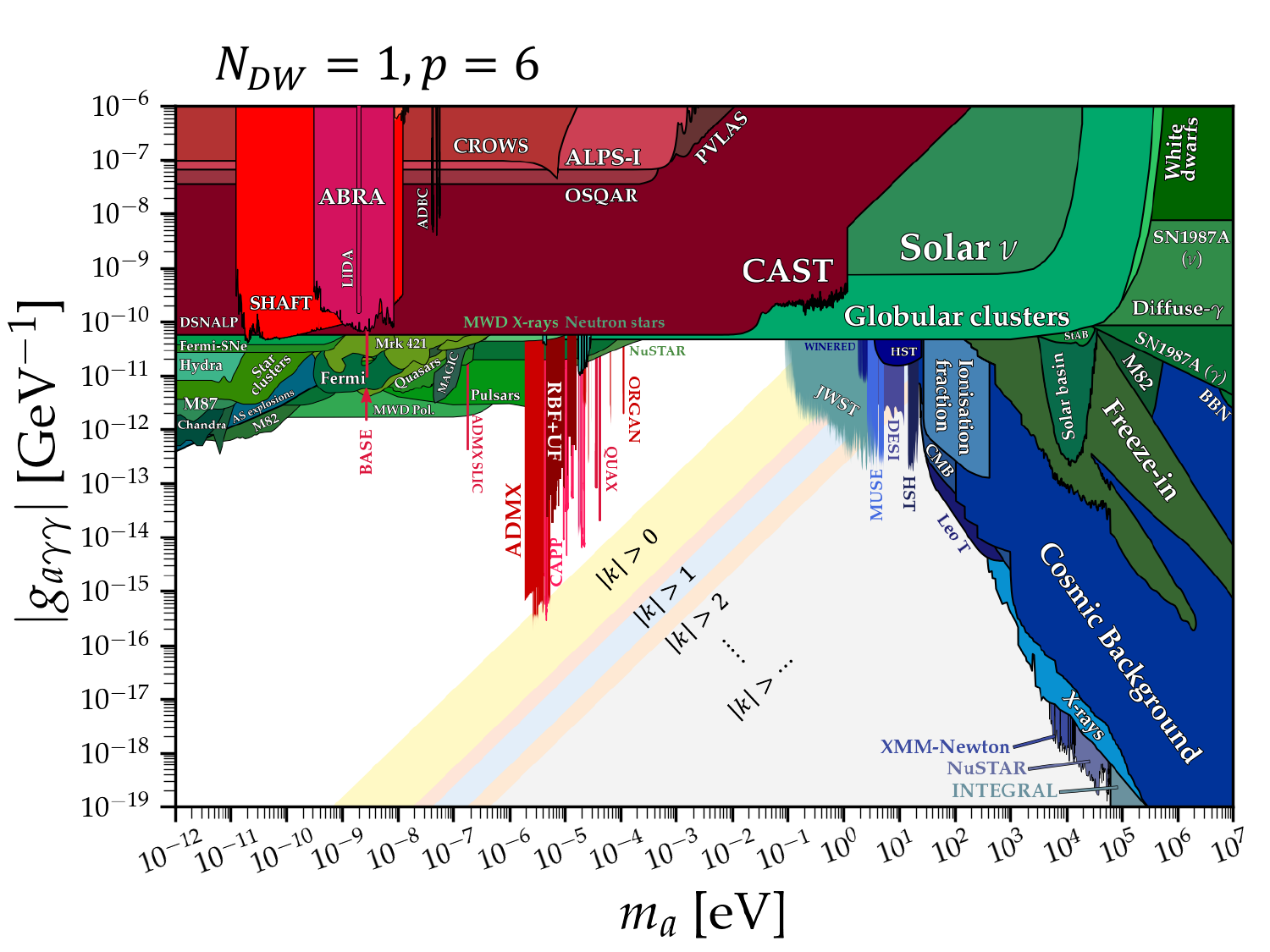}
  \caption{Current experimental bounds on the axion-photon coupling. We show the requirements on the compositeness degree $k$ if $N_{DW}=1$ and $p=6$. In the red region,we have $k \neq \{0,-1 \}$. In the orange region, we have $k \neq \{0,-1,1,-2 \}$. In the white region, $k$ can take any integer value. In this figure, we have neglected the uncertainty in the axion-pion mixing contribution. Taking this uncertainty into account, all but the white region are consistent with $g_{a \gamma \gamma}=0$. Figure modified from \cite{AxionLimits}.}
  \label{fig:Limits}
\end{figure}

The correlation between $E$ and $N$ couplings has consequences on the experimentally relevant photon coupling
\begin{equation}
g_{a \gamma \gamma}\simeq\frac{2 N\alpha_{e m}}{2 \pi f}\left(\frac{E}{N}-1.92(4)\right),
\label{eq:AxionPhotonCoupling}
\end{equation}
where $\alpha_{e m}$ is the fine structure constant and $f$ is the axion decay constant. The 2nd term in equation \eqref{eq:AxionPhotonCoupling} is the contribution from QCD and next-to-leading order, calculated with lattice simulations \cite{GrillidiCortona:2015jxo}. For compositeness degree $k=0$, it was shown in \cite{Choi:2023pdp} that for $G_6$, single axion models with $N_{DW} = 1$ must have $|g_{a \gamma \gamma}| \geq \frac{\alpha_{e m}}{2 \pi f}\left(\frac{8}{3}-1.92(4)\right) \simeq \frac{0.75 \alpha_{e m}}{2 \pi f}$. Allowing $k>0$ this constraint is relaxed since this allows for more finely grained quantised values of $E\sim \mathbb{Z}(\Qemin)^2$  with $\Qemin$ dependent on $k$ as in Eqs.~\eqref{eq:MinimumElectricCharge}, \eqref{eq:MinimumElectricCharge2}. For example, for $k=-1$ and $p=6$ we have $\Qemin = 1/15$ which allows $g_{a \gamma \gamma} \simeq \frac{0.03(4) \alpha_{e m}}{2 \pi f}$ for $N_{DW}=1$, which is already zero within the theoretical error. In this case, $E/N=146/75$.  As shown in Fig. \ref{fig:Limits}, higher $k$ theories would permit even finer tuning of $g_{a \gamma \gamma}$ as for $N_{\rm DW} = 1$ we have

\begin{equation}
{\rm min} \left| g_{a \gamma \gamma} \right| \,\simeq\, {\rm min} \left| \frac{\alpha_{e m}}{2 \pi f}\left(6(\Qemin)^2(3 \mathbb{Z} + 1)-1.92(4)\right) \right|.
\end{equation}

So the very general constraint of \cite{Choi:2023pdp} is relaxed for $k\neq0$, yet let us note that for a given axion model the $E/N$ value will not in general be the one that minimises the equation above, and higher compositeness degree does not imply worse observational prospects. Another possibility to relax the aforementioned constraint arises in multi-axion scenarios~\cite{Lee:2026umy}.

\subsection{Electromagnetic Axions}

As we have seen, the single axion solution to the strong CP problem requires that a non-zero $\theta_3$ is `shifted' by the axion to $\theta_{\rm em}^{\rm eff}$. As there is currently no experimental measurement of $\theta_{\rm em}$, this is an acceptable solution. Possible physical consequences of $\theta_{\rm em} \neq 0$ are discussed for example in \cite{GarciaGarcia:2025xqj}. Suppose we demand, out of theoretical interest, that the vacuum of nature be CP preserving. This can be achieved by adding a second axion field $a_{\rm em}$ coupling only to $F\tilde F$ to rotate away $\theta_{\rm em}$. Analogous to the QCD axion, we must add an additional heavy fermion charged both electromagnetically and under a new $U(1)$ symmetry $PQ_{\rm em}$, and electromagnetism must generate a potential for $a_{\rm em}$. Various possibilities for this have been discussed in \cite{Fan:2021ntg,GarciaGarcia:2025uub}.

Irrespective of the mechanism, the simplest allowed potentials from equation \eqref{eq:potentials}  not involving QCD instantons are

\begin{equation}
\begin{array}{cc}
G_3': \,\, & \cos(\theta_{\rm em}^{\rm eff'}/3(\Qemin)^2)\\
G_1': \,\, & \cos(\theta_{\rm em}^{\rm eff'}/(\Qemin)^2), \\
\end{array}
\end{equation}
with $\theta_{\rm em}^{\rm eff'} = \theta_{\rm em}^{\rm eff} + \frac{E^{\rm em} a_{\rm em}}{f_{\rm em}}$, taking into account that the vacuum $\theta_{\rm em}$ may also be shifted by the QCD axion. For $G_3^{\prime}$, the quantisation conditions in equation \eqref{eq:ENquant} require $E^{\rm em}_{DW} = 3\mathbb{Z}$ for $N = 0$, while for $G_1^{\prime}$ we still have $E^{\rm em}_{DW}=\mathbb{Z}$.
The identification of $E_{DW}$ as the relevant topological number does return the phenomenology straightforwardly, for $G_1'$ the choice $E_{DW}=1$ will provide a single minimum, but for $G_3'$ the smallest non-trivial choice is $E_{DW}=3$ and three distinct minima, each leading to one of the dyonic lattices of Eq.~\eqref{eq:thestG3p} and Fig.~\ref{fig:EMQCDtheta} and a potential electromagnetic domain wall problem.

\section{Summary}

This work has explored the generation of electric charges for monopoles produced by the $\theta$ angles of the Standard Model. The effect is closely and distinctively connected with the global structure of the Standard Model group and the focus was on the discrete and finite electric and magnetic $n$-alities $\hex,\mhex$ that most closely present the correlation. For each group, $\theta$ space was explored identifying the known periodicity and CP invariant points as well as presenting for the first time the $\theta$ values that lead to dyonic lattices, i.e. well defined and non-trivial magnetic and electric $n$-alitites $\hex,\mhex$. 

Unphysical $B+L$ shifts make one direction in the original $\theta$ three-torus unphysical; for the projection onto the physical two-torus care was taken to consider the topology and one-dimensional trajectories that collapse to a point to obtain an explicit formula. The projection in $\theta$ space has an associated projection in group space and selects $U(1)_{\rm em}$ from $SU(2)_{L}\times U(1)_Y$, remarkably before and independently of electro-weak symmetry breaking. The resulting $(\theta_{\rm em},\theta_3)$ physical space is only constrained experimentally to lie around the $\theta_3=0$ line with $\theta_{\rm em}$ unknown and  compatible with the values that lead to dyonic theories. A physical observable to determine $\theta_{\textrm{em}}$ is provided by the effect here studied: were a magnetic monopole to be detected, its electric charge will provide $\theta_{\textrm{em}}$. 

The axion couplings are quantised and parametrise the number of possible ways to wrap a circle $S^1$ round the $\theta$ two-torus $T^2$. These quantisation conditions depend on the global SM group and were known under the assumption of the LH quark doublet having the smallest possible hypercharge. Here they were generalised to the case in which another smaller hypercharge $\Qmin\neq\Qq$ exits. Study of the topology of the axion embedding in $\theta$ leads to the well-known result of $N_{DW}$ as the integer coupling that counts the number of vacua and domain walls of the QCD-induced axion potential; an analogous coupling $E_{DW}$ is identified that counts the number of minima of a possible QED-induced potential. For all intents and purposes however, the QCD axion potential is dominated by QCD, and the cosmologically favoured choice is $N_{DW}=1$ in order to avoid domain walls. We have shown that for $\Qmin\neq\Qq$, $N_{DW}=1$ is compatible with arbitrarily small axion-photon couplings for arbitrarily small $\Qmin$ and any SM gauge group. The axion solution to the strong CP problem will lead generically to $\langle\theta_{\textrm{em}}\rangle\neq 0$. To insure CP invariance of the full SM vacuum a second axion could be introduced, if coupling only to photons and for the group $G_3'$ such an axion would have necessarily multiple minima and potentially lead to electromagnetic domain walls.

\section*{Acknowledgments}
The authors are supported by STFC under ST/X000745/1.

\newpage

\appendix
\section{The Witten effect and $\theta$ angle space}
\label{App:W}

This Appendix presents an introduction to the Witten effect~\cite{Witten:1979ey} for Abelian and non-Abelian theories, highlighting how this effect can be used to deduce $\theta$-space properties.

\subsection{Abelian gauge theory}
\label{App:WU1}

 Consider the Lagrangian 
 \begin{align}
     \mathcal L=-\frac{1}{4e^2}F_{\mu\nu}F^{\mu\nu}+\frac{\theta}{16\pi^2}F_{\mu\nu}\tilde F^{\mu\nu}\,,
 \end{align}
where $\tilde F_{\mu\nu}=1/2\epsilon_{\mu\nu\alpha\beta} F^{\alpha\beta}$ and the covariant derivative reads $D_\mu =\partial_\mu+iA_\mu Q$ with $Q$ the charge operator. Note that, unlike the main text, gauge bosons here have non-canonically normalised kinetic terms, which one can however obtain by rescaling $A\to e A$.  Following the derivation by Wilczek~\cite{Wilczek:1987mv}, we proceed to consider a spacetime-dependent $\theta(x)$; such field would feature in the gauge boson EoM as
\begin{align}
    \frac{1}{e^2}\partial_\mu F^{\mu\nu}-\frac{1}{4\pi^2}\partial_\mu (\theta \tilde F^{\mu\nu})= J^\nu\,,
\end{align}
where the current is such that for a fermion one has $J_\mu=\bar\psi Q\gamma_\mu \psi$. If $\theta$ was a constant, the second term would be trivially vanishing since $\nabla B=\nabla\wedge\nabla A=0$.
The time-component of this equation in the presence of a point-like particle with electric charge $q$ and a monopole of magnetic flux $\Fl$ can be written as,
\begin{align}
   \frac{1}{e^2} \nabla (E)\,=\, q\,\delta^3(x) -\frac{1}{4\pi^2}\nabla( B\theta )\,.
\end{align}
If we now integrate over volume containing the monopole and the point-like particle and use Gauss's theorem, we find that the electric charge of the combined particle and monopole system is
\begin{align}
   q_{\rm \,eff}\,=\, q- \frac{\theta}{4\pi^2} \Fl\,.\label{eq:WitWil}
\end{align}
This implies that the monopole itself has acquired an effective electric charge 
$q_{\rm \, eff\, mono}=-\frac{\theta}{4\pi^2} \Fl$.

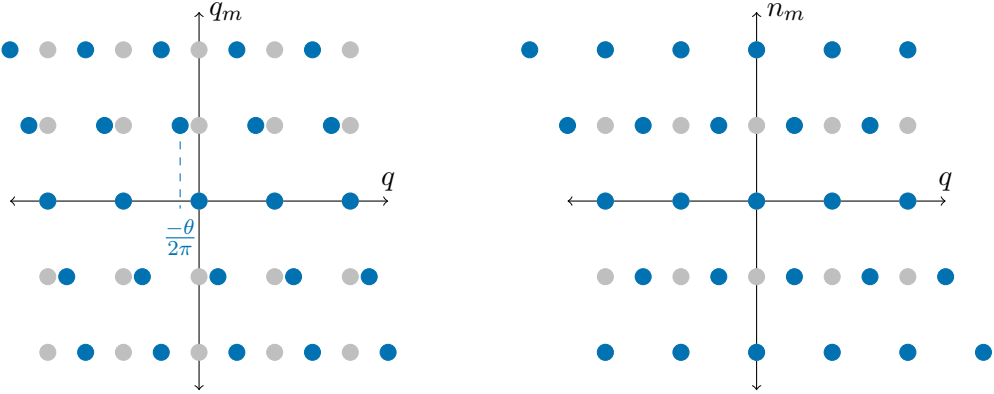
\begin{figure}[h!]
    \centering
    \begin{tikzpicture}
    \draw [<->] (-2.5,0) -- (2.5,0) node [above] {$q$};
    \draw [<->] (0,-2.5) -- (0,2.5) node[right] {$q_m$};    
    \foreach \x in {-2,...,2} {
    \foreach \y in {-2,...,2} {
    \filldraw[lightgray] (\x,\y) circle (3pt);
    }
    }
    \foreach \x in {-2,...,2} {
    \foreach \y in {-2,...,2} {
    \filldraw[OIblue] (\x-0.25*\y,\y) circle (3pt);
    }
    }
    \draw [dashed, OIblue] (-0.25,1) -- (-0.25,-0.1) node[below] {$\frac{-\theta}{2\pi}$};
    \end{tikzpicture}\qquad\qquad
      \begin{tikzpicture}
    \draw [<->] (-2.5,0) -- (2.5,0) node [above] {$q$};
    \draw [<->] (0,-2.5) -- (0,2.5) node[right] {$n_m$};    
    \foreach \x in {-2,...,2} {
    \foreach \y in {-2,...,2} {
    \filldraw[lightgray] (\x,\y) circle (3pt);
    }
    }
    \foreach \x in {-2,...,2} {
    \foreach \y in {-2,...,2} {
    \filldraw[OIblue] (\x-0.5*\y,\y) circle (3pt);
    }
    }
    \filldraw[OIblue] (2,2) circle (3pt);
    \filldraw[OIblue] (-2,-2) circle (3pt);
    \end{tikzpicture}
    \caption{Magneto-electric lattices for non-zero $\theta$ angle in $U(1)$ and $\Qmin=1$. In the LHS we have a generic $\theta$ value whereas in the RHS $\theta=\pi$ is shown to illustrate its CP invariant array of charges. The gray dots give the lattice for $\theta=0$.}
    \label{fig:U1dyons}
\end{figure}

The flux $\Fl$ itself takes discrete values, determined by Dirac quantisation conditions, 
\begin{align}
    e^{i\Fl \,Q}=1\,, \qquad\Fl=\frac{2\pi q_m}{\Qmin}\,,\quad q_m\in\mathbb{Z}\,,
\end{align}
where $\Qmin$ is the smallest quantum of electric charge. One could define $\Qmin$ to be $1$ with a rescaling of the coupling and obtain the original result of~\cite{Witten:1979ey}, but it is more convenient for our purposes to keep $\Qmin$ general.
An ordinary monopole (i.e. an electrically neutral state with magnetic charge $q_m$) therefore acquires a non-trivial electric charge proportional to $\theta/2\pi$ and the monopole's magnetic charge~\cite{Witten:1979ey},
\begin{align}
    q_{\rm \, eff\, mono}\, =\, -\,q_m\,\frac{\theta}{2\pi}\, \frac{1}{\Qmin}\,.\label{eq:appWit}
\end{align} 
A depiction of this `tilt' in the lattice of electric and magnetic charges is shown in Fig.~\ref{fig:U1dyons}. It follows from this figure that by increasing $\theta$ one would reach a stage in which the tilted dyonic lattice would coincide again with the original $\theta=0$ untilted spectrum. This defines the periodicity of the $\theta$-parameter since by travelling along the $\theta$-direction we have circled back to the original theory we started from. The condition is a shift of charge for the monopoles equal to the minimum electric quantum, i.e.
\begin{align}
\label{eq:A7}
    \theta&\sim \theta+\Delta \theta\,, \qquad \frac{\Delta \theta}{2\pi \Qmin}\equiv \Qmin\,,
\end{align}
and thus the period of the $\theta$-angle in QED is 
\begin{align}
\label{eq:A8}
\Delta \theta \,=\, 2\pi \,\Qmin^2\,.
\end{align}

As a cross-check of this periodicity, one can introduce a chiral fermion of smallest charge, rotate it as $\psi_L\to e^{i\alpha}\psi_L$ and, this being an anomalous global symmetry, find a shift,
\begin{align}
    \theta\to \theta+\alpha \Qmin^2\,.
\end{align}
The period of $\alpha$ is $2\pi$ which should be compatible with that of $\theta$. Compatible period means, to be explicit $\Qmin^2\Delta\alpha=\mathbb{Z}\Delta\theta$, where, since this is the smallest chiral representation, the proportionality-integer is $1$.
 
While the presence of this fermion does give us the right period, it does also turn the angle itself unphysical, since the phase of the fermion is unobservable. Having a `reference point' in the form of another parameter in the Lagrangian that shifts under the chiral rotation, provides an invariant linear combination. Take for example the mass term
\begin{align}
    \mathcal L\,\supset\, m \bar\psi_R\psi_L+m^*\bar\psi_L\psi_R\,,\qquad 
    \psi_L\,\to\, e^{i\alpha}\psi_L\,,\quad \arg(m)\,\to\, \arg(m)+\alpha\,,
\end{align}
so that the combination
\begin{align}
    \bar\theta=\theta-\Qmin^2\arg(m)\,,
\end{align}
is shift invariant and physical. In the main text we will often drop the bar over $\theta$ for brevity of notation.

\medskip

We further note that since the action of CP changes the sign of the electric charge, $q\to -q$, and does not modify the magnetic flux or charge, $\Fl\to\Fl$, it follows that the dyonic lattice of a CP-invariant theory must be invariant under axial symmetry around the $q=0$ axis. This is axial symmetry of the spectrum is realised for $\theta=0$ and also for $\theta=\pi$,  
and can be seen on the RHS panel in Fig.~\ref{fig:U1dyons}.

\subsection{SU(N) and quotients}
\label{App:WSUN}

Let us take special unitary groups as case-studies of non-Abelian theories. The set of groups that share the same Lie algebra and local properties are
\begin{align}
    \frac{SU(N)}{Z_p}\,,
\end{align}
with smallest $p=1$ for the $SU(N)$ case ($Z_1=1$) and largest $p=N$, with $Z_N$ generated by $n$-ality $n$ as $Z_N=\{1,e^{2\pi ni/N},e^{4\pi ni/N},\dots\}$, through intermediate values of $p$ for possible subgroups of $Z_N$. For reference a quark (fundamental) in $SU(3)$ has $n$-ality $1$, a symmetric representation made of two quarks has $n$-ality $2$ etc so that in Young tableaux language, $n$-ality counts the number of boxes modulo $N$.

In non-Abelian theories $n$-ality takes the place of charge in QED as the gauge-invariant representation-dependent observable; both are discrete but while for $U(1)$ charges are in one-to-one correspondence with the infinite set of integers $\mathbb{Z}$, for $n$-ality $N$ values suffice. 

Consider the gauge Lagrangian
\begin{align}
    \mathcal L=-\frac{1}{4\hat g^2} G_{\mu\nu}^aG^{\mu\nu}_a+\frac{\hat\theta}{32\pi^2}G_{\mu\nu}^a\tilde G^{\mu\nu}_a\,,
\end{align}
with
\begin{align}
    D_\mu=&\partial_\mu+G_\mu^a\tilde T_a\,, & \tilde T_{N^2-1}&=\mathrm{Diag}(1\,,\,1\,,\,1\,,\, \dots\,,\, -(N-1))\,.
\end{align}
The connection with the convention in which generators in the fundamental satisfy tr$(T_aT_b)=\delta_{ab}/2$ is given by a redefinition of the coupling which also allows to connect with $\theta$ as given in the main text, Eq.~\eqref{eqs:2.3}. The connection reads explicitly 
\begin{align}
    \hat g^2&=\frac{g^2}{2N(N-1)} \,, & \hat\theta&=2N(N-1)\theta\,,
\end{align}
and will be used in due course. One has that the group generated by $n$-ality reads $\exp(2\pi \tilde T_{N^2-1}i/N)$ which is the reason why we will single out this particular generator in the following.

The EoM for dynamical $\hat \theta$ is analogous to the Abelian case
\begin{align}
    \frac{1}{\hat g^2}\partial_\mu G^{\mu\nu}_a-\frac{1}{8\pi^2}\partial_\mu (\hat \theta \tilde G^{\mu\nu}_a)= J^\nu_a\,.
\end{align}
The monopole solution we embed solely along the $\tilde T_{N^2-1}$ direction; this being an Abelian subgroup of $SU(N)$ one can borrow the Abelian result for the field strength which only has non-zero component along the same direction. To obtain the $n$-ality of this monopole for $\theta\neq0$ one should put the monopole up against a consistent electric source that excites the $\tilde T_{N^2-1}$ direction only. A fermion current reads $J_{\mu,a}=\bar\psi \tilde T_a\gamma_\mu \psi$ and one can check that a fundamental $F$ with only the last component non-vanishing $\mathcal F=(0,0,0...,\mathcal F_N)$ would indeed have only non-zero current for the required generator, $a=N^2-1$. Equivalently any representation built out of a number $n_{\mathcal F}$ of such fundamental representations would only excite $\tilde T_{N^2-1}$ but its current would be $n_F$ times that of the fundamental. This result can be understood as arising from restricting ourselves to an Abelian subgroup of $SU(N)$ and indeed $n_{\mathcal F}$ is the analogous to $Q/\Qmin$, i.e. the number of elementary charges. 

All in all the outcome for a monopole along the $\tilde T_{N^2-1}$ direction and a representation built out of $n_{\mathcal F}$ fundamentals of the above specifications and point-like reads 
\begin{align}
   &\frac{1}{\hat g^2} \nabla E_G+\frac{1}{8\pi^2}\nabla (\hat \theta B_G)=-(N-1) n_{\mathcal F} \delta^3(x)\\
   &-(N-1)n^{\textrm{eff}}_{\mathcal F}=-\frac{\hat \theta}{8\pi^2}\Fl-(N-1)n_{\mathcal F}\,,
\end{align}
where we have performed the same volume integral of the Abelian case and have defined $n^{\textrm{eff}}_{\mathcal F}$ so that it agrees with $n_{\mathcal F}$ in the absence of monopoles. This equation leads to (dropping the eff index)
\begin{align}
    n_\mathcal{F}=\frac{\hat \theta\,\Phi}{(N-1)8\pi^2}=\frac{N\theta\Phi}{4\pi^2}\,,
\end{align}
where we note it has the opposite sign of the Abelian case in Eq.~\eqref{eq:WitWil} but otherwise $n_\mathcal{F}, \Phi$ are the analogous of $Q/\Qmin, \Phi$ in the Abelian case.

For the possible values of the flux, let us give the gauge transformation required by the monopole in the Wu-Yang formulation~\cite{Wu:1975es} and its quantisation condition
\begin{align}
    U(\phi)&=e^{i (\phi/2\pi) \Fl \tilde T_{N^2-1} }\,, & A^+d\phi-A^-d\phi&=U^\dagger dU\,, & e^{i \Fl \,\tilde T_{N^2-1} }&=1\,,
\end{align}
where $\phi$ is the azimuthal angle.
The flux quantisation condition does depend on the quotient since it determines the allowed representations in which to evaluate $\tilde T$, in particular one can (cannot) have the fundamental in $SU(N)$ ($SU(N)/Z_N$) and so
\begin{align}
    \Fl=&\frac{2\pi q_m}{N}\,, & q_m=&\left\{\begin{array}{cc}
         SU(N)&  q_m=N\times \mathbb{Z}\\
         SU(N)/Z_N&  q_m= \mathbb{Z}
    \end{array}\right.\,.\label{eq:Flqm}
\end{align}
An equivalent way to write the quantisation condition reads
\begin{align}
    e^{i \Fl \tilde T_{N^2-1} }&=\left[\left(e^{2\pi i\tilde T_{N^2-1}/N}\right)^{N/p}\right]^\mathbb{Z}\,,  &\Fl=2\pi \frac{\mathbb{Z}} {p}  \,,
\end{align}
in this case the RHS is still the identity: for $p=1$ ($SU(N)$) one has powers of exp$(2\pi i\tilde T_{N^2-1})$ which given the eigenvalues of $\tilde T_{N^2-1}$ being integers is the identity; for $p=N$ we can have the $N$'th root of exp$(2\pi i\tilde T_{N^2-1})$ because $\tilde T_{N^2-1}$ is only  evaluated in representations with eigenvalues in $N\mathbb{Z}$ (the centre is absent i.e. degenerate with the identity). 

It is at this point that the main difference with the Abelian case resurfaces; 
for an unbroken non-Abelian theory,
the flux (or equivalently $q_m$) is not a conserved quantity.\footnote{
If instead, $SU(N)$ is broken spontaneously to an Abelian subgroup $U(1)^r$ then for each of the physical $U(1)$ factors in the IR there exists a topologically conserved integer magnetic flux governed by $\pi_1(U(1))=\mathbb{Z}$. For example, in the Georgi-Glashow $SU(2)\to U(1)$ model 't~Hooft-Polyakov monopoles carry topologically conserved $U(1)$ magnetic charges, $q_m\in Z$; this also applies to generic non-Abelian embeddings of the SM, see e.g.~Refs.~\cite{Khoze:2024hlb,Alonso:2024pmq}, where monopoles carry topologically conserved magnetic fluxes of QED. In this case, monopoles with $U(1)$ magnetic charges $q_m$ and $q_m+N$ are topologically distinct configurations, though, for notational convenience, they can still be grouped into the same universality class~\eqref{eq:nm-qmodN} labelled by $n_m$.} 
In $U(1)$ flux was conserved due to its topological nature based on $\pi_1(U(1))=\mathbb{Z}$, now one has $\pi_1(SU(N)/Z_p)=Z_p$ and fluxes are conserved modulo $p$.  The conserved quantity can be labelled as
\begin{align}
\label{eq:nm-qmodN}
    n_m=q_m\mod N\,.
\end{align}
Together with the quantisation conditions this means that in the case of $SU(N)$ there is no non-trivial flux (the only allowed value of $n_m$ is $0$) whereas in $SU(N)/Z_N$ there are in addition $N-1$ non-trivial fluxes ($n_m=1,..,N-1$).  Let us recall the analogous electric relation
\begin{align}
    n=n_\mathcal{F}\mod N\,.
\end{align}
It is relevant to highlight that $n,n_m$ are labels and the actual monopole that a $q_m$-flux solution will decay into is given by a semi-classical stability analysis (though it is generically the smallest compatible $q_m$). Understood as defining an equivalence class, the indices $n,n_m$ can be visualised in Fig.~\ref{fig:CPbrokenP} in the case of $SU(3)/Z_3$ theory, where they are labelling states inside the dashed square.
The point marked in red in the LHS graph of coordinates $(n_\mathcal{F},q_m)=(3,-3)$ is taken to $(0,0)$ by the action of one vertical and one horizontal green arrows that define the equivalence class. With such procedure any other point in the plane can be taken to within the dashed square were $n,n_m$ lie. The introduction of $\theta\neq0$ tilts the lattice but does not change the equivalence relation; the same point in the RHS panel of Fig:~\ref{fig:CPbrokenP} as changed coordinates but the green arrows that define the equivalence have also tilted so that the point can be taken from $(3-3\theta/2\pi,-3)$ to within-the-dashed-square point $(0,0)$. 

Figure~\ref{fig:CPbroken} illustrates the $\theta$-tilting effect of the $SU(3)$ lattice and complements the discussion above of this effect in the $SU(3)/Z_3$ theory in Fig.~\ref{fig:CPbrokenP}. The only difference is the interchange of the allowed values of $n$ and $n_m$ between the two theories, 
$n=0\,,\, n_m=1,2,\, \leftrightarrow \, n=1,2\,,\, n_m=0.$

\medskip

The summary of this discussion and central result of this subsection is that
for all $SU(N)$ and $SU(N)/Z_p$ theories, the Witten effect reads
\begin{align}
\label{eq:A.25}
    n=&\frac{\theta }{2\pi}\left( q_m \mod N\right) \mod N =\frac{\theta }{2\pi} n_m \mod N\,,
\end{align}
in the appropriate for each theory $(n,n_m)$ subset.

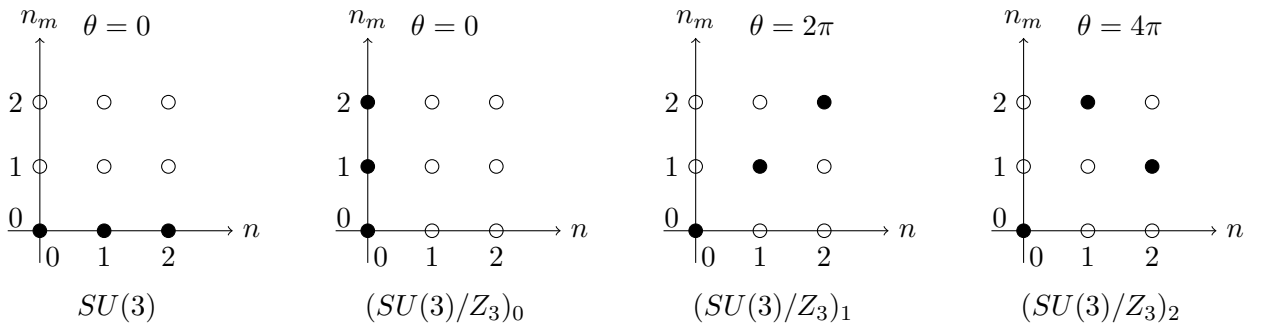
\begin{figure}[h!]
\centering
\hspace{5mm}
\begin{tikzpicture}[scale=0.85]
  \node at (1.2,3.2) {$\theta=0$};
  \draw[->] (-0.5,0) -- (3,0) node[right] {$n$};
  \draw[->] (0,-0.5) -- (0,3) node[above] {$n_m$};
  \node[left] at (-0.1,0.2) {$0$};
  \node[left] at (-0.1,1) {$1$};
  \node[left] at (-0.1,2) {$2$};
  \foreach \x in {1,...,2}
    \node[below] at (\x,-0.1) {\x};
  \node[below] at (0.2,-0.1) {0};
  \foreach \x in {0,...,2} {
    \foreach \y in {0,...,2} {
      \draw (\x,\y) circle (3pt);
    }
  }
  \fill (0,0) circle (3pt);
  \fill (1,0) circle (3pt);
  \fill (2,0) circle (3pt);
  \node at (1.2,-1.2) {$SU(3)$};
\end{tikzpicture}
\hspace{5mm}
\begin{tikzpicture}[scale=0.85]
  \node at (1.2,3.2) {$\theta=0$};
  \draw[->] (-0.5,0) -- (3,0) node[right] {$n$};
  \draw[->] (0,-0.5) -- (0,3) node[above] {$n_m$};
  \node[left] at (-0.1,0.2) {$0$};
  \node[left] at (-0.1,1) {$1$};
  \node[left] at (-0.1,2) {$2$};
  \foreach \x in {1,...,2}
    \node[below] at (\x,-0.1) {\x};
  \node[below] at (0.2,-0.1) {0};
  \foreach \x in {0,...,2} {
    \foreach \y in {0,...,2} {
      \draw (\x,\y) circle (3pt);
    }
  }
  \fill (0,0) circle (3pt);
  \fill (0,1) circle (3pt);
  \fill (0,2) circle (3pt);
  \node at (1.2,-1.2) {$(SU(3)/Z_3)_0$};
\end{tikzpicture}
\hspace{5mm}
\begin{tikzpicture}[scale=0.85]  
\node at (1.5,3.2) {$\theta=2\pi$};
  \draw[->] (-0.5,0) -- (3,0) node[right] {$n$};
  \draw[->] (0,-0.5) -- (0,3) node[above] {$n_m$};
  \node[left] at (-0.1,0.2) {$0$};
  \node[left] at (-0.1,1) {$1$};
  \node[left] at (-0.1,2) {$2$};
  \foreach \x in {1,...,2}
    \node[below] at (\x,-0.1) {\x};
  \node[below] at (0.2,-0.1) {0};
  \foreach \x in {0,...,2} {
    \foreach \y in {0,...,2} {
      \draw (\x,\y) circle (3pt);
    }
  }
  \fill (0,0) circle (3pt);
  \fill (1,1) circle (3pt);
  \fill (2,2) circle (3pt);
  \node at (1.2,-1.2) {$(SU(3)/Z_3)_1$};
\end{tikzpicture}
\hspace{5mm}
\begin{tikzpicture}[scale=0.85]
\node at (1.5,3.2) {$\theta=4\pi$};
  \draw[->] (-0.5,0) -- (3,0) node[right] {$n$};
  \draw[->] (0,-0.5) -- (0,3) node[above] {$n_m$};
  \node[left] at (-0.1,0.2) {$0$};
  \node[left] at (-0.1,1) {$1$};
  \node[left] at (-0.1,2) {$2$};
  \foreach \x in {1,...,2}
    \node[below] at (\x,-0.1) {\x};
  \node[below] at (0.2,-0.1) {0};
  \foreach \x in {0,...,2} {
    \foreach \y in {0,...,2} {
      \draw (\x,\y) circle (3pt);
    }
  }
  \fill (0,0) circle (3pt);
  \fill (2,1) circle (3pt);
  \fill (1,2) circle (3pt);
  \node at (1.2,-1.2) {$(SU(3)/Z_3)_2$};
\end{tikzpicture}
\caption{\label{figCPvacSU3} The four different theories arising from a gauge theory with the same $su(3)$ Lie algebra.}
\end{figure}

\medskip

The study of $n,n_m$ and the extended lattice of $n_{\mathcal{F}},q_m$ can now also be used to determine and compare the periodicity structure of the $\theta$-angle for these models. Let us note that the equivalence relation that gives $n,n_m$ from $n_{\mathcal{F}},q_m$ reads, for generic $\theta$
\begin{align}
    (n_{\mathcal{F}},q_m)\sim(n_{\mathcal{F}},q_m)+\mathbb{Z}(N\theta/2\pi,N)+\mathbb{Z}'(N,0)\,,
\end{align}
one recovers the $\theta=0$ orthogonal equivalence vectors for $\theta=2\pi\mathbb{Z}$ together with well defined $n$-ality based group actions. For $SU(N)$ in particular for $\theta=2\pi$ the extended lattice is also the same as for $\theta=0$ and one indeed has that the period of $\theta$ in $SU(N)$ is $2\pi$. For $SU(N)/Z_N$ on the other hand one obtains a diagonal-like lattice shown in Fig.~\ref{fig:CPbrokenP}. This is not the same theory but one of the Dyonic theories that the main text expands upon. For the full period one requires $N$ times that shift, i.e. for $SU(N)/Z_N$, $\theta\sim\theta+2\pi N$. CP invariant points can also be obtained from the extended lattice demanding a mirror self-reflected lattice as shown in the Abelian case. Results of our Witten-effect-based derivation of $\theta$-periods agree with other known constructions in the literature, be it chiral rotations, fractional instantons, or other effects, see Refs.~\cite{Choi:2023pdp,Cordova:2023her}.

Let us collect our results for $SU(N)$ and $SU(N)/Z_p$ groups here
\begin{align}
    n&=\frac{n_m \theta}{2\pi}\mod N \,, &&\left\{
    \begin{array}{ccc}
        SU(N) & n_m=0& 0\leq\theta<2\pi \\
        SU(N)/Z_N & \quad n_m=\left\{0,1,\dots,N-1\right\} & \quad \,0\leq\theta<2\pi N
    \end{array}\right\}\,.
\end{align}

\begin{figure}[h!]
    \centering
    \begin{tikzpicture}
        \draw [<->] (0,-3.5) -- (0,4.5) node[left] {$q_m$};
        \draw [<->] (-1.5,0) -- (3.5,0) node[right] {$n_\mathcal{F}$};
    \foreach \y in {-3,...,4}
            \node[below] at (-0.25,\y) {\y};
    \foreach \y in {-3,...,4} 
      \filldraw[OIblue] (0,\y) circle (3pt);
    \foreach \y in {-3,...,4} 
      \filldraw[OIblue] (3,\y) circle (3pt);
    \foreach \x in {0,1,2}
    \draw (\x,0) circle (3pt);
    \draw [<->,OIgreen,very thick] (0,4) -- (0,1);
    \draw [<->,OIgreen,very thick] (3,-3) -- (0,-3);
    \draw [<->,OIgreen,very thick] (0,0) -- (0,-3);
    \filldraw[OIvermilion] (3,-3) circle (3pt);
    \filldraw[OIorange] (0,4) circle (3pt);
    \draw [dashed] (-0.5,-0.5) -- (2.5,-0.5) -- (2.5,2.5) -- (-0.5,2.5) --(-0.5,-0.5);
    \end{tikzpicture}\qquad
    \begin{tikzpicture}
        \draw [<->] (0,-3.5) -- (0,4.5)  node[left] {$q_m$};
        \draw [<->] (-1.5,0) -- (3.5,0) node[right] {$n_\mathcal{F}$};
    \draw [OIblue,dashed] (0.2,1) -- (0.2,-0.1)node[below] {$\frac{\theta}{2\pi}$};
    \foreach \y in {-3,...,4}
            \node[below] at (-0.25,\y) {\y};
    \foreach \y in {-3,...,4} 
      \filldraw[lightgray] (0,\y) circle (3pt);
    \foreach \y in {-3,...,4} 
      \filldraw[OIblue] (0.2*\y,\y) circle (3pt);
    \foreach \y in {-3,...,4} 
      \filldraw[lightgray] (0+3,\y) circle (3pt);
    \foreach \y in {-3,...,4} 
      \filldraw[OIblue] (0.2*\y+3,\y) circle (3pt);
      \filldraw[OIorange] (0.8,4) circle (3pt);
    \filldraw[OIvermilion] (-0.6+3,-3) circle (3pt);
    \foreach \x in {0,1,2}
    \draw (\x,0) circle (3pt);
    \draw [dashed] (-0.5,-0.5) -- (2.5,-0.5) -- (2.5,2.5) -- (-0.5,2.5) --(-0.5,-0.5);
    \draw [<->,OIgreen,very thick] (0.78,3.9) -- (0.21,1.1);
    \draw [<->,OIgreen,very thick] (0.78-0.8,3.9-4) -- (0.21-0.8,1.1-4);
    \node[right,OIgreen] at (1.15,3.5) {$\left(\Delta n_{\mathcal{F}}=\frac{3\theta}{2\pi},\Delta q_m=3\right)$};
    \draw [<->,OIgreen,very thick] (3-0.6,-3) -- (0-0.6,-3);
    \end{tikzpicture}
     \caption{Lattices of electric and magnetic charges for $SU(3)/Z_3$ for $\theta=0$ (LHS) and for non-vanishing $\theta$ (RHS). These charges are conserved modulo the equivalence relation given by the green arrows such that any allowed charge combination (blue points) can be brought within the dashed square and assigned a $(n,n_m)$ label.} 
    \label{fig:CPbrokenP}
\end{figure}
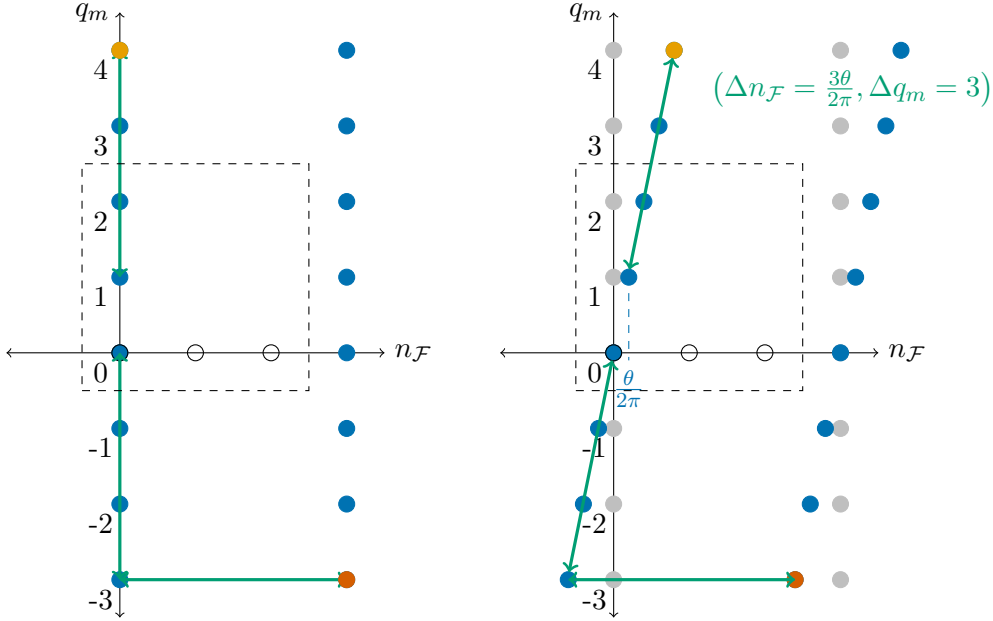
\begin{figure}[h!]
    \centering
    \begin{tikzpicture}
          \draw [<->] (0,-3.5) -- (0,4.5) node[left] {$q_m$};
        \draw [<->] (-1.5,0) -- (3,0)  node[right] {$n_\mathcal{F}$};
    \foreach \y in {-3,...,4}
            \node[below] at (-0.25,\y) {\y};
    \foreach \y in {-3,...,4} 
      \draw (0,\y) circle (3pt);
    \foreach \x in {0,...,2} 
      \filldraw[OIblue] (\x,-3) circle (3pt);
      \foreach \x in {0,...,2}
      \filldraw[OIblue] (\x,3) circle (3pt);
    \foreach \x in {0,1,2}
    \filldraw[OIblue] (\x,0) circle (3pt);
    \draw [<->, OIgreen, very thick] (1 , 2.9) -- (1,0.1);
      \draw [dashed] (-0.5,-0.5) -- (2.5,-0.5) -- (2.5,2.5) -- (-0.5,2.5) --(-0.5,-0.5);
    \end{tikzpicture}\qquad\quad
    \begin{tikzpicture}
      \draw [dashed] (-0.5,-0.5) -- (2.5,-0.5) -- (2.5,2.5) -- (-0.5,2.5) --(-0.5,-0.5);
        \draw [<->] (0,-3.5) -- (0,4.5) node[left] {$q_m$};
        \draw [<->] (-1.5,0) -- (3,0) node[right] {$n_\mathcal{F}$};
    \draw [OIblue,dashed] (0.6,3) -- (0.6,-0.1)node[below] {$\frac{3\theta}{2\pi}$};
    \foreach \y in {-3,...,4}
            \node[below] at (-0.25,\y) {\y};
    \foreach \y in {-3,...,4} 
      \draw (0,\y) circle (3pt);
    \foreach \x in {0,...,2} 
      \filldraw[OIblue] (\x+0.6,3) circle (3pt);
      \foreach \x in {0,...,2}
      \filldraw[OIblue] (\x-0.6,-3) circle (3pt);
      \foreach \x in {0,...,2}
      \filldraw[lightgray] (\x,-3) circle (3pt);
      \foreach \x in {0,...,2}
      \filldraw[lightgray] (\x,3) circle (3pt);
    \foreach \x in {0,1,2}
    \filldraw[OIblue] (\x,0) circle (3pt);
    \draw [<->, OIgreen, very thick] (1.58 , 2.9) -- (1,0.1);
    \node[right, OIgreen] at (1.25 , 1.5) {$\left(\Delta n_\mathcal{F}=\frac{3\theta}{2\pi},\Delta q_m=3\right)$};
    \end{tikzpicture}
   \caption{Lattices of electric and magnetic charges for $SU(3)$ for $\theta=0$ (LHS) and for non-vanishing $\theta$ (RHS). These charges are conserved modulo the equivalence relation given by the green arrows such that any allowed charge combination (blue points) can be brought within the dashed square and assigned a $(n,n_m)$ label.} 
    \label{fig:CPbroken} 
\end{figure}
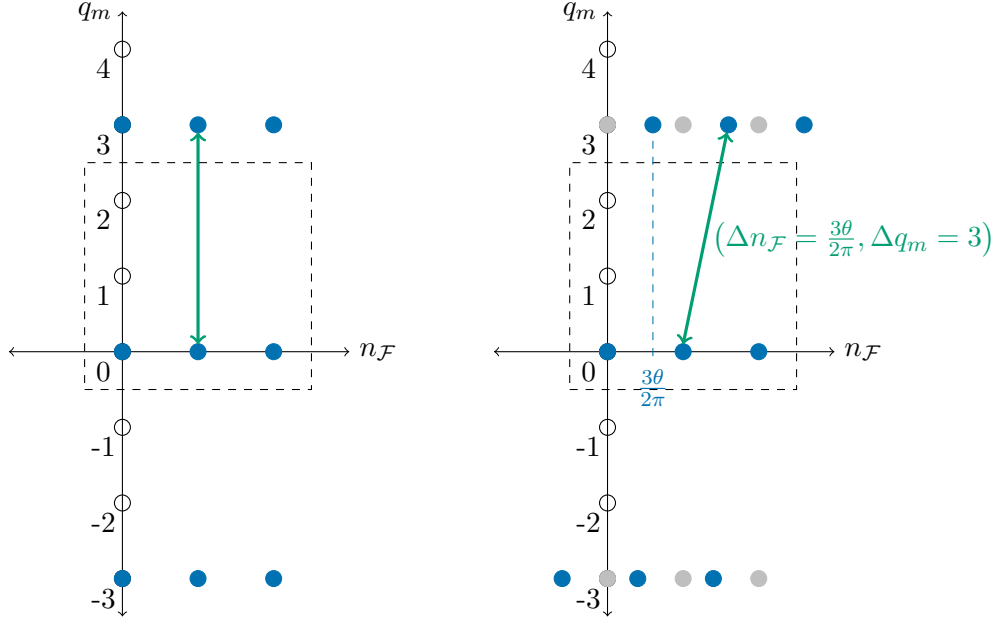

\subsection{U(N) and its universal cover}\label{sec:UN}
The last type of groups required for a study aimed at the Standard Model contains both Abelian and non-Abelian components, specifically
\begin{align}
    G_p\,=\,\frac{SU(N)\times U(1)}{Z_p}\,,\quad Z_p\,=\,\left\{1,e^{2\pi i(\tilde T-Q/Q_{\min})/N},e^{4\pi i(\tilde T-Q/Q_{\min})/N},\dots\right\}.
\end{align}
with the case $G_p=U(N)$. This discreet group that straddles both sectors prompts the definition of a modified $n$-ality as
\begin{align}
    &\exp\left(\frac{2\pi i}{N}n\right)\equiv\exp\left(\frac{2\pi i}{N}\left(\tilde T-\frac{Q}{Q_{\textrm{min}}}\right) \right)\,, \quad \,n\,=\, n_{SU(N)}-\frac{Q}{\Qmin} \mod N\,.
\end{align}
The monopole quantisation condition returns different solutions depending on the quotient, here as above we define
\begin{align}
    &e^{i\Fl Q+i\Fl_G \tilde T}=
    \left[e^{2\pi i (\tilde T-Q/Q_{\min})/N}\right]^{n_m}\,, &
     &\left\{\begin{array}{cc}
         U(N)&  \qquad n_m=\left\{0,1,\ldots,N-1\right\}\\
         SU(N)\times U(1)&  n_m=0
    \end{array}\right.\,,
\end{align}
and one has an extended lattice connected to $n,n_m$ as
\begin{align}
    n_\mathcal{F}\,=\, n_{\mathcal{F}}^{SU(N)}-\frac{Q}{\Qmin} \,,
    \qquad n_m \,=\,q_m \mod m,
\end{align}
with $q_m$ connected to fluxes as
\begin{align}
    \Fl\,=\,-\frac{2\pi }{N Q_{\mathrm{min}}}q_m\,, \quad  
    \Fl_G\,=\,\frac{2\pi }{N}(q_m+N \phi_G) \,: \qquad
    \left\{\begin{array}{cc}
         U(N)&  q_m=\mathbb{Z}\\
         SU(N)\times U(1)&  q_m=N\mathbb{Z}
    \end{array}\right.,
\end{align}
and $\phi_G\in\mathbb{Z}$. Application of the results obtained in the previous sections allows us to write the Witten effect for the extended lattice as
\begin{align}
    n_\mathcal{F}=\frac{N\Fl_G\theta_G}{4\pi^2}+\frac{\Phi\theta}{4\pi^2\Qmin}=\frac{(q_m+N\phi_G)\theta_G}{2\pi}-\frac{ q_m \theta}{2\pi N\Qmin^2}
\end{align}
while applied to the $n,n_m$ subset
\begin{align}
    n=\frac{n_m}{2\pi}\left(\theta_G-\frac{\theta}{N\Qmin^2}\right)\mod N.
\end{align}

\begin{figure}[h!]
     \centering
     \begin{tikzpicture}     
         \draw [->] (-0.1,0) -- (3.25,0) node[right] {$\theta$} ;
         \draw [->] (0,-0.1) -- (0,3.25) node[above] {$\theta_G$};
         \draw (3,0.1) -- (3,-0.1) node[below] {$2\pi \Qmin^2$};
         \draw (0.1,3) -- (-0.1,3) node[left] {$2\pi $};
         \draw [gray,dashed] (0,0) -- (3,0) -- (3,3) -- (0,3) -- (0,0);
         \draw [OIgreen,<->] (0,.5) -- (3,0.5);
         \draw [OIgreen,<->] (.5,0) -- (0.5,3);
         \draw [OIvermilion, very thick] (0,0) -- (3,0);
         \draw [OIblue,thick] (1.8,0) -- (1.8,3);
     \end{tikzpicture}\quad\quad\qquad\qquad
     \raisebox{12pt}{\begin{tikzpicture}
         \draw [->] (-0.1,0) -- (6.25,0) node[right] {$\theta$} ;
         \draw [->] (0,-0.1) -- (0,2.25) node[above] {$\theta_G$} ;
         \draw [OIgreen,<->] (0,.5) -- (6,0.5);
         \draw [OIgreen,<->] (0,0) -- (2,2);
           \draw (2,0.1) -- (2,-0.1) node[below] {$2\pi N \Qmin^2$};
           \draw (6,0.1) -- (6,-0.1) node[below] {$2\pi N^2 \Qmin^2$};
         \draw (0.1,2) -- (-0.1,2) node[left] {$2\pi $};
           \draw [gray,dashed] (0,0) -- (0,2) -- (6,2) -- (6,0) -- (0,0);
           \draw [OIblue,thick] (1.5,0) -- (1.5,2);
           \draw [OIblue,thick] (3.5,0) -- (3.5,2);
           \draw [OIblue,thick] (5.5,0) -- (5.5,2);
           \draw [OIvermilion, very thick] (0,0) -- (6,0);
     \end{tikzpicture}}
     \includegraphics[width=0.45\textwidth]{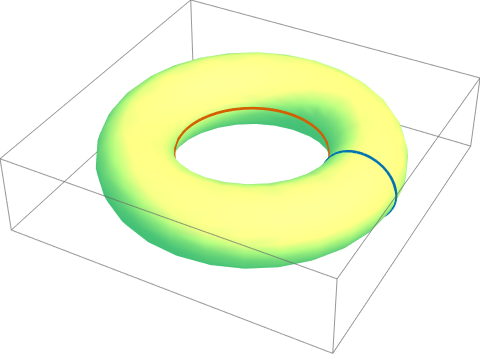}\qquad \qquad
     \includegraphics[width=0.45\textwidth]{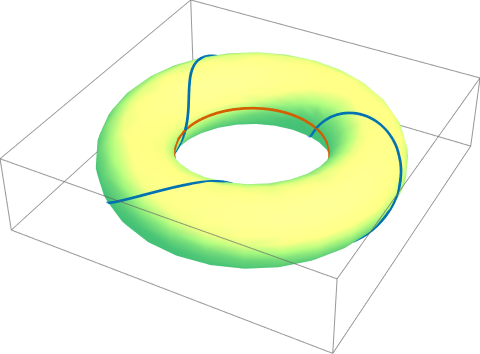}     \caption{The $(\theta,\theta_G)$ plane for $SU(N)\times U(1)$ [LHS] and $U(N)$ [RHS] for $N=3$. Green double-pointed arrows give the equivalence relations and they turn the plane into the torus immediately below with the same thick black and cyan lines displayed in both the plane and torus.}
     \label{fig:UNthetatorus}
 \end{figure}
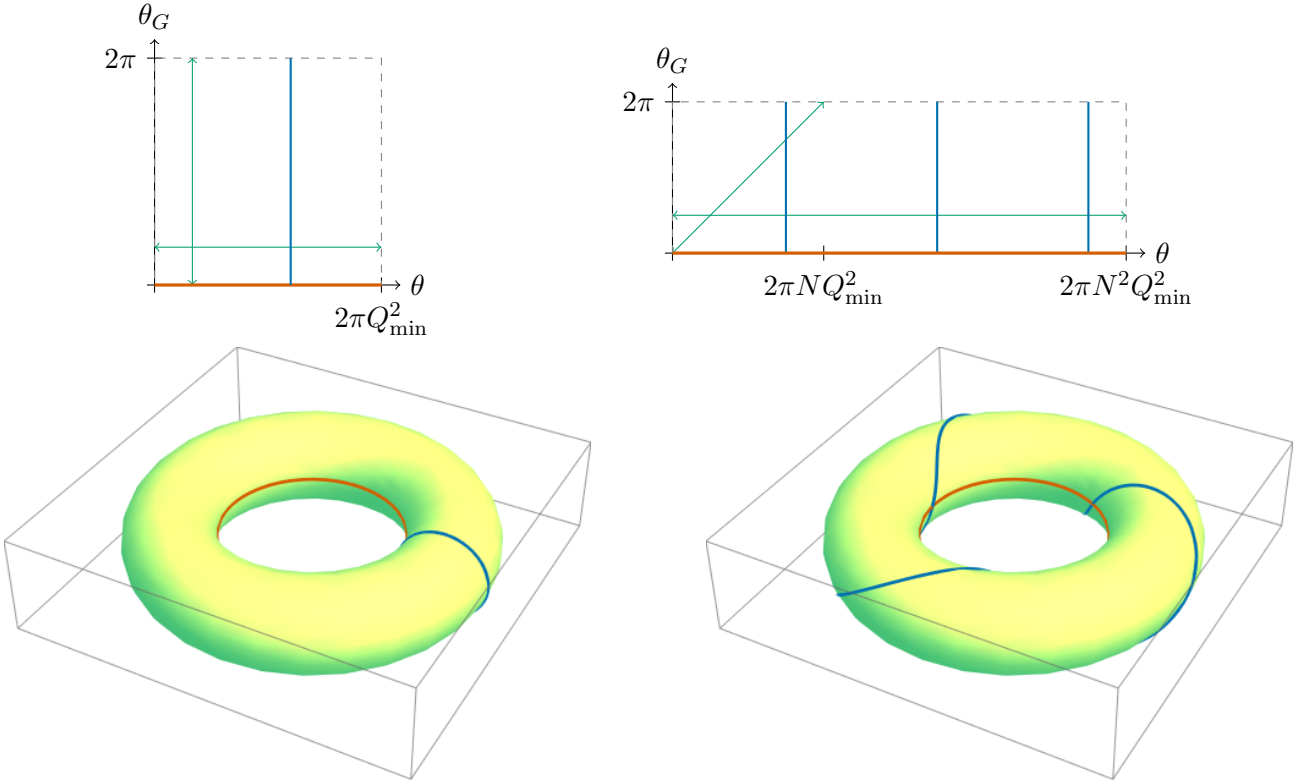

These results can in turn be used to study $\theta$ space, as detailed in the main text; here we simply note that for $U(N)$ the smallest periodicity given by $\delta n=N$ is now  $(\theta,\theta_G)\sim (\theta,\theta_G)+2\pi(N\Qmin^2,1)$ which can be contrasted with the result of a chiral rotation of the smallest fermion
 \begin{align}
     (\Delta\theta,\Delta\theta_G)\sim(\alpha  N Q_{\min}^2, \alpha  )\,.
 \end{align}
This means in the two-torus of $\theta$ as depicted in Fig.~\ref{fig:UNthetatorus}, the axes $\theta=0$, $\theta_3=0$ are not at 90 degrees in case of $U(N)$.

\clearpage
\bibliographystyle{JHEP}
\bibliography{bibliography}
\end{document}